# Luminous Fast Blue Optical Transients as "Failed" Gravitational-wave Sources: Helium Core–Black Hole Mergers Following Delayed Dynamical Instability

Jakub Klencki[1] and Brian D. Metzger[2,3]
[1] Max Planck Institute for Astrophysics, Karl-Schwarzschild-Strasse 1, 85748 Garching, Germany; jklencki@mpa-garching.mpg.de
[2] Department of Physics and Columbia Astrophysics Laboratory, Columbia University, Pupin Hall, New York, NY 10027, USA; bdm2129@columbia.edu
[3] Center for Computational Astrophysics, Flatiron Institute, 162 5th Avenue, New York, NY 10010, USA
*Received 2025 October 22; revised 2026 April 30; accepted 2026 May 11; published 2026 June 19*

## Abstract

Binaries in which a massive donor star undergoes an extended ($\gtrsim$kyr) phase of stable mass transfer onto a black hole (BH) accretor offer a promising channel for creating LIGO gravitational-wave sources. However, in many systems the mass transfer terminates prematurely in a delayed dynamical instability (DDI), culminating in the BH plunging into the partially stripped donor and potentially disrupting and accreting its helium core (HeC) at super-Eddington rates. Combining binary evolution models with analytic estimates and population synthesis, we predict the population of luminous transients from DDI and attribute them to the "luminous" class of fast blue optical transients (LFBOTs). The initial BH plunge typically ejects $\sim 10\,M_\odot$ of H/He-enriched material at $\sim 10^2$–$10^3$ km s$^{-1}$, generating a compact circumstellar medium (CSM) of $\lesssim 1000\,R_\odot$ before the BH tidally disrupts the HeC. Rapid BH accretion generates a highly aspherical wind-driven explosion into the (already aspherical) environment, powering UV/optical emission via CSM interaction and X-ray reprocessing, peaking after several days at $\sim 10^{44}$–$10^{45}$ erg s$^{-1}$ before fading as the disk spreads and the accretion rate drops. If the BH is rapidly spinning, its relativistic jet may power an ultralong gamma-ray transient for on-axis observers. Luminous radio/submillimeter emission arises over months as the jet collides with the slow quasi-spherical outflow generated by the stable mass transfer preceding DDI, extending to $\sim 10^{17}$ cm, consistent with the remarkable CSM of LFBOTs. We estimate event rates of 5–300 Gpc$^{-3}$ yr$^{-1}$, with a preference for low metallicity ($\lesssim 0.4\,Z_\odot$), consistent with LFBOT demographics. Taken together, our results support LFBOTs as luminous signposts of "failed" gravitational-wave sources.

## 1. Introduction

A majority of massive stars reside in binary or higher-order systems that interact through one or more phases of mass transfer during their evolution (e.g., H. Sana et al. 2012; M. Moe & R. Di Stefano 2017). This has implications for myriad observable properties of stellar populations (e.g., O. R. Pols 1994; S. E. de Mink et al. 2013; E. Glebbeek et al. 2013; F. R. N. Schneider et al. 2015; Y. Götberg et al. 2018; C. Wang et al. 2020; M. Renzo & Y. Götberg 2021), as well as the final fates of these systems (e.g., F. R. N. Schneider et al. 2021; T. M. Tauris & E. P. J. van den Heuvel 2023; E. Laplace et al. 2025). Of particular recent interest are those binaries that retain a massive compact object after the first core-collapse event (e.g., E. P. J. van den Heuvel 1976; I. J. Iben & A. V. Tutukov 1984; P. Podsiadlowski et al. 1992; N. Sravan et al. 2019; E. Zapartas et al. 2019), which serve as progenitor systems for the populations of merging binary black holes (BHs) and neutron stars (NSs) reported by the LIGO–Virgo–Kagra collaboration (K. Belczynski et al. 2016; B. P. Abbott et al. 2017; A. Vigna-Gómez et al. 2018; F. S. Broekgaarden et al. 2019; The LIGO Scientific Collaboration et al. 2025).

A chief requirement for producing a compact object merger is a sufficiently tight orbital separation for the binary to coalesce through gravitational waves in a Hubble time. In the usually considered "common envelope" channel (e.g., M. Dominik et al. 2012; M. U. Kruckow et al. 2016; J. Klencki et al. 2021), this is achieved through a rapid contraction of a wide binary resulting from the compact object spiraling into and ejecting the donor's envelope (B. Paczynski 1976; N. Ivanova et al. 2013). The latter occurs following the onset of unstable mass transfer and is traditionally envisioned to occur soon after a highly evolved convective donor star overflows its Roche lobe onto the compact object. However, several challenges afflict the common envelope mechanism, particularly regarding its prevalence (K. Pavlovskii et al. 2017; P. Marchant et al. 2021; J. Klencki et al. 2026, hereafter K26) and uncertainties about the final outcome (N. Ivanova & J. L. A. Nandez 2016; T. Fragos et al. 2019; M. Y. M. Lau et al. 2022; S. Tuna & B. D. Metzger 2023; D. Wei et al. 2024; M. Y. M. Lau et al. 2025). These issues, which lead to orders-of-magnitude uncertainty in the predicted LIGO detection rates (e.g., M. Chruslinska et al. 2018; N. Giacobbo & M. Mapelli 2018; M. U. Kruckow et al. 2018; K. Breivik et al. 2020; A. Olejak et al. 2021; I. Mandel & F. S. Broekgaarden 2022), have motivated investigations of alternative or complementary merger channels.

One such alternative invokes tight initial binaries in which both stars undergo chemically homogeneous evolution due to rotational mixing (A. Maeder 1987; N. Langer 1992); by preventing the stars from expanding off the main sequence (MS) and undergoing mass transfer, this can lead to massive BH mergers at low metallicities (S. E. de Mink &

I. Mandel 2016; I. Mandel & S. E. de Mink 2016; P. Marchant et al. 2016; H. F. Song et al. 2016).

Another alternative, relevant also for wider initial systems, are cases in which the donor instead undergoes a long-lived phase of *stable mass transfer* (E. P. J. van den Heuvel 2017; S. S. Bavera et al. 2021; M. Gallegos-Garcia et al. 2021; P. Marchant et al. 2021; C. J. Neijssel et al. 2021; L. A. C. van Son et al. 2022; A. Olejak et al. 2024; K26). In this scenario, the binary tightens comparatively slowly as the donor is gradually stripped of mass and contracts in tandem with the binary separation, over timescales of ~10 kyr or even ~1 Myr (P. Marchant et al. 2017; J. Klencki et al. 2022). The orbital shrinkage is a natural outcome of angular momentum conservation and may be further enhanced via slow outflows from the outer Lagrangian point (W. Lu et al. 2023; K26) or through interaction with a circumbinary disk (S. Tuna & B. D. Metzger 2023; D. Wei et al. 2024).

A fraction of stable mass-transferring systems will detach at short enough separations to later evolve to produce double compact object gravitational-wave sources. However, this is not the only or necessarily even the most likely outcome. In particular, many binaries that start out transferring mass stably eventually become dynamically unstable (K. Pavlovskii & N. Ivanova 2015; K. Pavlovskii et al. 2017; H. Ge et al. 2020; N. Blagorodnova et al. 2021; K. D. Temmink et al. 2025). For donor stars with a radiative envelope (the most common case among massive stars), the instability is particularly delayed, often occurring only once the donor has been stripped down to its inner regions, where the entropy profile flattens (M. S. Hjellming & R. F. Webbink 1987; H. Ge et al. 2015; K26). By that point, the donor has transferred a significant fraction of its H-rich envelope (~20%–60%) and contracted to a typical radius of ~10–30 $R_\odot$, depending on its evolutionary state and internal structure ( K26).

The processes that follow dynamical instability are complex (e.g., C. J. Lombardi James et al. 2002; M. MacLeod et al. 2018; F. K. Röpke & O. De Marco 2023), but its final outcome involves the BH rapidly plunging into the donor envelope over just a few orbits. In the standard common envelope paradigms, if the liberated energy is sufficient to remove most of the remaining envelope, the outcome could be a tighter stable binary composed of the donor's evolved helium core (HeC) and the BH. However, if during this process the BH migrates all the way inward to reach the HeC (especially likely for radiative envelopes), a more cataclysmic outcome becomes possible. In particular, if the HeC is tidally disrupted by the BH, the resulting extremely super-Eddington accretion episode onto the BH likely triggers a "merger-driven explosion" (e.g., R. A. Chevalier 2012; N. Soker et al. 2019; S. L. Schrøder et al. 2020) with associated high-energy jetted emission (e.g., C. L. Fryer & S. E. Woosley 1998).

Building on an earlier suggestion by N. Soker et al. (2019), B. D. Metzger (2022, hereafter M22) argued that such binary merger events between HeCs ("Wolf–Rayet" stars) and BHs (or NSs) offer a promising model for "fast blue optical transients" (FBOTs; M. R. Drout et al. 2014; I. Arcavi et al. 2016; M. Pursiainen et al. 2018; A. Y. Q. Ho et al. 2023b; M. Nicholl et al. 2023). FBOTs are a class of stellar explosions characterized by very short UV/optical rise times of a few days and that can reach peak luminosities $\gtrsim 10^{44}$ erg s$^{-1}$, similar to or exceeding those of the most luminous supernovae (SNe) known (e.g., C. Inserra 2019). As a whole, FBOTs form a heterogeneous class with several distinct origins (e.g., A. Y. Q. Ho et al. 2023b; Z. Y. Wang et al. 2025). Of particular interest here are the most luminous FBOTs ("LFBOTs"), which form a rare subpopulation with unique multiwavelength properties implicating a central engine, as reviewed below.

The first and best-studied LFBOT is AT2018cow (S. J. Prentice et al. 2018; L. E. Rivera Sandoval et al. 2018; N. P. M. Kuin et al. 2019; R. Margutti et al. 2019; D. A. Perley et al. 2019; A. Y. Q. Ho et al. 2019; A. J. Nayana & P. Chandra 2021; Y.-X. Chen et al. 2023), but the sample has now expanded to include a handful of other events: AT2018lug (ZTF18abvkwla; A. Y. Q. Ho et al. 2020), AT2020xnd (D. A. Perley et al. 2021; A. Y. Q. Ho et al. 2022b; J. S. Bright et al. 2022), AT2020mrf (Y. Yao et al. 2022b), AT2022tsd (D. Matthews et al. 2023; A. Y. Q. Ho et al. 2023a), CSS161010 (D. L. Coppejans et al. 2020; C. P. Gutiérrez et al. 2024), AT2023fhn (A. A. Chrimes et al. 2024b; A. A. Chrimes et al. 2024a), AT2024wpp (M. Pursiainen et al. 2025; N. LeBaron et al. 2026; A. J. Nayana et al. 2025; D. A. Perley et al. 2026; J. Liu et al. 2026), AT2024qfm (M. Fulton et al. 2024), and (possibly) AT2024puz (J. J. Somalwar et al. 2025). LFBOTs are rare, with a volumetric rate ~0.3–400 Gpc$^{-3}$ yr$^{-1}$ in the local Universe, corresponding to $\lesssim$0.6% of the core-collapse rate (e.g., D. L. Coppejans et al. 2020; A. Y. Q. Ho et al. 2023b).

Near their discovery close to peak light, the optical spectra of LFBOTs are largely featureless (e.g., S. J. Prentice et al. 2018; D. A. Perley et al. 2019), indicating a rapidly expanding photosphere velocity $v \gtrsim 0.1c$–$0.3c$. Spectra taken a week or two later sometimes reveal the emergence of narrower H and He emission features (e.g., S. J. Prentice et al. 2018), down to much lower velocities $v \sim$ 3000–6000 km s$^{-1}$, even while the color evolution shows minimal evidence for ejecta cooling (e.g., D. A. Perley et al. 2019; R. Margutti et al. 2019; D. Xiang et al. 2021; N. LeBaron et al. 2026; however, see M. Nicholl et al. 2023; J. Vinkó et al. 2015). Narrow He emission lines indicating a ~300 km s$^{-1}$ velocity component seen in AT2018cow support interaction between the explosion ejecta and He-rich circumstellar medium (CSM) on radial scales $\lesssim 10^{14}$ cm surrounding the progenitor system (e.g., O. D. Fox & N. Smith 2019; L. Dessart et al. 2021; C. Pellegrino et al. 2022). LFBOTs exhibit highly variable nonthermal X-ray emission (e.g., L. E. Rivera Sandoval et al. 2018; N. P. M. Kuin et al. 2019; R. Margutti et al. 2019; A. J. Nayana et al. 2025), which fades in tandem with the optical emission, as well as with a complex light curve and spectral evolution.[4] The jetted compact object remains active to late times, as evidenced by extremely luminous optical/infrared flares seen several months into the evolution of AT2022tsd (A. Y. Q. Ho et al. 2023a; but not yet from AT2024wpp; E. O. Ofek et al. 2025).

LFBOTs are also accompanied by luminous radio/submillimeter synchrotron emission lasting months (A. Y. Q. Ho et al. 2019; R. Margutti et al. 2019; D. L. Coppejans et al. 2020; J. S. Bright et al. 2022; A. J. Nayana & P. Chandra 2021; Z. Liu et al. 2023; A. J. Nayana et al. 2025; C. Sevilla et al. 2026), created by the shock interaction of transrelativistic outflow from the compact object with a dense CSM shell with a sharp outer edge starting around $\sim 3 \times 10^{16}$ cm (B. Margalit et al. 2022). Although the sample size is modest, the presence and inferred properties of this extended CSM appear remarkably uniform across the LFBOT sample (C. Sevilla et al. 2026), suggesting that a robust explanation is required. Finally, in the two LFBOTs where such observations are possible, a separate component of slowly decaying infrared emission is observed (D. A. Perley et al. 2019; M. Pursiainen et al. 2025; N. LeBaron et al. 2026); this could represent a "dust echo" generated by reprocessing of the transient's UV light by dust existing in the extended CSM prior to the explosion that arrives to Earth following a light-travel delay (B. D. Metzger & D. A. Perley 2023; J.-Y. Li et al. 2025), though other explanations have been suggested (e.g., C. Chen & R.-F. Shen 2022).

[4] Distinct components of hard $\gtrsim$10 keV X-ray emission were seen over a brief epoch in AT2018cow (R. Margutti et al. 2019), with a distinct spectrum characterized by broadened Fe emission lines and a Compton-hump-like feature around 30 keV, similar to accretion disk reflection spectra (see also A. J. Nayana et al. 2025).

Many progenitor models have been proposed for LFBOTs, but most can be divided into two broad classes: (1) stellar core-collapse events giving birth to a compact object (S. J. Prentice et al. 2018; D. A. Perley et al. 2019; S.-C. Leung et al. 2020; R. Ouyed 2025; D. Lazzati et al. 2024; J. N. D. van Dalen et al. 2025; A. A. Chrimes et al. 2026), and (2) the tidal disruption of stars by stellar- or intermediate-mass BHs (e.g., H. B. Perets et al. 2016; K. Kremer et al. 2021; D. R. Pasham et al. 2022; W. Zhang et al. 2022; M22; C. P. Gutiérrez et al. 2024; A. A. Chrimes et al. 2024a; I. Linial & E. Quataert 2024; D. Tsuna & W. Lu 2025). Tidal disruption scenarios are favored by UV (N.-C. Sun et al. 2022, 2023; Y. Chen et al. 2023; A. Inkenhaag et al. 2023) and X-ray (G. Migliori et al. 2024) observations of AT2018cow taken several years after the explosion, which reveal a large accretion disk with a higher angular momentum than expected from the collapse of even a rapidly spinning star (G. Migliori et al. 2024; M. Winter-Granic & E. Quataert 2025; however, see A. A. Chrimes et al. 2026).

Unlike in most tidal disruption scenarios, a dense extended CSM around the explosion site is naturally expected in the abovementioned scenario of HeC tidal disruption by a BH or NS orbiting companion (N. Soker et al. 2019; M22; A. Grichener 2023, 2025). M22 envisioned the CSM to arise from a combination of a bound relic disk left over from an earlier common envelope event (S. Tuna & B. D. Metzger 2023; D. Gagnier & O. Pejcha 2023) and He-rich mass loss from the HeC/BH binary during the earliest stages of the merger, mainly through the L2 Lagrange point. However, as emphasized above, dynamical instability in massive-star binaries is often delayed and presaged by a long-lived stable mass transfer phase (e.g., N. Blagorodnova et al. 2021). Nonconservative mass transfer that takes place during this long-lived phase (most likely also through L2; O. Pejcha et al. 2016a; W. Lu et al. 2023; P. Scherbak et al. 2025) therefore provides a distinct source of radially extended CSM, and one that can arguably be predicted through binary stellar evolution modeling with greater certainty.

Motivated to explore this scenario further, here we combine detailed binary evolution models (K26) with analytic estimates (M22) to predict population-level properties of the explosive transients produced from delayed dynamical instability (DDI) mergers. This paper is organized as follows: In Section 2 we summarize the grid of binary evolution models employed and key properties of the systems at the onset of dynamical instability. In Section 3 we present analytic estimates for key properties of the system at different phases in the DDI transients using our stellar models as input. The model stages and associated schematic light curves are illustrated in Figure 1. In Section 4 we discuss implications of our results, particularly for the progenitors of LFBOTs and the connections of these events to other measurable binary outcomes such as LIGO gravitational-wave sources. In Section 5 we summarize our conclusions.

## 2. Binary Evolution Models

### 2.1. Model Details

We use detailed binary evolution models from K26 calculated with the MESA code (B. Paxton et al. 2011, 2013, 2015, 2019; A. S. Jermyn et al. 2023). These models follow the evolution of systems comprised of a BH and an O- or B-type star companion (BH+OB star systems) BH+OB star systems beginning from the zero-age MS (ZAMS) of the star until central carbon depletion or mass transfer instability.

*Grid setup.* We extend the BH+OB grids of K26 to five metallicities $Z/Z_{\odot} = \{0.04, 0.1, 0.2, 0.4, 1.0\}$ and two BH masses $M_{\bullet} = \{5, 10\}\, M_{\odot}$, yielding 10 fiducial grids. In addition, we compute 10 variation grids with reduced convective core overshooting, $\sigma_{\rm ov} = 0.18$ (fiducial: $\sigma_{\rm ov} = 0.35$), for a total of 20 grids. For each ($Z$, $M_{\bullet}$) we cover orbital periods $\log_{10}(P_{\rm ini}/{\rm day}) \in [0.2, 3.1]$, i.e., the full range relevant for binary interaction, and a sufficiently wide range in initial mass ratios to capture the complete parameter space for BH+HeC mergers that lead to LFBOTs. Wide noninteracting binaries are not considered. The left panel of Figure 2 shows an overview for one representative grid; outcome maps for all grids are provided in Figures 15 and 16.

*Stellar physics assumptions.* The fiducial grids adopt the same microphysics, wind mass loss, convection, and internal mixing prescriptions as K26 (see their Section 2 for details) and in particular assume $\sigma_{\rm ov} = 0.35$ for core overshooting (I. Brott et al. 2011). The secondary set of grids adopts a smaller overshooting $\sigma_{\rm ov} = 0.18$. This ad hoc choice is to emulate the possible internal structure of mass-gainer stars that have accreted mass without fully rejuvenating the core, leading to a core-to-envelope mass ratio that is smaller compared to normal stars (H. Braun & N. Langer 1995; M. Renzo & Y. Götberg 2021; T. Wagg et al. 2024; A. Miszuda et al. 2025; X.-T. Xu et al. 2025). The choice of $\sigma_{\rm ov}$ has little effect on the transient model and predicted LFBOT properties. Its main impact is on the radii of MS stars and therefore on the LFBOT rate and its metallicity dependence (Section 3.7).

*Mass transfer ingredients.* Following Roche lobe overflow (RLOF), the mass transfer rate through L1 $\dot{M}_{\rm L1}$ is computed for every time step using the scheme of P. Marchant et al. (2021). The accretion rate is Eddington limited, yielding a typical accretion efficiency of $\lesssim 1\%$. In the fiducial grids ($\sigma_{\rm ov} = 0.35$), we assume that a fraction $f_{\rm L2} = 0.25$ of the nonaccreted mass leaves the system with the specific angular momentum of the outer Lagrangian point. The $\sigma_{\rm ov} = 0.18$ grids are evolved without this L2-specific angular momentum loss term. Any particular choice of $f_{\rm L2}$ (including its possible time dependence or $\dot{M}$ dependence) plays only a secondary role in shaping the detailed time history of $\dot{M}(t)$. The crucial bulk quantities for our model, such as the overall duration of the pre-instability mass transfer and the donor's mass and radius at the onset of instability, are primarily determined by the donor star (K26).

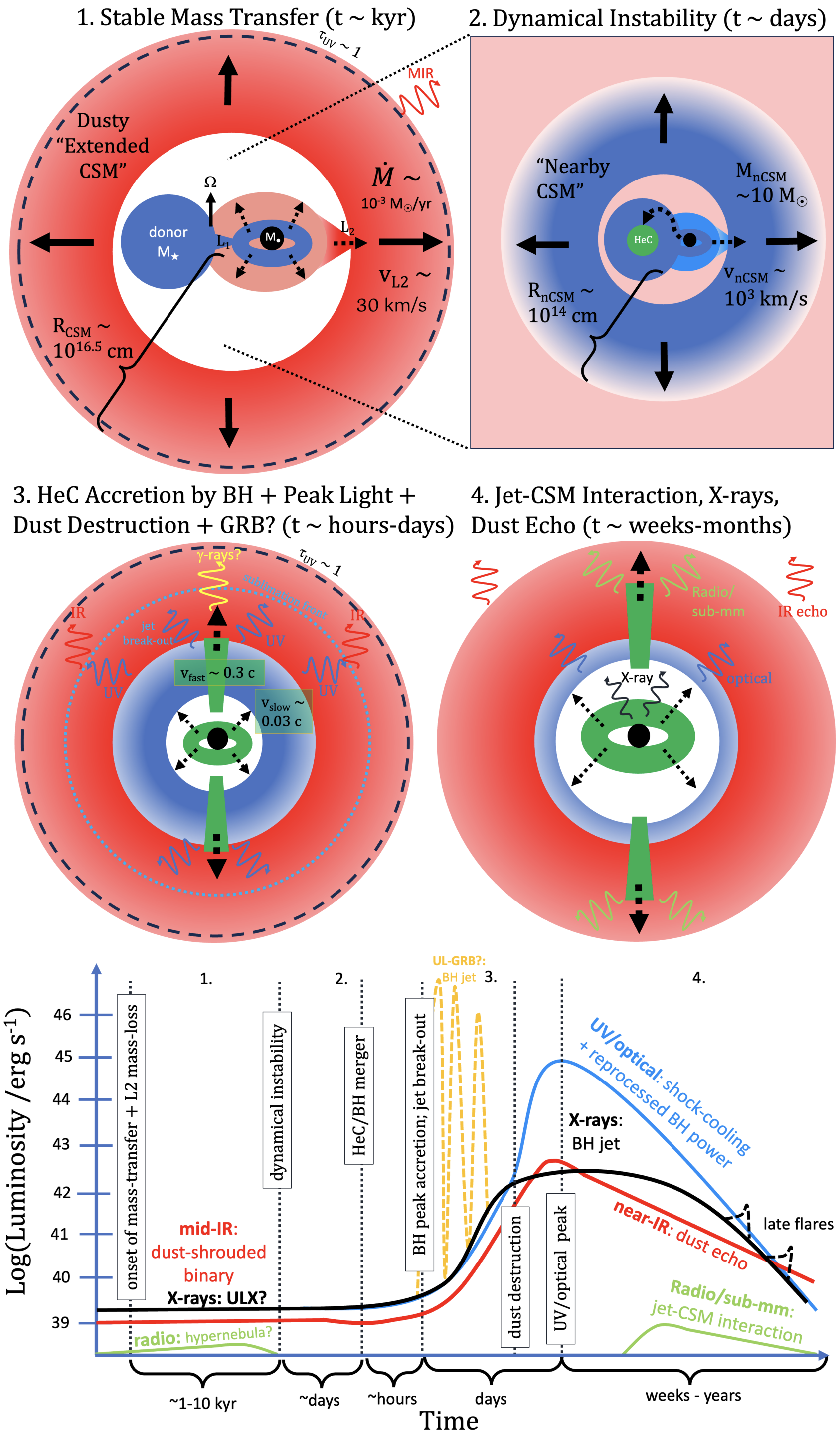

**Figure 1.** Stages of DDI transients and associated multiwavelength light curve: (1) A massive star transfers mass stably onto a BH companion, for thousands of years. A slow outflow from the L2 point tightens the binary, creating a dense, dusty, and opaque CSM extending to large radii $R_{\rm CSM} \sim 10^{16}$–$10^{17}$ cm (Figure 4). (2) once enough of the donor envelope has been removed, the mass transfer becomes dynamically unstable; the BH plunges into the star over just a few orbital periods (typically days), ejecting the envelope and creating "nearby" He-rich CSM extending to radii $R_{\rm nCSM} \sim 10^{13}$–$10^{14}$ cm. (3) The BH tidally disrupts the star's helium ("Wolf–Rayet") core, creating a massive disk around the BH that feeds it at highly super-Eddington rates. Accretion peaks over hours to a day, thereafter decaying as a power law $\dot{M}_\bullet \propto t^{-\beta}$ with $\beta \sim 2$–$3$. The disk produces both fast bipolar outflows ("jet") and wider-angle slow winds, which initially collide with the nearby CSM. If the BH is rapidly spinning, it may power a GRB-like ultrarelativistic jet visible for on-axis viewers (Section 4.4). The jet breaks out of the nearby CSM within a few hours, but the bulk of the UV emission occurs over the longer peak diffusion time of $t_d \sim$ days (Equation (23)). A portion of the early UV emission is absorbed and reprocessed by dust in the extended CSM before the dust is sublimated prior to peak light. (4) The growing BH accretion funnel allows X-rays from the jet base to reach the distant observer. Time variability in the jet or its orientation may result in late flares and associated nonthermal afterglow (A. Y. Q. Ho et al. 2023a). Shock interaction of the fast jet with the extended CSM produces synchrotron radio/submillimeter emission. The reprocessed dust emission from stage 3 reaches the observer as an "echo" over the several weeks required for light to travel across the extended CSM.

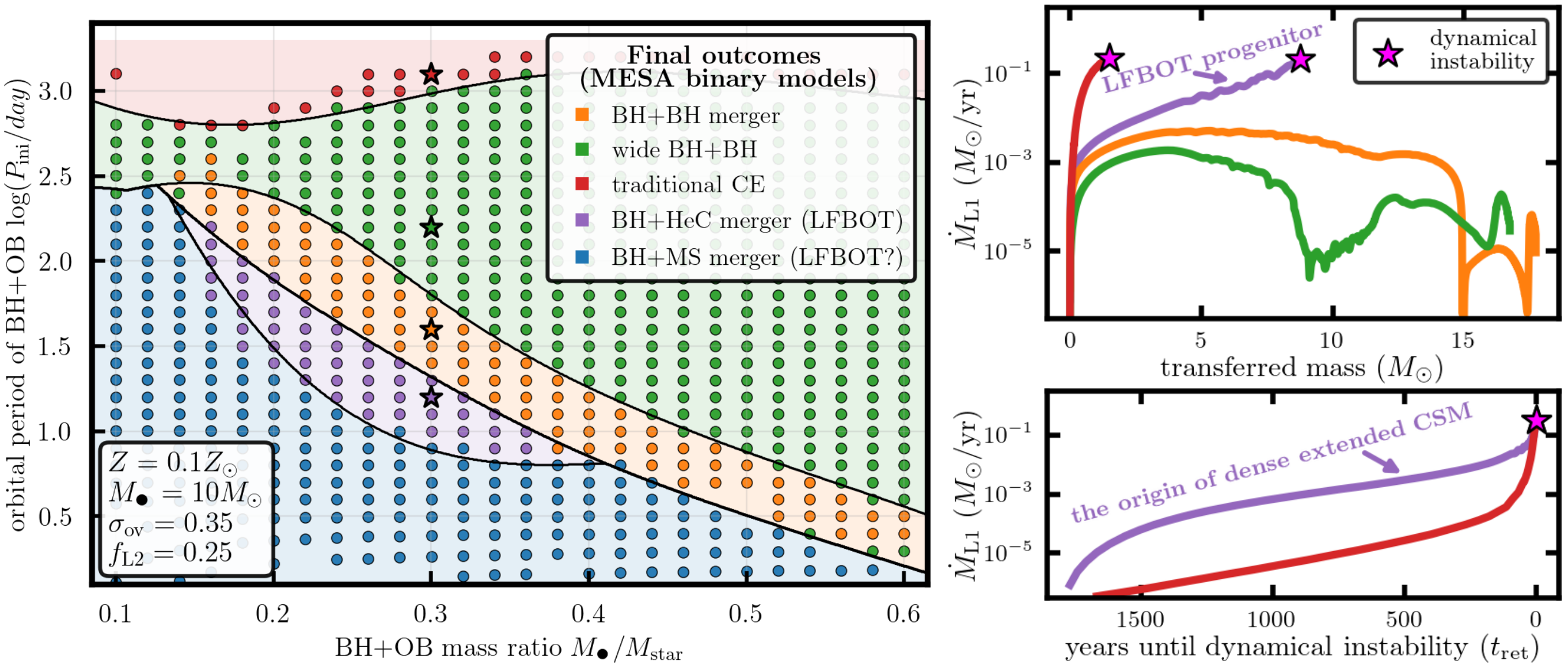

**Figure 2.** An overview of MESA binary evolution models and their final outcomes. Left: 1 of the 19 binary model grids of BH+OB star systems employed in this study. The grid covers BH–OB systems with varying initial mass ratios and orbital periods for a fixed initial BH mass $M_\bullet = 10\,M_\odot$ and metallicity $Z = 0.1\,Z_\odot$, assuming fiducial core overshooting ($\sigma_{\rm ov} = 0.35$) and L2 outflows of $f_{\rm L2} = 0.25$. See Figures 15 and 16 for the other grids. Each binary model (circle) is color-coded according to its final outcome (see text). Solid lines mark approximate boundaries between different final outcomes. The LFBOT progenitors (marked in purple as BH+HeC mergers) are systems in which the mass transfer from a radiative post-MS donor encounters DDI. Star symbols mark models shown in the right panels. Top right: mass transfer rate as a function of the total mass transferred since the onset of RLOF, shown for a few example binaries of different final outcomes. Bottom right: mass transfer rate in the final 1,750 years before dynamical instability for an LFBOT progenitor (purple) and a system with a convective donor, leading to traditional CE (red).

To predict the extended CSM profiles of LFBOT progenitors (Section 3.1), we treat the mass loss through the L2 point more carefully in post-processing following the model of W. Lu et al. (2023). We define the onset of mass transfer instability as the point when $\dot{M}_{\rm L1} > 10^{-0.5}\,M_\odot\,{\rm yr}^{-1}$. This threshold is motivated by the thermal relaxation timescale of the steep-entropy outer layers in radiative envelopes (for an in-depth discussion, see K. D. Temmink et al. 2023; K26). Although somewhat arbitrary, adopting, e.g., $0.1\,M_\odot\,{\rm yr}^{-1}$ instead would not significantly change our predicted CSM profiles or event rates. Unstable systems with convective donors enter a traditional common envelope phase, which we do not model. Unstable systems with radiative donors experience DDI and are assumed to always lead to mergers (M. U. Kruckow et al. 2016; J. Klencki et al. 2021; P. Marchant et al. 2021). We distinguish between mergers with MS and post-MS donors with a HeC, which likely produce distinct transients (Sections 4.1 and 4.5).

*Final outcomes.* Each BH+OB binary model is assigned one of five final outcomes (color-coded in Figure 2):

1. *BH+BH merger.* Stable mass transfer leads to close binaries that become BH–BH mergers owing to gravitational-wave-driven inspiral.
2. *Wide BH+BH.* Binaries that terminate mass transfer while too wide to become BH–BH mergers.
3. *Traditional Common Envelope (CE).* Systems with convective red supergiant donors that become unstable at wide orbits; although some of these systems may also end up producing HeC/BH mergers and LFBOTs (M22), they are not the main focus of this paper.
4. *BH+HeC merger (LFBOT).* DDI mergers from evolved donors with compact HeCs, which we associate with LFBOTs.
5. *BH+MS mergers (LFBOT?).* DDI mergers from MS donors, which may be associated with LFBOTs or related transients.

The right panels of Figure 2 illustrate typical mass transfer rate histories for examples of these outcomes.

### 2.2. Overview of Systems Leading to DDI and Mergers

Across our binary grids, the donor stars span $M_{\star,0} \sim$ 10–100 $M_\odot$, with the BH masses $M_\bullet = 5$, $10\,M_\odot$. Here we summarize the sequences that end in unstable mass transfer. We focus on systems with radiative post-MS donors, which experience DDI and lead to BH+HeC mergers (blue in Figure 3). For comparison we also show unstable events with convective donors (orange), leading to traditional CE. The key difference between these two classes is in how much mass is lost before the instability ($\Delta M$) and in the physical size of the donor at instability ($R_{\star,f}$). Convective donors interact in a wide orbit ($a_0 \sim 10^3\,R_\odot$), compatible with red supergiant radii. The instability develops quickly, and the total pre-instability mass loss is modest, $\Delta M \sim$ 0.5–1 $M_\odot$ (top panel of Figure 3). By comparison, radiative donors begin the interaction in a wider range of orbital separations ($a \sim$ 20–$10^3\,R_\odot$ for post-MS donors). The onset of instability is delayed: systems experience prolonged, nearly steady mass transfer at $\dot{M} \sim 10^{-3}\,M_\odot\,{\rm yr}^{-1}$ for $\sim 10^3$ yr before DDI (see purple example in the right panels of Figure 2), removing far more mass from the donor: $\Delta M \sim$ 2–20 $M_\odot$. This mass is predominantly lost from the system, feeding L2 outflows and building a dense extended CSM (Section 3.1). By the time radiative donors reach DDI, the remaining envelope is (i) helium enriched, with an average $X_{\rm He} \sim 0.6$; (ii) reduced in mass, $M_{\rm env} \sim$ 4–40 $M_\odot$; and (iii) compact in size, with $R_{\rm env} \sim$ 10–20 $R_\odot$ (middle panel of Figure 3). In contrast, convective donors encounter instability while still very large, $R \sim 500$–$1000\,R_\odot$, and with little prior stripping. For models that undergo DDI and proceed to BH+HeC mergers (our LFBOT progenitors), the population trends in Figure 3 (bottom panel) are well summarized by $M_{\rm env} \sim M_{\rm He} \sim 3\,\Delta M$. These relationships provide convenient inputs for the analytic estimates in Section 3. We note that the precise ratios between $M_{\rm env}$, $M_{\rm He}$, and $\Delta M$ are subject to uncertainties in the amount of internal mixing in massive stars. This is especially relevant for stellar companions to BHs, as they have likely accreted mass in the past, affecting their internal structure. Mass gainers that have not fully rejuvenated, with an elevated envelope-to-core mass ratio (S. Justham et al. 2014; F. R. N. Schneider et al. 2024) and a large intermediate convective zone (M. Renzo & Y. Götberg 2021; A. Miszuda et al. 2025), may lead to somewhat lower values of $M_{\rm He}/M_{\rm env}$ and $\Delta M/M_{\rm env}$ than those predicted in Figure 3 (see K26). An opposite trend could be expected if some mass gainers develop unusually massive HeCs owing to efficient rejuvenation and rotationally induced mixing.

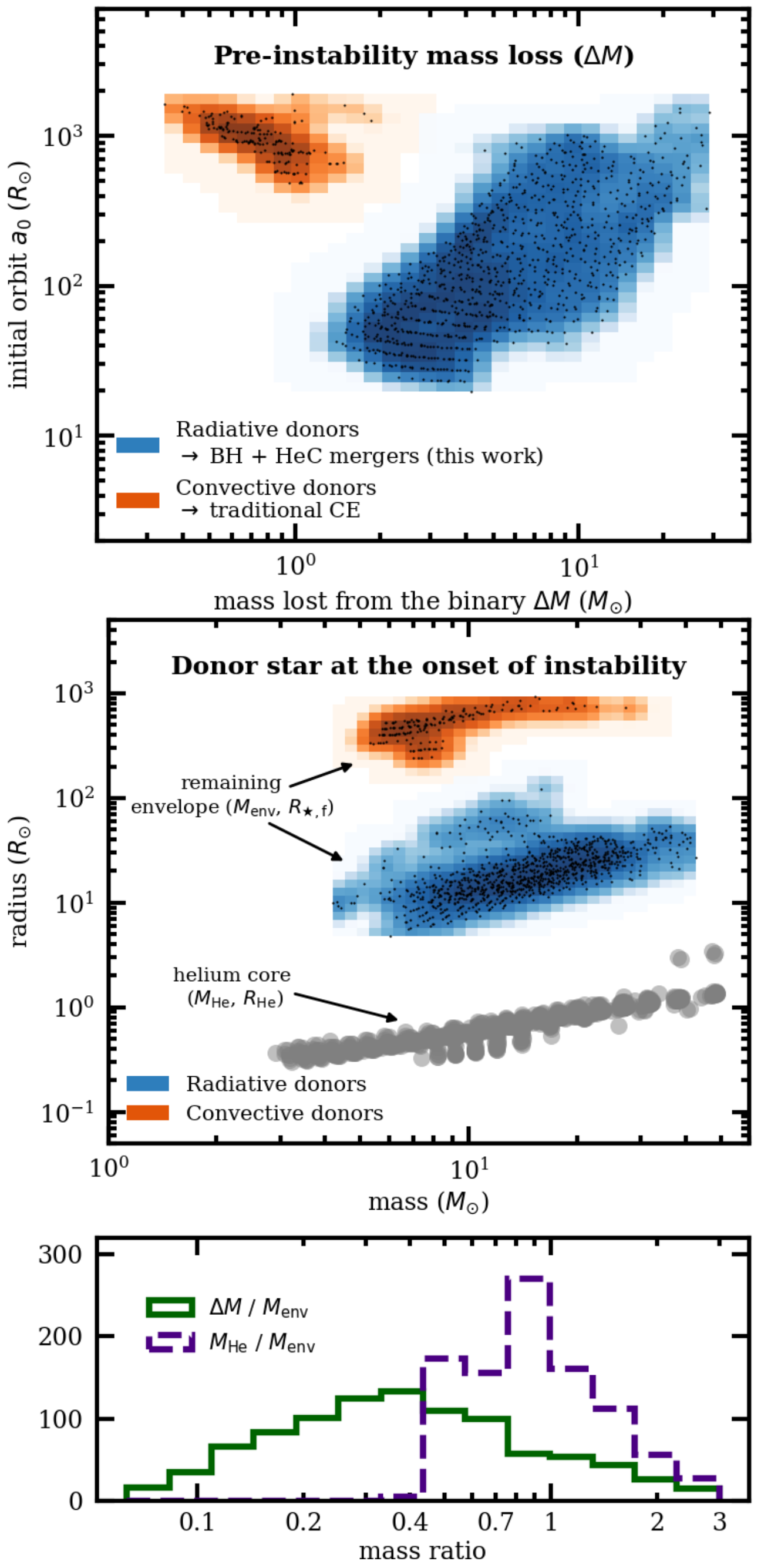

**Figure 3.** Top: mass loss preceding dynamical instability ($\Delta M$) as a function of initial semimajor axis of the BH+OB orbit ($a_0$), shown separately for binary models that terminate in unstable mass transfer by a convective donor (orange, leading to traditional CE) or with a radiative post-MS donor (blue, leading to LFBOTs, as explored in this work). Middle: properties of donor stars at the onset of dynamical instability. Show in color are the mass and radius of the remaining envelope, following the same convection as the top panel. Shown in gray are the mass and radius of the core. Bottom: fractional mass loss $\Delta M/M_{\rm env}$ and core mass $M_{\rm He}/M_{\rm env}$ for all the binary models that lead to LFBOTs (i.e., radiative donors from the other panels).

## 3. Transient Model

We consider a model for the creation of luminous transients from DDI, which follows a modified version of the picture laid out in M22. As we proceed, we provide analytic estimates in parallel with refined distributions derived from our simulated binary population.

### 3.1. Stable Mass Transfer Generates Extended CSM

Because the mass transfer rates of the binaries of interest are highly super-Eddington, the BH cannot accept most of the mass (e.g., P. Podsiadlowski et al. 2002), which is instead lost from the system, mostly through the outer L2 point (e.g., W. Lu et al. 2023). Hydrodynamic simulations that follow the fate of mass loss through L2 find that the ejected material either remains bound to the binary (likely forming a circumbinary disk; e.g., R. E. Taam & H. C. Spruit 2001) or becomes part of a slow unbound outflow (e.g., O. Pejcha et al. 2016a, 2016b, 2017; M. MacLeod & A. Loeb 2020; P. Scherbak et al. 2025). However, even if a circumbinary disk forms, its high feeding rate implies high densities and long radiative cooling timescales, rendering the disk susceptible to outflows (e.g., O. Pejcha et al. 2016a).

Motivated thus, we assume that a large fraction of the transferred mass goes into an unbound outflow, which expands away from the binary at a small fraction $\chi \ll 1$ of the characteristic orbital velocity,

$$v_{\rm L2} = \chi\left(\frac{GM_{\rm tot}}{a}\right)^{1/2} \approx 30\ {\rm km\ s^{-1}} \times \left(\frac{\chi}{0.1}\right)\left(\frac{M_{\rm tot}}{50\,M_\odot}\right)^{1/2}\left(\frac{a}{100\,R_\odot}\right)^{-1/2}, \quad (1)$$

where $a$ is the semimajor axis, $M_{\rm tot} = M_\bullet + M_\star$ is the total mass of the binary, and we expect $\chi \approx 0.1$–$0.2$ across a range of binary mass ratios (see Figure 3 of O. Pejcha et al. 2016b, though note their different normalization convention, and Figure 8 of P. Scherbak et al. 2025). Over the duration of the stable mass transfer phase $t_{\rm ST} \gtrsim 10^3$ yr, the L2 mass loss will expand to radial scales

$$R_{\rm CSM} \sim v_{\rm L2}(a_0)t_{\rm ST} \sim 10^{17}\ {\rm cm}, \quad (2)$$

where $a_0$ is the orbital separation at the onset of RLOF. The composition of this extended CSM could be slightly helium enhanced if the donor star had accreted a substantial amount of mass in the past (e.g., He/H number ratio enhanced by ∼10%–50% in mass-gainer models by C. Martínez-Sebastián et al. 2025). Although the L2 outflow is concentrated in the equatorial binary plane, it can become quasi-spherical on larger scales from pressure forces if radiative cooling is inefficient (O. Pejcha et al. 2016b). This is expected to occur if the expansion timescale, $t_{\rm exp} \sim r/v_{\rm L2}$, near the base of the outflow on a radial scale $r \sim a$ greatly exceeds the photon diffusion time $t_{\rm diff} \sim \tau(r/3c)$, where $\tau \approx \kappa\dot{M}/4\pi r c v_{\rm L2}$ is the optical depth through the outflow of opacity $\kappa$. We define $\dot{M}_{\rm iso}$ as a lower limit on the mass-loss rate above which the binary outflow becomes quasi-spherical on large scales. The rough condition of $\tau \gtrsim (c/3v_{\rm L2})$ can be rewritten to obtain

$$\dot{M}_{\rm iso} \equiv \frac{4\pi ac}{3\kappa} \approx 4 \times 10^{-3}\, M_\odot\ {\rm yr}^{-1}\left(\frac{a}{10\, R_\odot}\right), \tag{3}$$

where we have taken $\kappa \approx 0.3\ {\rm cm}^2\ {\rm g}^{-1}$ as an estimate of the electron scattering opacity. Approximating the mass-loss rate using its average value $\dot{M} \sim \Delta M/t_{\rm ST}$ during the stable transfer phase, and for typical values $\Delta M \sim 10\, M_\odot$, $t_{\rm ST} \sim 10^3$ yr, $a \sim 10{-}100\, R_\odot$, we find $\dot{M} \sim \dot{M}_{\rm iso}$, close to the limit for a quasi-spherical wind (see also Figure 2).

In general, the CSM density will not follow the $\rho \propto r^{-2}$ profile expected of a steady wind, because neither the mass-loss rate nor the wind speed $v_{\rm L2} \propto a^{-1/2}$ (Equation (1)) remain constant as the binary evolves. For mass ejected at a retarded time $t_{\rm ret}$ at the rate $\dot{M}(t_{\rm ret})$ with a radial velocity $v_{\rm L2}(t_{\rm ret})$, we estimate its distance from the binary at the time of dynamical instability as $r = t_{\rm ret} v_{\rm L2}(t_{\rm ret})$, and we calculate its contribution to the CSM density at that radius as

$$dn_{\rm CSM} = \frac{1}{4\pi r^2 m_p}\frac{\dot{M}(t_{\rm ret})}{v_{\rm L2}(t_{\rm ret})}dt_{\rm ret}. \tag{4}$$

Here $n_{\rm CSM}$ is the number density of free electrons, which is the relevant quantity for radio emission and one that can be inferred empirically. The $1/m_p$ factor (with $m_p$ being the proton mass) makes the usual assumption that the ejecta is dominated by ionized hydrogen. Accounting for helium, with its mass fraction possibly enhanced up to ~0.5 in the envelopes of mass-gainer stars (C. Martínez-Sebastián et al. 2025), would reduce $n_{\rm CSM}$ by no more than a factor of 2.

For each of our binary models, we obtain the radial CSM density profile at DDI by integrating over the entire past history of mass loss:

$$n_{\rm CSM}(r) = \frac{1}{4\pi r^2 m_p} \times \int_0^{t_{\rm age}} \frac{\dot{M}(t_{\rm ret})}{v_{\rm L2}(t_{\rm ret})}\delta[t_{\rm ret} - r/v_{\rm L2}(t_{\rm ret})]\,dt_{\rm ret}, \tag{5}$$

where $t_{\rm age}$ is the age of the binary and the delta Dirac term $\delta[t_{\rm ret} - r/v_{\rm L2}(t_{\rm ret})]$ selects only those mass ejection episodes that contribute the CSM density at distance $r$.[5] During the mass transfer phase, we calculate $\dot{M}$ from the fraction of the binary mass transfer lost through the L2 point, following the model of W. Lu et al. (2023, their Figure 3, top panel). During detachment, $\dot{M}$ is simply the stellar wind, for which we assume $v_{\rm L2} = 1000\ {\rm km\ s}^{-1}$.

Figure 4 shows $R_{\rm CSM}$, $n(R_{\rm CSM})$ and profiles of $n_{\rm CSM}(r)$, across our grid of models. For comparison we show the density profiles surrounding LFBOTs based on radio/submillimeter modeling, for different assumptions about the microphysical parameters of the shock (e.g., J. S. Bright et al. 2022; A. A. Chrimes et al. 2024a).[6]

A sharp cutoff in the predicted CSM density is seen to occur on a radial scale $\sim 10^{17}$ cm matching the estimate in Equation (2), where the relatively low density stellar wind prior to the onset of mass transfer ($r > R_{\rm CSM}$) is replaced by the L2 outflow following RLOF ($r < R_{\rm CSM}$), the mass-loss rate from which grows as the binary tightens (W. Lu et al. 2023), leading to the sharp rise in $n(r)$ toward smaller radii. Small-scale bumps in the CSM profiles of some of the binary models in Figure 4 (gray lines) are related to fluctuations in the mass-loss rate, as well as rare cases when a faster ejecta catches up with a prior, slower outflow (details in Section A.1).

The CSM structure predicted by our models also broadly matches, if not modestly exceeds, the normalization implied by the radio data. However, there are several uncertainties to note, related not only to microphysical parameters of the shock but also to the relative importance of slow L2 outflows from the binary (related to the mass transfer efficiency we have assumed), versus faster winds from the BH accretion disk, particularly along the predominantly polar directions likely probed by the radio/millimeter data.

### 3.2. Nearby CSM from Dynamical Envelope Removal

In contrast to the extended CSM produced gradually over the long stable mass transfer phase, a more compact "nearby" CSM will be produced by mass loss that accompanies the dynamical plunge of the BH into the stellar envelope. Here we estimate the asymptotic velocity of this nearby ejecta ($v_{\rm nCSM}$) and its distance from the central source at the time when the BH reaches the HeC and powers a transient ($R_{\rm nCSM}$). Let $a_{\rm DDI}$ be the orbital separation at the onset of DDI, when the BH begins to spiral in through its companion, and let $a_{\rm HeC}$ be the separation when the BH reaches the HeC and tidally disrupts it. The quasi-circular inspiral of the BH will release an amount of gravitational energy approximately given by

$$\Delta E \approx \frac{GM_\bullet M_{\rm He}}{2a_{\rm HeC}}, \tag{6}$$

where we have assumed $a_{\rm DDI} \gg a_{\rm HeC}$ and

$$a_{\rm HeC} \approx R_{\rm He}\frac{0.6q^{2/3} + \ln(1 + q^{1/3})}{0.49q^{2/3}} \underset{q\approx 3}{\approx} 2.1R_{\rm He} \tag{7}$$

is the semimajor axis of RLOF for mass ratio $q \equiv M_{\rm He}/M_\bullet$ (P. P. Eggleton 1983). Assuming that an order-unity fraction of $\Delta E$ ultimately goes into the kinetic energy of the ejected envelope, the unbound envelope will achieve an asymptotic

[5] In practice, we calculate the integral in Equation (6) as a discrete sum over evolutionary time steps of a binary model, with $\Delta t_{\rm ret}$ of the order of 10–50 yr during a mass transfer phase. We evaluate $n_{\rm CSM}(r)$ at distances $r$ ranging from $10^{14}$ to $10^{19}$cm in steps of 0.04 dex.

[6] Rather than making the usual equipartition assumption, first-principles simulations of magnetic field amplification (e.g., D. Caprioli & A. Spitkovsky 2014; P. C. Duffell & A. I. MacFadyen 2014) and nonthermal electron acceleration in nonrelativistic (e.g., J. Park et al. 2015) and ultrarelativistic (e.g., A. Spitkovsky 2006) shocks motivate our adoption of modest values for $\epsilon_e \lesssim 0.1$, $\epsilon_B \lesssim 10^{-2}$.

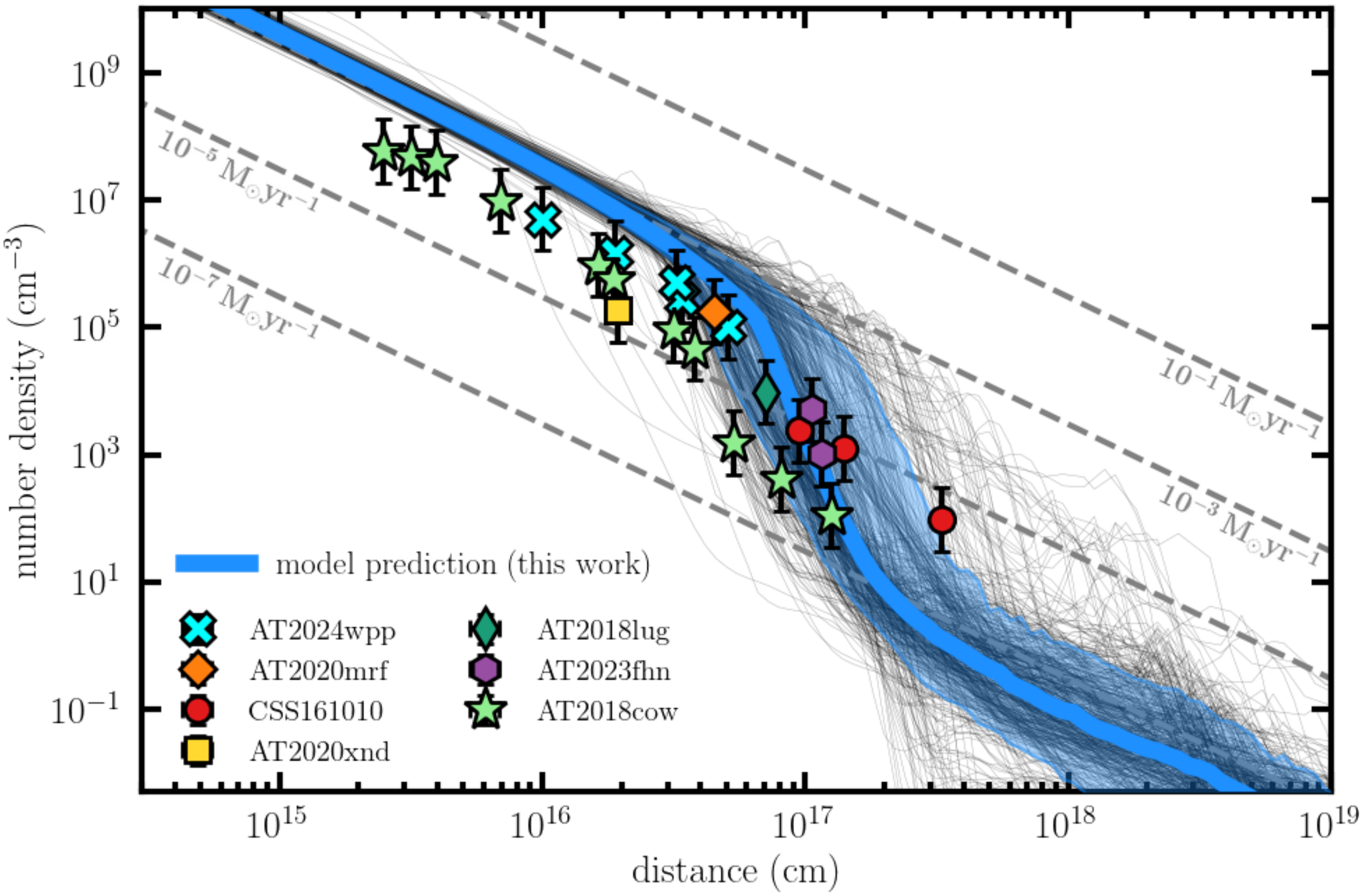

**Figure 4.** Radial CSM density profiles $n_{\rm CSM}(r)$ (Equation (6)) at the time of dynamical instability, due to the combination of stellar winds and L2 mass loss from the preceding stable mass transfer phase, calculated from our binary evolution models. For the L2 mass loss we assume an outflow velocity $v_w = v_{\rm L2}$ (Equation (1)) for $\chi = 0.1$ and mass-loss rate $\dot{M}$ following W. Lu et al. (2023) capped at 50% $\dot{M}_{\rm L1}$. Gray lines show the density profiles of individual binary evolution models, the thick blue line shows the mean density profile, and the blue shading indicates the 10th–90th percentiles. For comparison, symbols show measured CSM density profiles of individual LFBOTs obtained by modeling the radio/submillimeter synchrotron emission (J. S. Bright et al. 2022; A. A. Chrimes et al. 2024a; D. A. Perley et al. 2026). The shock microphysics is uncertain, in particular the fraction of internal energy placed into relativistic electrons ($\epsilon_e$) and amplified magnetic fields ($\epsilon_B$). The error bars on $n_{\rm CSM}$ therefore span the ranges $\epsilon_e = 0.01-0.1$ and $\epsilon_B = 0.001-0.01$, adopting the scaling $n_{\rm CSM} \propto \epsilon_e^{-6/19}\epsilon_B^{-13/19}$ (R. A. Chevalier 1998). For comparison, diagonal dashed lines show the density profiles for steady wind mass loss at $\dot{M} = 0.1$, $10^{-3}$, $10^{-5}$, $10^{-7}\,M_\odot$ yr$^{-1}$ for an assumed outflow speed $v_w = 10$ km s$^{-1}$. To the extent that L2 mass loss is preferentially concentrated in the binary plane, while the polar jets in LFBOTs interact with gas along the rotational axis, our predicted spherically averaged CSM profiles may somewhat overpredict the afterglow-inferred densities at $r \lesssim 10^{17}$ cm.

velocity:

$$v_{\rm nCSM} \approx \left(\frac{2\Delta E}{M_{\rm nCSM}}\right)^{1/2} \approx \left(\frac{GM_\bullet M_{\rm He}}{a_{\rm HeC} M_{\rm env}}\right)^{1/2} \approx 10^3\,{\rm km\,s^{-1}} \times \left(\frac{M_\bullet}{10\,M_\odot}\right)^{1/2}\left(\frac{a_{\rm HeC}}{2R_{\rm He}}\right)^{-1/2}\left(\frac{R_{\rm He}}{R_\odot}\right)^{-1/2}\left(\frac{M_{\rm He}}{M_{\rm env}}\right)^{1/2}. \quad (8)$$

This estimate neglects the initial gravitational binding energy of the envelope. In Section A.2 we show that this is a reasonable approximation for the outer parts of the hydrogen envelope but that the binding energy cannot be neglected for matter closer to the HeC, which will likely emerge with a lower speed ∼100 km s$^{-1}$. Ejection of the outer envelope is expected to take place over a timescale comparable to the orbital period at DDI, $P_{\rm orb,DDI}$ (e.g., M. MacLeod et al. 2018),

$$P_{\rm orb,DDI} = \sqrt{\frac{4\pi^2 a_{\rm DDI}^3}{GM_{{\rm tot},f}}} \approx 2.3\ {\rm days}\left(\frac{a_{\rm DDI}}{30\,R_\odot}\right)^3\left(\frac{M_{{\rm tot},f}}{30\,M_\odot}\right)^{-1/2}, \quad (9)$$

of typically a few days, where $M_{{\rm tot},f}$ is the total binary mass at DDI. The end result is the creation of a "nearby" CSM of mixed H/He composition, which by the time the BH spirals into and tidally disrupts the HeC will have expanded to radii

$$R_{\rm nCSM} \approx P_{\rm orb,DDI} v_{\rm nCSM} \sim 10^{13} - 10^{14}\ {\rm cm}. \quad (10)$$

This large CSM shell, comparable in size to a giant star, has implications for the optical emission from the merger-driven explosion, as discussed below. Figure 5 shows the distributions of ejected envelope mass $M_{\rm nCSM}$, velocity $v_{\rm nCSM}$, and characteristic radial extent $R_{\rm nCSM}$ from our binary model suite, following the above estimates. The mean helium abundance of this nearby CSM is typically $X_{\rm He} \approx 0.6$, over twice solar and substantially higher than the extended CSM (Figure 4). In practice, the nearby ejecta will most likely be highly aspherical and with components both slower and faster than $v_{\rm nCSM}$ estimated above (for 3D hydrodynamic simulations of dynamical inspiral, see P. A. Ondratschek et al. 2022; M. Vetter et al. 2024, 2025, although limited to much larger scales of convective envelopes).

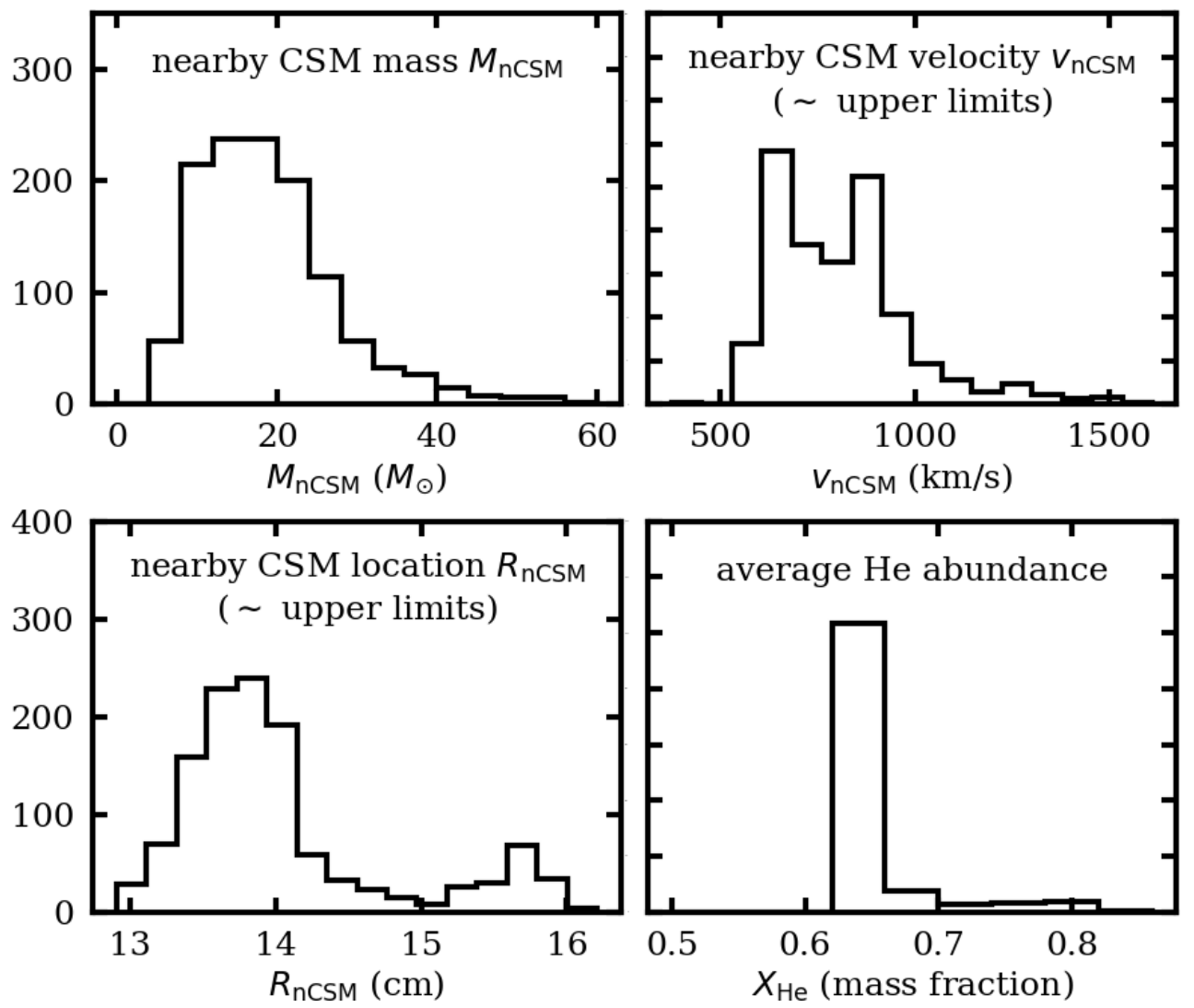

**Figure 5.** Properties of the "nearby CSM" ejected when the BH dynamically plunges into the envelope of the stripped donor star. Histograms compiled from our suite of binary simulations show estimates of the total mass of the nCSM (envelope mass), as well as its average radial velocity $v_{\rm nCSM}$ (Equation (6)), radial extent at the time of the HeC–BH merger $R_{\rm nCSM}$ (Equation (10)), and mean He abundance. The estimated $v_{\rm nCSM}$ and $R_{\rm nCSM}$ values are likely upper limits (see text and Section A.2).

### 3.3. Tidal Disruption and Accretion of the He Core

Unlike in a star with a large convective envelope, there is no steep core–envelope gradient to halt the inspiral of the BH after plunging into the donor envelope (Section A.2). The HeC will begin to RLOF onto the BH at an orbital period,

$$P_{\rm orb,HeC} = 2\pi\left(\frac{a_{\rm HeC}^3}{GM_{\rm bin}}\right)^{1/2} \underset{q\approx 3}{\approx} 6.1\times 10^3\,{\rm s}\left(\frac{a_{\rm HeC}}{2\,R_\odot}\right)^{3/2}\left(\frac{M_{\rm bin}}{20\,M_\odot}\right)^{-1/2}, \quad (11)$$

of typically 1–2 hr, where $M_{\rm bin} = M_\bullet + M_{\rm He}$ is the total mass of the BH+HeC system. We assume that the mass transfer process is unstable, such that the HeC is tidally disrupted by the BH on a timescale comparable to $P_{\rm orb,HeC}$. Though uncertain, the likelihood of this is greater owing to the dense rotating envelope surrounding the binary, the drag from which accelerates the mass transfer rate and whose presence may prevent any angular momentum from being transferred back into the orbit (D. Gagnier & O. Pejcha 2023; S. Tuna & B. D. Metzger 2023). The tidally disrupted HeC will form a rotationally supported disk around the BH with a characteristic radius (e.g., B. Margalit & B. D. Metzger 2016),

$$R_{d,0} \simeq a_{\rm HeC}(1+q)^{-1} \underset{q=3}{\approx} 0.5 R_{\rm He}, \quad (12)$$

where the second equality makes use of Equation (7).

After forming, the disk will accrete onto the BH as a result of angular momentum transport driven by the magnetorotational instability (S. A. Balbus & J. F. Hawley 1998) and/or gravitational instabilities (C. F. Gammie 2001). The peak accretion rate occurs on the "viscous" timescale (e.g., J. Frank et al. 2002),

$$t_{\rm visc,0} \sim \frac{R_{d,0}^2}{\nu} \sim \frac{1}{\alpha}\frac{1}{\theta^2}\left(\frac{R_{d,0}^3}{GM_\bullet}\right)^{1/2} \approx 0.92\ {\rm days}\ \alpha_{0.1}^{-1}\left(\frac{M_\bullet}{10\,M_\odot}\right)^{-1/2}\left(\frac{R_{d,0}}{R_\odot}\right)^{3/2}, \quad (13)$$

where $\nu = \alpha c_s H = \alpha r^2\Omega_{\rm K}\theta^2$ is the effective kinematic viscosity, with $\Omega_{\rm K} \equiv (GM_\bullet/R_{d,0}^3)^{1/2}$, $c_s \approx H\Omega_{\rm K}$ the midplane sound speed, and $\alpha = 0.1\alpha_{0.1}$ the viscosity parameter (N. I. Shakura & R. A. Sunyaev 1973) scaled to a typical value (e.g., A. R. King et al. 2007). In the above we have assumed that the disk is hot and geometrically thick after forming, with a vertical scale height $H$ and aspect ratio $\theta \equiv H/R_{d,0} \approx 1/3$.

On timescales $t \gtrsim t_{\rm visc,0}$, the disk will establish a steady flow onto the BH. The peak inflow rate near the outer disk $\sim R_{d,0}$,

$$\dot M_0 \approx \frac{M_{\rm He}}{t_{\rm visc,0}} \approx 4\times 10^{29}\left(\frac{M_{\rm He}}{20\,M_\odot}\right)\left(\frac{t_{\rm visc,0}}{1\ {\rm day}}\right)^{-1}\ {\rm g\ s^{-1}}, \quad (14)$$

is typically $\gtrsim$10 orders of magnitude larger than the BH Eddington rate $\dot M_{\rm Edd} \equiv L_{\rm Edd}/(0.1c^2) \sim 10^{19}(M_\bullet/10\,M_\odot)$ g s$^{-1}$, justifying our earlier assumption of a thick disk.

### 3.4. Disk Outflows

For the high mass inflow rates $\dot M \gg \dot M_{\rm trap} \equiv \dot M_{\rm Edd}(R_{d,0}/R_{\rm in}) \sim 10^4 \dot M_{\rm Edd}$ of interest (see Equation (14)) photons are trapped and advected inward through the disk at radii $\lesssim R_{d,0}$ (e.g., M. C. Begelman 1979), where $R_{\rm in} = 6GM_\bullet/c^2 \approx 10^7$ cm corresponds to the inner edge of the disk, taken here to be the innermost stable orbit of a slowly spinning BH. Outflows from the radiatively inefficient accretion flow (e.g., R. Narayan & I. Yi 1995; R. D. Blandford & M. C. Begelman 1999; T. Kitaki et al. 2021) decrease the mass inflow rate $\dot M$ approaching the BH as a power law in disk radius $r$,

$$\dot M(r) \approx \dot M_0\left(\frac{r}{R_{d,0}}\right)^p, \quad (15)$$

where the power-law index obeys $p \leqslant 1$.

Equation (15) predicts that the outermost disk radii $\sim 10^{10}$–$10^{11}$ cm dominate the total mass carried away by outflows, while the inner radii $\sim 10^7$–$10^8$ cm dominate their energy budget.[7] The total mass-loss rate,

$$\dot M_w = \dot M(R_{d,0}) - \dot M(R_{\rm in}) = \dot M_0\left[1 - \left(\frac{R_{\rm in}}{R_{d,0}}\right)^p\right], \quad (16)$$

almost equals the total inflow rate $\approx \dot M_0$ because $R_{d,0} \gg R_{\rm in}$. Most of the disrupted HeC becomes unbound, with only a

[7] Here we estimate mass and energy loss per logarithmic radius interval. From Equation (15), $d\dot M_w/d\ln r \propto r^p$, while $d\dot E_w/d\ln r \sim v_{\rm K}^2\, d\dot M_w/d\ln r \propto r^{p-1}$.

small fraction accreting onto the BH:

$$\frac{M_{\rm acc}}{M_{\rm He}} \approx \frac{\dot{M}(R_{\rm in})}{\dot{M}(R_{d,0})} \sim \left(\frac{R_{\rm in}}{R_{d,0}}\right)^{p} \approx 4 \times 10^{-3}\left(\frac{M_{\bullet}}{10\,M_{\odot}}\frac{R_{\odot}}{R_{d,0}}\right)^{0.6}$$
$$\Rightarrow M_{\rm acc} \approx 0.1\,M_{\odot}\left(\frac{M_{\rm He}}{20\,M_{\odot}}\right)\left(\frac{M_{\bullet}}{10\,M_{\odot}}\frac{R_{\odot}}{R_{d,0}}\right)^{0.6}, \quad (17)$$

where in the final numerical estimate we take $p = 0.6$ motivated by numerical simulations of radiatively inefficient accretion (e.g., F. Yuan & R. Narayan 2014).

The disk outflows possess a wide range of speeds, with the wind velocity $v_w$ at radius $r$ scaling with the Keplerian orbital velocity $v_{\rm K} = r\Omega_{\rm K}$ according to $v_w \approx 1.2 v_{\rm K}$, where the prefactor is for $p = 0.6$ following B. Margalit & B. D. Metzger (2016, their Figure 3). Roughly half of the wind material emerges from $r > R_w \approx R_{d,0}/2^{1/p} \sim 0.3 R_{d,0}$ with an average velocity

$$v_{\rm slow} \approx 1.2 v_{\rm K}|_{R_w} \approx 1.2\left(\frac{GM_{\rm bin}}{R_{\rm w}}\right)^{1/2}$$
$$\approx 5300\ {\rm km\ s^{-1}}\left(\frac{M_{\rm bin}}{30\,M_{\odot}}\right)^{1/2}\left(\frac{R_{d,0}}{R_{\odot}}\right)^{-1/2}, \quad (18)$$

carrying a kinetic energy

$$E_{\rm slow} \approx \frac{1}{2} M_{\rm He} v_{\rm slow}^2 \sim 3 \times 10^{51} {\rm erg}. \quad (19)$$

This "slow" ejecta, comprising the bulk of the mass, expands with velocities $\lesssim$8000 km s$^{-1}$. After sweeping up the nearby CSM (Section 3.2), final outflow speeds of several thousand kilometers per second are expected, broadly similar to the lowest ejecta speeds implied by LFBOT spectral line widths (e.g., D. A. Perley et al. 2019; R. Margutti et al. 2019; D. Xiang et al. 2021; N. LeBaron et al. 2026).

A smaller fraction of the disk-wind ejecta mass $\gtrsim M_{\rm acc}$ (Equation (17)) emerges close to the BH with much higher velocities $v_{\rm fast} \gg v_{\rm slow}$ as a collimated "jet." Radiation general relativistic MHD simulations of super-Eddington accretion find the generation of transrelativistic outflows ($v_{\rm fast} \gtrsim 0.3c$) from the innermost polar region above the BH, carrying a combined radiative and kinetic luminosity $L_{\rm acc} \approx \eta \dot{M}_{\bullet} c^2$ (e.g., A. Sadowski & R. Narayan 2016), where $\dot{M}_{\bullet} = \dot{M}(R_{\rm in})$ is the accretion rate reaching the BH and $\eta \approx 0.01-0.1$ is an efficiency factor that depends weakly on the accretion rate but more strongly on the magnetic field threading the disk (e.g., A. Sadowski & R. Narayan 2016).

Over the initial accretion timescale $t_{\rm visc,0} \sim 1$ day (Equation (13)), the fast jet thus carries a total energy,

$$E_{\rm fast} \approx \eta M_{\rm acc} c^2 \approx 10^{51} {\rm erg}\ \eta_{-2}\left(\frac{M_{\rm acc}}{0.1\,M_{\odot}}\right), \quad (20)$$

comparable or exceeding $E_{\rm slow}$, depending on $\eta = 0.01\eta_{-2}$, and a peak luminosity

$$L_{j,0} \approx \frac{E_{\rm fast}}{t_{\rm visc,0}} \approx 2 \times 10^{46}\ {\rm erg\ s^{-1}}\ \eta_{-2}\left(\frac{M_{\rm acc}}{0.1\,M_{\odot}}\right)\left(\frac{t_{\rm visc,0}}{1\ {\rm day}}\right)^{-1}. \quad (21)$$

A jetlike outflow of velocity $v_{\rm fast} \gtrsim 0.1c-0.5c$ is consistent with the highest outflow speeds from LFBOTs based on the early photosphere expansion rate and bright radio synchrotron emission (e.g., D. L. Coppejans et al. 2020; A. Y. Q. Ho et al. 2020). If the BH is rapidly spinning, it may also power a more tightly collimated relativistic jet and associated nonthermal emission (Section 4.4).

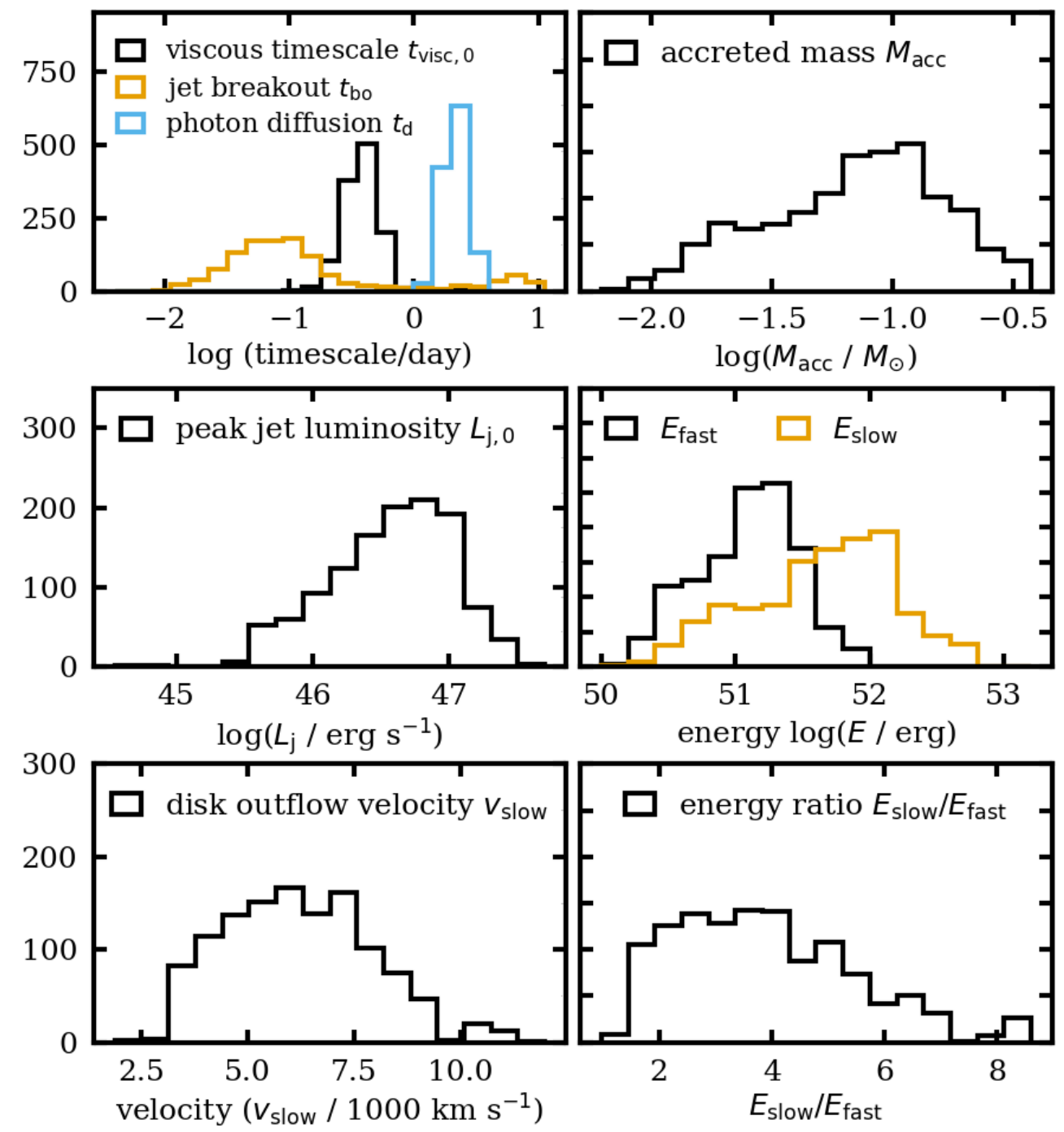

**Figure 6.** Quantities related to the BH engine after disruption of the HeC: initial disk accretion timescale $t_{\rm visc,0}$ (Equation (13) for $\alpha = 0.1$); jet breakout time (Equation (22) for $v_{\rm fast} = 0.3c$); diffusion time through polar ejecta (Equation (23) for $f_j = 0.03$); mass accreted by BH (Equation (17)); peak jet luminosity, $L_{\rm jet}$ (Equation (21) for $\eta = 10^{-2}$); energy in the fast (Equation (20) for $\eta = 10^{-2}$) and slow (Equation (19)) disk outflows; and slow outflow velocity (Equation (18)).

Figure 6 summarizes several of the above key properties of the HeC/BH accretion engine across our grid of binary evolution models. The energies in the fast and slow disk-wind ejecta are generally comparable for fiducial parameters and can span $\sim 10^{50.5}-10^{52}$ erg. The peak luminosity of the jet spans a typical range $L_j \sim 10^{46}-10^{47}$ erg s$^{-1}$, over several hours.

### 3.5. Transient Electromagnetic Emission

#### 3.5.1. Jet Breakout and Cooling Envelope Emission

Before reaching the extended CSM (Section 3.1), the fast BH accretion-powered jet will first interact with the nearby CSM produced from the dynamical envelope ejection (Section 3.2). Assuming that the jet propagates through the nearby CSM at a velocity $\sim v_{\rm fast}$, it will break out of the photosphere on a timescale

$$t_{\rm bo} \approx \frac{R_{\rm nCSM}}{v_{\rm fast}} \sim 0.1\ {\rm day}\left(\frac{R_{\rm nCSM}}{10^{14}\ {\rm cm}}\right)\left(\frac{v_{\rm fast}}{0.3c}\right)^{-1}. \quad (22)$$

As the portion of the nearby CSM freshly shocked by the jet expands, it gradually becomes transparent to radiation, powering so-called "shock cooling emission" (SCE; E. K. Grasberg & D. K. Nadezhin 1976; S. W. Falk & W. D. Arnett 1977; R. A. Chevalier 1992; E. Nakar & R. Sari 2010; A. L. Piro 2015; O. Gottlieb et al. 2022; B. Margalit 2022). We assume that a fraction $f_j \equiv M_{\rm sh}/M_{\rm env} \lesssim 0.1$ of the envelope mass is shocked by the fast polar jet. Because $t_{\rm bo}$ can be comparable to the jet duration ($t_j \sim t_{\rm visc,0}$; Figure 6), the energy imparted to the shocked CSM is a significant fraction of $E_{\rm fast}$ (Equation (20)). Following standard estimates (e.g., A. L. Piro 2015), the SCE emission will peak on a timescale set by photon diffusion,

$$t_d \approx \left(\frac{M_{\rm sh}\kappa}{4\pi v_{\rm fast}c}\right)^{1/2} \approx 2.7\ {\rm days}\left(\frac{v_{\rm fast}}{0.3c}\right)^{-1/2} \times \left(\frac{f_j}{0.03}\right)^{1/2}\left(\frac{M_{\rm env}}{10\,M_\odot}\right)^{1/2}, \quad (23)$$

at a luminosity

$$L_{\rm SCE} \approx \frac{E_{\rm fast}R_{\rm nCSM}}{v_{\rm fast}t_d^2} \approx 2 \times 10^{43}\ {\rm erg\ s^{-1}}\left(\frac{E_{\rm fast}}{10^{51}\ {\rm erg}}\right) \times \left(\frac{R_{\rm nCSM}}{10^{14}\ {\rm cm}}\right)\left(\frac{v_{\rm fast}}{0.3c}\right)^{-1}\left(\frac{f_{\rm j}}{0.03}\right)\left(\frac{M_{\rm env}}{10\,M_\odot}\right)^{-1}, \quad (24)$$

where we again take $\kappa \approx 0.30\ {\rm cm^2\ g^{-1}}$ for the electron scattering opacity. These estimates are crude and should be calibrated with environment-specific radiation hydrodynamic simulations for better accuracy.

Our estimates for $t_d$ and $L_{\rm SCE}$ are comparable to rise times and peak UV/optical luminosities of some FBOTs. However, the persistently high temperatures of LFBOT emission over several weeks ($t \gg t_d$; e.g., D. A. Perley et al. 2019) are inconsistent with the predicted reddening of SCE (e.g., A. L. Piro 2015). This suggests that CSM interaction is not the only source powering the UV/optical light curves of LFBOTs (e.g., R. Margutti et al. 2019).

### *3.5.2. Reprocessed Engine Power*

An additional source of longer-lasting optical UV/optical emission is *continuous* shock heating of the environment by the jet (O. Gottlieb & B. D. Metzger 2024) or reprocessing of its X-ray emission (e.g., R. Margutti et al. 2019; A. L. Piro & W. Lu 2020; K. Uno & K. Maeda 2020; D. Calderón et al. 2021; M22; C. Chen & R.-F. Shen 2022).

At times $t \gg t_{\rm visc,0}$ (Equation (13)) the disk mass drops owing to accretion and its outer edge viscously spreads outward (e.g., B. D. Metzger et al. 2008). This results in a power-law decay of the BH accretion rate and jet luminosity $L_{\rm jet} \propto \dot{M}_\bullet \propto t^{-\beta}$, where $\beta \approx 2-2.7$ for $p = 0.5-1$ (Equation (A12) in Section A.3). Spreading of the disk to radii $\gg R_{d,0}$ is supported by the large outer disk radii $\approx 40\,R_\odot$ inferred from UV observations of AT2018cow taken several years after the explosion (N.-C. Sun et al. 2022, 2023; Y. Chen et al. 2023; G. Migliori et al. 2024; M. Winter-Granic & E. Quataert 2025). Recent modeling of this process favors a BH accretor of mass $\sim 10-100\,M_\odot$ (M. Winter-Granic & E. Quataert 2025).

Because $t_d \gtrsim t_{\rm visc,0}$, the rise of the engine-powered light curve still occurs on the diffusion time through the shocked CSM (Equation (23)). Assuming that an order-unity fraction of $L_{\rm jet}$ goes into reprocessed emission, and substituting $t = t_d$ (Equation (23)) into $L_{\rm jet}(t)$ (Equation (A12)), the reprocessed UV/optical light curve roughly obeys

$$L_{\rm rep}(t) \approx L_{\rm pk}\left(\frac{t}{t_{\rm pk}}\right)^{-2.1}, \quad (25)$$

where the peak luminosity is given by

$$L_{\rm pk} \approx L_{\rm rep}(t_d) \approx 1.3 \times 10^{45}\ {\rm erg\ s^{-1}}\eta_{-2}\alpha_{-1}^{-1.13} \times \left(\frac{M_\bullet}{10\,M_\odot}\right)^{0.03}\left(\frac{M_{\rm He}}{20\,M_\odot}\right)\left(\frac{R_{d,0}}{R_\odot}\right)^{1.1} \times \left(\frac{0.3c}{v_{\rm fast}}\frac{f_j}{0.03}\frac{M_{\rm env}}{10\,M_\odot}\right)^{-1.07}, \quad (26)$$

and we have again assumed $p = 0.6$, $\theta = 1/3$.

Figure 7 shows both luminosity sources, $L_{\rm pk}$ and $L_{\rm SCE}$, as a function of $t_d$ for our grid of models, in comparison to the observed peak luminosities and rise time of LFBOTs. Although our estimates depend on several uncertain parameters ($v_{\rm fast}$, $f_j$, $\alpha$, $\eta$), broadly speaking our models show substantial overlap with the LFBOT population. Furthermore, the transients with the largest nCSM masses and longer durations that we predict might be classified as superluminous SNe (SLSNe) or jetted tidal disruption events (TDEs), instead of FBOTs (e.g., AT2024puz; J. J. Somalwar et al. 2025; see also Section 4.5).

As the fast polar ejecta clears out, slower CSM and disk-wind ejecta from progressively larger angles off the jet axis will take over as the dominant reprocessing region (M22), eventually dominating the photosphere emission and generating lower observed spectral velocities. A torus photosphere geometry is consistent with the asymmetric spectral line shapes (R. Margutti et al. 2019) and polarimetry (J. R. Maund et al. 2023) seen after several weeks in AT2018cow for viewing angles off the binary/jet axis.

Nonthermal X-rays generated internal to the BH jet, which are not absorbed by the disk-wind ejecta or CSM, are directly visible to the external observer, with a luminosity and spectrum depending sensitively on viewing angle and associated complex scattering geometry (R. Margutti et al. 2019; A. J. Nayana et al. 2025). On top of the smoothly declining accretion rate, fluctuations in the accretion rate, magnetic flux, or jet orientation with respect to the observer site line (as postulated in other jetted TDEs; A. Tchekhovskoy et al. 2014) could generate late-time flares, somewhat akin to microquasar eruptions, with an associated brief nonthermal afterglow (A. Y. Q. Ho et al. 2023a). If the BH is rapidly spinning and its direction need not be aligned with the angular momentum of the disk defined by the binary orbit, precession of the jet may also play a role (e.g., A. Tchekhovskoy et al. 2014; O. Teboul & B. D. Metzger 2023).

### *3.6. Infrared Dust Echo*

A final source of electromagnetic emission is an infrared echo due to reprocessing of the earliest phases of the transient light by dust in the extended CSM.

The slow wind from the binary during the stable mass transfer phase (Section 3.1; Figure 4) is sufficiently dense and cool for efficient dust formation to occur (B. D. Metzger & D. A. Perley 2023), potentially shrouding the binary in an opaque medium prior to the final dynamical transient (e.g., O. Pejcha et al. 2016a). As the fast polar jet expands to large radii $\gg R_{nCSM}$, it will begin to shock this extended CSM, powering synchrotron radio/submillimeter emission lasting several months to years (A. Y. Q. Ho et al. 2019; B. Margalit et al. 2022).

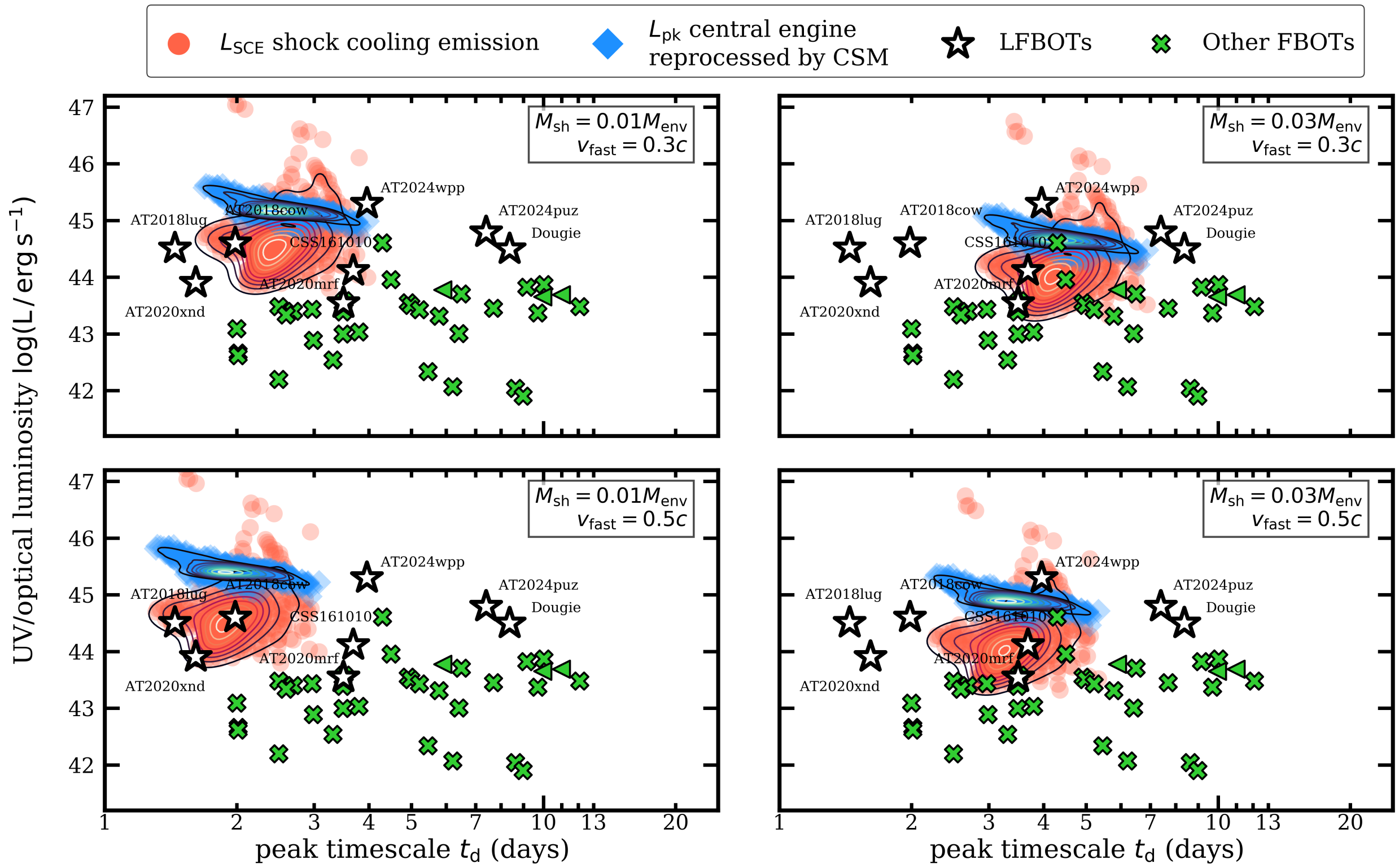

**Figure 7.** Distribution of peak optical/UV luminosity vs. peak timescale ($t_d$; Equation (23)) of BH jet-powered emission for our modeled binary population. We show separately the luminosities associated with (a) passive SCE, $L_{SCE}$ (Equation (24)), from the jet of velocity $v_{fast}$ breaking out of the nearby CSM (Figure 5); (b) continuous heating of the CSM by the jet power, $L_{pk}$ (Equation (26)). For comparison, we also show measured rise timescales and pseudo-bolometric luminosities of LFBOTs (black stars/triangles) and the broader FBOT population (green crosses) from N. LeBaron et al. (2026, their Figure 4). The values shown in each panel are calculated for different assumed values of $f_j = 0.01$, 0.03 and $v_{fast}/c = 0.3$, 0.5 as marked, while all cases assume $\alpha = 0.1$, $\eta = 10^{-2}$.

However, even prior to the arrival of the jet, the earliest UV photons emitted by the transient will reach the CSM and be absorbed by the dust. This will redden and attenuate the earliest phases of the transient, before the dust is heated sufficiently to be sublimated by the rising UV luminosity. However, before the dust is destroyed, the absorbed transient's light is reemitted as infrared radiation (B. D. Metzger & D. A. Perley 2023; S. Tuna et al. 2025). This IR emission signal is referred to as an "echo" because it reaches the observer later than the direct UV light from the transient due to the light propagation time from the absorbing dusty CSM (i.e., the dust has "already" been destroyed by the time we see the echo). Section A.4 reviews the duration and luminosity ($t_{echo}$, $L_{echo}$) of the dust echo from B. D. Metzger & D. A. Perley (2023) in terms of the assumed properties of the dust and the density profile of the extended CSM.

Figure 8 shows ($t_{echo}$, $L_{echo}$) for each of our models, as determined from their CSM density profiles (Figure 4), for fiducial assumptions about the dust grain size and sublimation temperature ($a = 1\ \mu$m; $T_s = 1700$ K). The modest predicted echo luminosities $\sim 10^{40}$–$10^{42}$ erg s$^{-1}$ will be buried initially by the much brighter UV/optical light of the transient (Figure 7); however, the echo's longer duration of several weeks to months and cooler temperature can permit its detection. Indeed, the excess IR emission component seen in AT2018cow (D. A. Perley et al. 2019) and AT2024wpp (N. LeBaron et al. 2026) falls in the center of our predicted distribution, consistent with being a dust echo (B. D. Metzger & D. A. Perley 2023; S. Tuna et al. 2025). Spherical dust distributions generate relatively flat light curves, which become more bell shaped for equatorially concentrated dust shells (S. Tuna et al. 2025).

The photosphere of the pre-transient dust shell, which controls the echo duration, roughly coincides with the drop in the CSM profile around $r_{ph} \sim 3 \times 10^{16}$ cm. The cool photosphere has implications for the observed appearance of LFBOT progenitors leading up to dynamical instability (Section 4). We also show the distribution of $L_{thin}$, the transient luminosity sufficient to sublimate dust out to the photosphere. The fact that $L_{thin} \sim 10^{42}$–$10^{43}$ erg s$^{-1}$ is much less than the peak luminosity of the transient (Figure 7) shows that the dust photosphere is destroyed during the rise phase of the transient, consistent with the lack of observed dust reddening in LFBOT spectra taken near and after peak light. Indeed, the LFBOT candidate MUSSES2020J exhibited red colors prior to maximum light (J.-a. Jiang et al. 2022), which might arise from dust absorption prior to dust destruction.

### 3.7. Rates

In order to assess the predicted rate of DDI transients, we developed a rapid binary population synthesis code (Appendix B). Although our code neglects several secondary effects (e.g., no binary tides or eccentric orbits), it is built on modern MESA stellar tracks up to $150\,M_\odot$ and includes an up-to-date treatment of mass transfer stability following C. Schürmann & N. Langer (2024) and K26 and the donor's core response to stripping following C. Schürmann et al. (2024, 2025). Starting from ZAMS binaries, we synthesize BH+OB populations at five metallicities $Z/Z_\odot = \{0.04, 0.1, 0.2, 0.4, 1.0\}$. To gauge the uncertainty of population predictions, we explore five accretion efficiencies $\beta = \{0.1, 0.3, 0.5, 0.7, 0.9\}$ (which also affect stability) and two SN engines for BH formation (delayed C. L. Fryer et al. 2012 and K. Maltsev et al. 2025), i.e., 10 variants per metallicity. We assume that BHs receive natal kicks with magnitudes drawn from the Hobbs distribution (Maxwellian with $\sigma = 265$ km s$^{-1}$; G. Hobbs et al. 2005) but scaled down by a $(1 - f_{\rm fb})$ factor, where $f_{\rm fb}$ (determined by the SN engine prescription) is the fraction of mass that ultimately falls back onto the BH (see also Section B.2.3). This results in the typical BH kick magnitudes ranging from $\sim$10 km s$^{-1}$ to a few hundreds of kilometers per second.

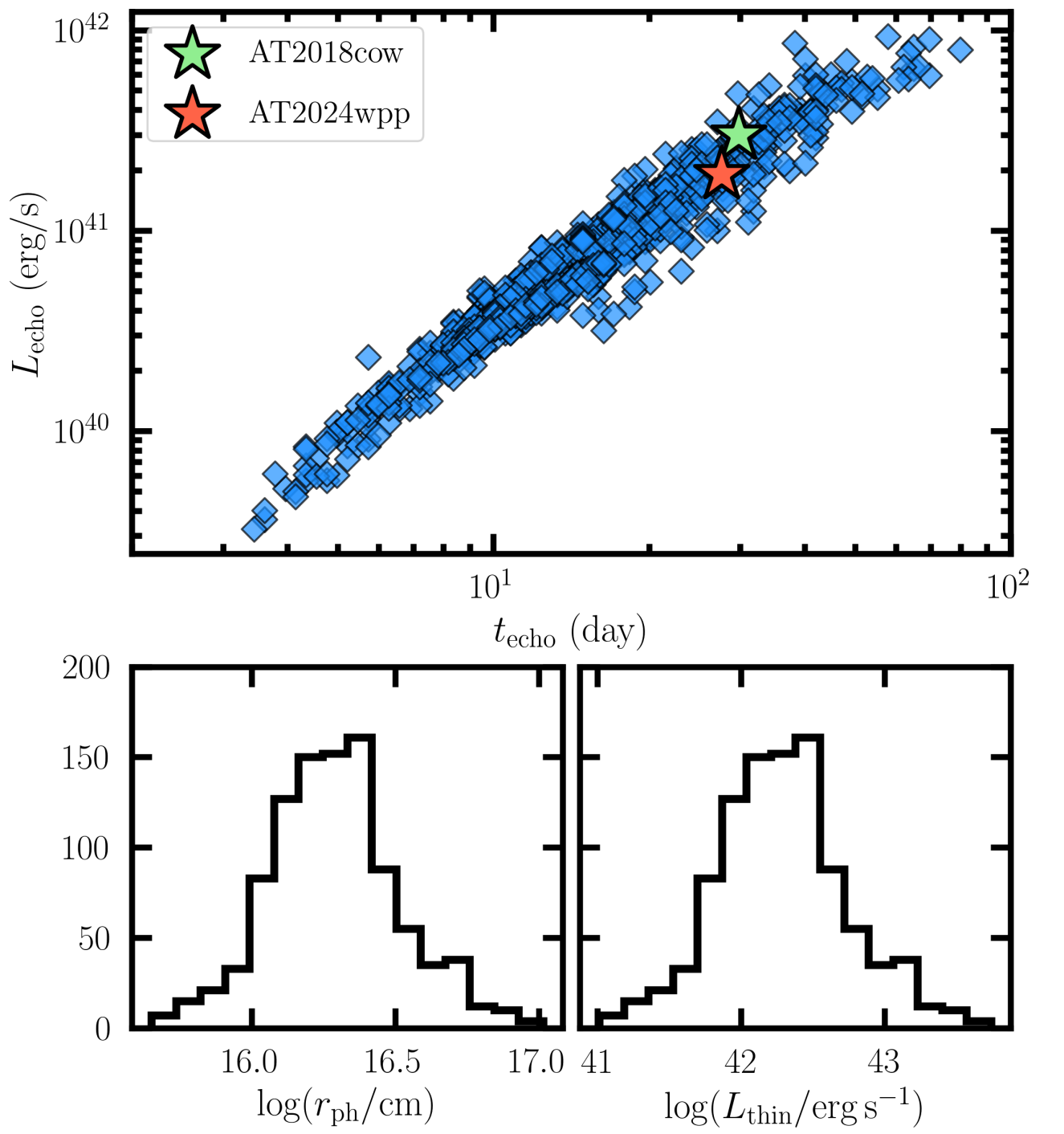

**Figure 8.** Top panel: approximate luminosity $L_{\rm echo}$ and duration $t_{\rm echo}$ (Equation (A16)) of the infrared echo produced by reprocessing of UV emission by dust within the extended CSM (Figure 4), for an assumed dust grain size $a = 1$ $\mu$m and sublimation temperature $T_s = 1700$ K. Shown for comparison with stars are the measured infrared emission component in AT2018cow (D. A. Perley et al. 2019) and AT2024wpp (N. LeBaron et al. 2026). Bottom panels: the dust photosphere radius of extended CSM $r_{\rm ph}$ (Equation (A13)) and UV luminosity $L_{\rm thin}$ at which dust is sublimated to the photosphere (see Section A.4 for details).

For each BH+OB system produced by the rapid code, we assign a final fate informed by our MESA binary grid of the same metallicity (Section 2.1). These detailed binary models begin at the BH+OB stage and allow us to map the parameter space of mass ratios/orbital periods of BH+OBs into five final outcomes: BH+BH mergers, wide BH+BH, traditional CE, BH+HeC mergers (our LFBOT channel), and BH+MS mergers, as illustrated in Figure 2. We consider two variations of MESA binary grids: fiducial ($\sigma_{\rm ov} = 0.35$) and with reduced core overshooting ($\sigma_{\rm ov} = 0.18$), which yield somewhat different predictions for the BH+OB outcomes (for details, see Section B.3, Figures 15 and 16).

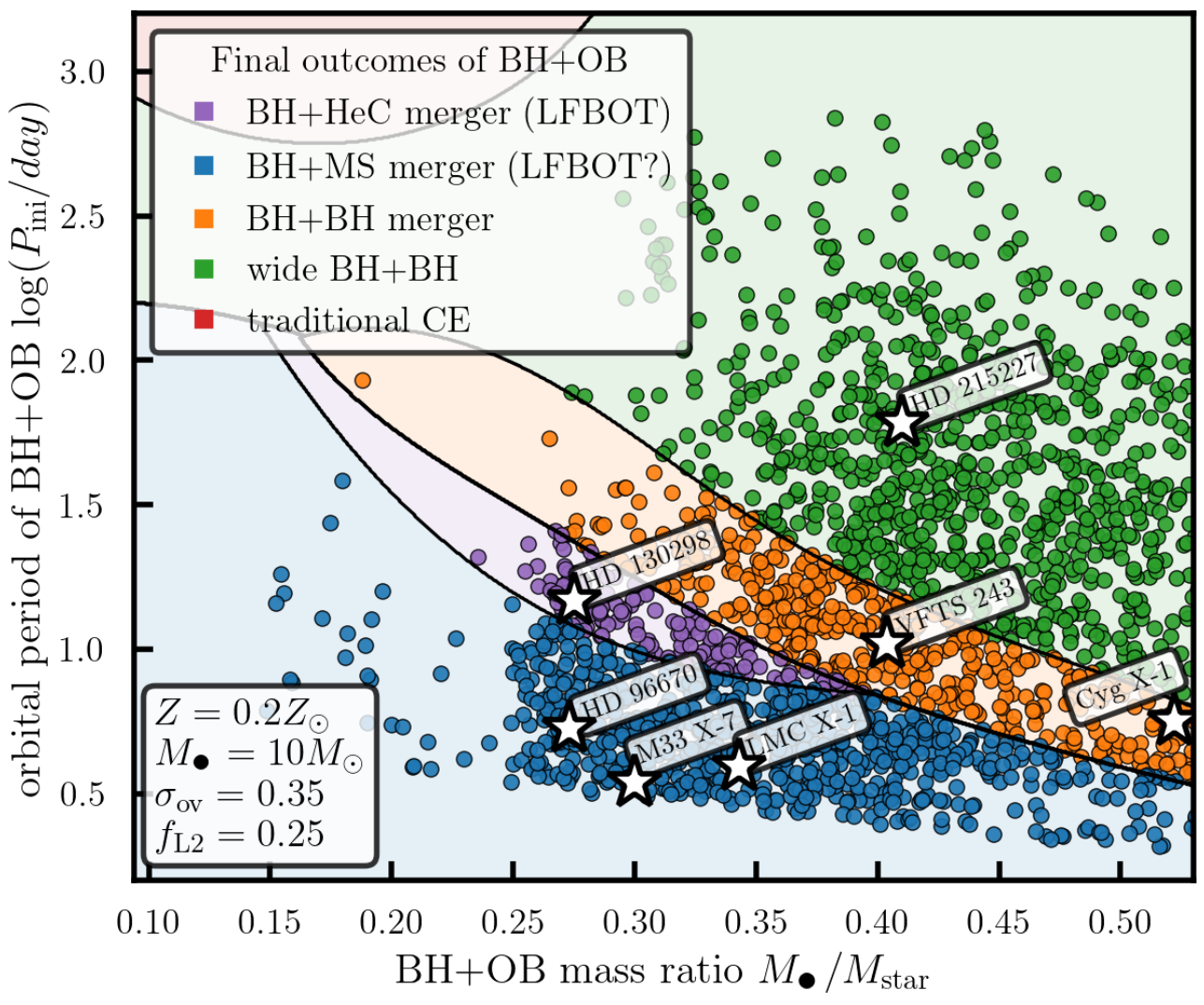

**Figure 9.** Population of BH+OB systems predicted by our rapid population synthesis code at $Z = 0.2\,Z_\odot$ (circles), mapped onto the final outcomes informed by our detailed binary MESA models of BH+OBs. The population shown here assumes K. Maltsev et al. (2025) criteria for BH formation and an equal contribution from five accretion efficiencies $\beta = \{0.1, 0.3, 0.5, 0.7, 0.9\}$. Out of all synthesized BH+OBs, we predict $\sim$18% to evolve to become BH+star mergers and possibly LFBOT-related transients, with $\sim$5% leading to BH+HeC mergers (our LFBOT channel). By comparison, $\sim$6.5% of BH+OB systems generate BH+BH mergers and gravitational-wave sources. Star symbols mark several known BH+OB systems, both X-ray bright and dormant. The full population model includes five metallicities to arrive at volumetric LFBOT rates in Figure 10.

An example synthetic population of BH+OB systems is shown in Figure 9, generated with the rapid code at $Z = 0.2\,Z_\odot$ metallicity for K. Maltsev et al. (2025) explodability criteria. The population shown combines equal contributions from all five choices of the accretion efficiency $\beta$ in order to reproduce the observed BH+OB systems (star symbols; see below). Each synthetic BH+OB system (circles) is color-coded by their final outcome derived from a MESA binary grid (variant with fiducial overshooting and $M_\bullet = 10\,M_\odot$). The entire population of synthetic BH+OBs also includes mass ratios not shown in the figure ($q > 0.5$). Systems with strongly unequal mass ratios ($q < 0.25$) are rarely predicted: these originate primarily from variants with efficient mass accretion in the first mass transfer ($\beta = 0.7$ or 0.9), for which our assumed stability criteria are quite stringent (C. Schürmann et al. 2024, 2025). For comparison, we plot all the well-characterized systems with a BH and a massive-star companion, both those interacting and X-ray bright (Cyg X-1, J. C. A. Miller-Jones et al. 2021; LMC X-1, J. A. Orosz et al. 2009; M33 X-7, V. Ramachandran et al. 2022) and those X-ray quiet (HD 130298, L. Mahy et al. 2022; VFTS 243, T. Shenar et al. 2022; HD 96670, S. Gomez & J. E. Grindlay 2021; HD 215227, J. Casares et al. 2014; although see S. Janssens et al. 2023). Notably, we find that no single choice of $\beta$ is able to reproduce all the observed systems and choose to instead combine all five variants in Figure 9. This may be in line with evidence for cases of both conservative and nonconservative mass transfer evolution among massive binaries (e.g., T. Lechien et al. 2025; K. Sen et al. 2026).

Combining rapid population synthesis of BH+OBs with their final outcomes from MESA binary models yields the total fraction of stars that evolve to become BH+HeC mergers. For the $Z = 0.2\,Z_\odot$ case shown in Figure 9, out of the initial star-formed population of $2 \times 10^8\,M_\odot$, $\sim$12,800 BH+OB systems were formed. We see that BH+star mergers from DDI account for ∼18% of their final outcomes (blue and purple combined). This number drops to ≈5% if we consider only evolved giant donors with a well-defined core/envelope structure (BH+HeC mergers), as may be necessary to produce the short accretion timescales needed to explain LFBOTs. This corresponds to one LFBOT formed per $\sim 3.3 \times 10^5\,M_\odot$ of star formation (or an LFBOT formation efficiency of $3 \times 10^{-6}\,M_\odot^{-1}$ at $Z = 0.2\,Z_\odot$). By contrast, we find that ∼6.5% of BH+OB systems generate tight binaries capable of evolving into BH+BH mergers (though with a possible strong metallicity dependence; Figures 15 and 16).[8]

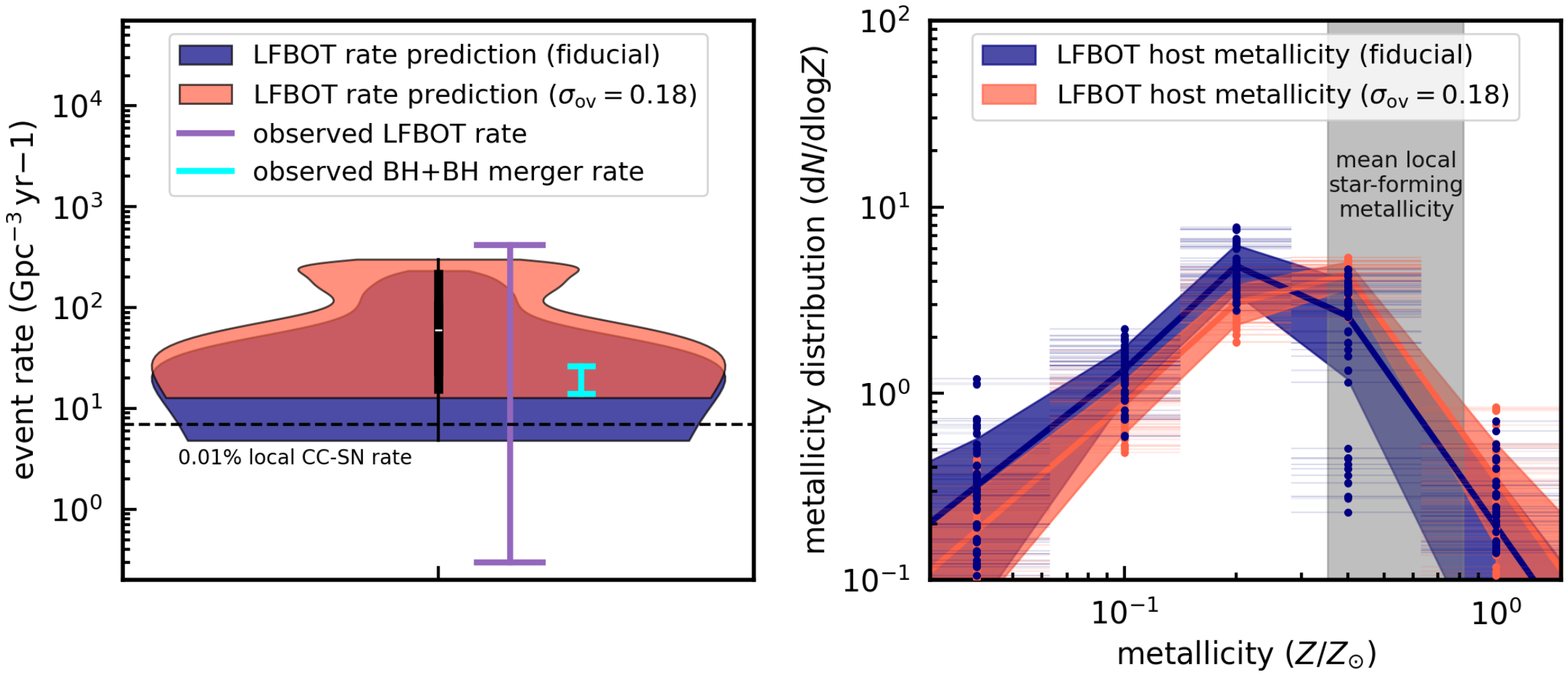

**Figure 10.** Estimate of the event rate of LFBOTs from BH+HeC mergers. Left: volumetric rates of LFBOTs predicted across our population model variations (violin plots), calculated for two variants of MESA binary grids: fiducial ($\sigma_{ov} = 0.35$), yielding LFBOT rates $R_{LFBOT} \sim$ 5–200 Gpc$^{-3}$ yr$^{-1}$, and with reduced core overshooting ($\sigma_{ov} = 0.18$), yielding $R_{LFBOT} \sim$ 15–300 Gpc$^{-3}$ yr$^{-1}$. These model predictions are consistent with the empirically estimated LFBOT rate 0.3–400 Gpc$^{-3}$ yr$^{-1}$ (D. L. Coppejans et al. 2020; A. Y. Q. Ho et al. 2023b) and also overlap with the BH+BH merger rate, supporting our proposal that these two types of events may share a near-twin evolutionary pathway. Right: prediction for the metallicity distribution of LFBOT hosts, derived from our population synthesis of BH+HeC mergers. The preference for subsolar metallicity is broadly consistent with LFBOT hosts that tend to be star-forming dwarfs of $Z \sim$ 0.2–0.8 $Z_\odot$; see Section 4.2.

Finally, these fractions can be transformed into volumetric rate estimates in the local volume $z < 0.3$ where LFBOTs are observed. We convolve the LFBOT formation efficiency at each metallicity with a suite of 10 metallicity-specific star formation histories from M. Chruślińska et al. (2019, 2021, 2025), normalized to the local star formation rate (SFR) density $\rho_{SFR} = 10^{-1.8}\,M_\odot\,\mathrm{yr}^{-1}\,\mathrm{Mpc}^{-3}$ (P. Madau & M. Dickinson 2014). The 10 variants explore uncertainties in the contribution from low-mass galaxies, the absolute abundance scale of gas-phase metallicity, and the oxygen-over-iron specific SFR (sSFR) relation (M. Chruślińska et al. 2024). Together with five variants of accretion efficiency $\beta$ and two remnant prescriptions, this yields a total of 100 population synthesis realizations, each producing a prediction for the LFBOT rate. We show the resulting distribution of LFBOT rate estimates in the form of violin plots in the left panel of Figure 10, one for fiducial BH+OB binary grids (blue) and another for grids adopting reduced overshooting (orange). For the fiducial grids, we obtain $R_{LFBOT} \sim$ 5–200 Gpc$^{-3}$ yr$^{-1}$; adopting reduced overshooting ($\sigma_{ov} = 0.18$) yields $R_{LFBOT} \sim$ 15–300 Gpc$^{-3}$ yr$^{-1}$. These ranges are consistent with the broad observational estimates of ∼0.3–400 Gpc$^{-3}$ yr$^{-1}$ (D. L. Coppejans et al. 2020; A. Y. Q. Ho et al. 2023b). For context, the empirically inferred BH+BH merger rate is ∼14–26 Gpc$^{-3}$ yr$^{-1}$ (The LIGO Scientific Collaboration et al. 2025), consistent with our picture that the origin of LFBOTs may be closely linked to the stable mass transfer pathway to BH+BH.[9] We do not include BH+MS mergers in our LFBOT rate estimate; some may nevertheless yield non-FBOT engine-powered transients (see Section 4).

The right panel of Figure 10 shows the predicted metallicity distribution of LFBOT progenitors. To construct the plot, for each of the five metallicities $Z/Z_\odot = \{0.04, 0.1, 0.2, 0.4, 1.0\}$ we calculate the metallicity-specific LFBOT event rate $dN/d\log Z$ for each of the 100 population synthesis variants (scatter points, with thin horizontal lines marking the $d\log Z$ bins). Solid lines connect the mean $dN/d\log Z$ values obtained, and the colored contours mark a $1\sigma$ deviation. Subsolar hosts are favored, peaking near $0.2\,Z_\odot$ (fiducial) or $0.4\,Z_\odot$ ($\sigma_{ov} = 0.18$). The preference for lower $Z$ stems from the larger parameter space for mass transfer from radiative envelope post-MS donors, because MS stars become more compact as metallicity decreases (e.g., J. Klencki et al. 2020).[10] Although a comprehensive analysis of LFBOT metallicities is lacking, the predicted low-$Z$ preference of the DDI scenario is broadly compatible with direct spectroscopic measurements and masses/SFRs of LFBOT host galaxies (D. A. Perley et al. 2019; J. D. Lyman et al. 2020; A. Y. Q. Ho et al. 2020; Y. Yao et al. 2022b; C. P. Gutiérrez et al. 2024; A. A. Chrimes et al. 2024a), as we discuss in Section 4.2.

8 We note that the LFBOT formation yield is mildly sensitive to the assumed BH natal kicks. For example, we estimate that assuming all the BHs to receive a natal kick of only 30 km s$^{-1}$ (small compared to orbital velocities of LFBOT progenitors, which are typically >100 km s$^{-1}$) would increase the predicted LFBOT rate by a factor of ∼3–5 compared to our default fallback-reduced Hobbs distribution (Section B.2.3).

9 We do not quote a BH+BH merger rate from our population model here; doing so would require calculating the signal-to-noise ratios for detectability and accounting for long delay times.

10 This trend might reverse at very low metallicities $\lesssim 0.02\,Z_\odot$ not explored here, where the maximum post-MS sizes of massive stars could be severely reduced to blue supergiant radii (∼100 $R_\odot$) rather than red supergiant radii (∼1000 $R_\odot$) (see, e.g., J. H. Groh et al. 2019; A. Tanikawa et al. 2020).

## 4. Discussion

### 4.1. Origin of LFBOTs

Three observations in particular strongly constrain progenitor scenarios for LFBOTs: (1) their star-forming, typically dwarf host galaxies (e.g., S. J. Prentice et al. 2018; D. A. Perley et al. 2019; A. Y. Q. Ho et al. 2020; D. L. Coppejans et al. 2020); (2) the seemingly ubiquitous presence of dense CSM with a density profile that steepens at radii $\gtrsim 3 \times 10^{16}$ cm; if interpreted as a progenitor outflow, this requires mass-loss rates $\gtrsim 10^{-3}\,M_\odot$ yr$^{-1}$, much higher than ordinary stellar winds (A. Y. Q. Ho et al. 2019; R. Margutti et al. 2019); and (3) the large size $\gtrsim 40\,R_\odot$ (high angular momentum) of the remnant accretion disk evidenced by late-time UV/optical observations of AT2018cow (N.-C. Sun et al. 2022, 2023; Y. Chen et al. 2023; A. Inkenhaag et al. 2023; G. Migliori et al. 2024; M. Winter-Granic & E. Quataert 2025).

The final requirement disfavors most core-collapse scenarios (e.g., G. Migliori et al. 2024; M. Winter-Granic & E. Quataert 2025; however, see A. A. Chrimes et al. 2026). Jetted transients with long durations can arise from the collapse of large progenitor stars, such as blue supergiants, for which the long freefall timescale of the stellar envelope feeds the BH over several days or longer (e.g., E. Quataert & D. Kasen 2012; R. Perna et al. 2018). However, at least LFBOTs likely arise from more compact partially stripped progenitors (based on their sometimes H-depleted spectra; e.g., D. A. Perley et al. 2019) with short freefall times, for which generating a long-duration transient thus requires a long viscous timescale and hence large initial disk radius, comparable in size to the stellar radius. This in turn may require the star to be very rapidly rotating at collapse, which is difficult to obtain in standard evolutionary channels (J. Fuller et al. 2019) but could point toward binary scenarios involving chemically homogeneous evolution or tidal spin-up in tight orbits (e.g., S. C. Yoon & N. Langer 2005; R. G. Detmers et al. 2008; S. E. de Mink & I. Mandel 2016; J. Fuller & W. Lu 2022).

Such high angular momentum debris arises naturally from the tidal disruption of an MS star by either a low- or intermediate-mass BH (e.g., K. Kremer et al. 2021; A. Inkenhaag et al. 2023); however, requirements (1) and (2) suggest the presence of a massive star and hence are more naturally explained by the HeC−BH "TDE" in a massive binary scenario explored here. Moreover, most TDEs arise from the disruption of stars on initially unbound parabolic orbits,[11] and such events are challenged by the low efficiency with which the weakly bound stellar debris ultimately reaches the central BH in the face of powerful outflows that unbind most of the mass (e.g., L. E. Strubbe & E. Quataert 2009; B. D. Metzger & N. C. Stone 2016). Disk outflows are a key part of the HeC −BH merger scenario (Section 3.4), but their effects in reducing the liberated gravitational energy are mitigated by the more tightly bound orbit and high compactness of the HeC relative to the MS victims of traditional TDE scenarios.

A massive-star association and large accretion disk are also potentially consistent with the scenario proposed by D. Tsuna & W. Lu (2025), in which an NS formed in a core-collapse SN occurring in a tight binary collides with its companion star after receiving a fortuitously directed kick (see also E. S. Phinney & B. M. S. Hansen 1993; R. Hirai & P. Podsiadlowski 2022). In this scenario, an extended CSM may also be expected from L2 mass loss during the case BB mass transfer that is predicted to occur in a wide range of orbits, particularly at low metallicity (E. Laplace et al. 2020; J. Klencki et al. 2022).[12] The amount of mass lost during case BB/BC mass transfer is likely to be significantly less ($\lesssim 1\,M_\odot$) than in the models considered in this paper. It might also occur over a somewhat longer timescale (∼10 kyr) prior to core collapse (A. Ercolino et al. 2024, 2025), unless the L2 outflows only become significant around the peak of $\dot{M}$ of case BB/BC interaction ($\lesssim$1 kyr). Traditional signatures of the coincident stripped-envelope SN would also need to be effectively "hidden" in this scenario. For example, $^{56}$Ni generated by the core collapse places a floor on the optical light curve around 100 days from $^{56}$Co decay; by comparison, little or no radioactive material is expected to be synthesized in the HeC merger outflows (M22). More broadly, there are general reasons to favor BH accretors as the engines of LFBOTs (Section 4.6).

### 4.2. LFBOT Environments

Existing metallicity estimates for LFBOT host galaxies place them broadly in the subsolar regime, $Z \simeq 0.2$–$0.8\,Z_\odot$, with no strong evidence for extremely metal-poor environments ($Z < 0.1\,Z_\odot$). For example, the hosts of AT2018cow (∼$0.5\,Z_\odot$; D. A. Perley et al. 2019; J. D. Lyman et al. 2020), AT2018lug (∼$0.5\,Z_\odot$; A. Y. Q. Ho et al. 2020), CSS161010 (∼0.3–$0.6\,Z_\odot$; D. L. Coppejans et al. 2020; C. P. Gutiérrez et al. 2024), AT2020mrf (∼0.3–$0.8\,Z_\odot$; Y. Yao et al. 2022b), and AT2023fhn (∼0.3–$0.6\,Z_\odot$; A. A. Chrimes et al. 2024b) all indicate moderately subsolar values. These measurements are based on a mix of emission-line diagnostics, stellar mass–metallicity scaling, and SED modeling, so systematic uncertainties are nonuniform across the sample. More recently, A. E. Nugent et al. (2026) reanalyzed LFBOT host galaxies in a homogeneous way, finding that they are more strongly star-forming than the hosts of ordinary core-collapse SNe but less extreme than those of hydrogen-poor SLSNe, as well as more metal-poor than Type II and Ibc SN hosts while remaining more metal-rich than the hosts of Type I SLSNe and long gamma-ray bursts (GRBs). Overall, these results place LFBOT environments in an intermediate regime in both star formation activity and metallicity among massive-star transients.

The tendency for subsolar metallicities among LFBOT hosts is consistent with our population model predictions (right panel of Figure 10). The preference for low-$Z$ environments in the DDI channel arises primarily from two effects. First, metal-poor massive stars more commonly form BHs than NSs owing to their weaker stellar winds. Second, the parameter space for BH +star systems that evolve into BH+HeC mergers expands the lower the metallicity (Figures 15 and 16). The latter trend follows from detailed binary MESA models and reflects the fact that low-metallicity MS stars are more compact, which favors binary interaction with post-MS donors that possess a HeC (e.g., J. Klencki et al. 2020). The size of massive MS stars remains uncertain in 1D stellar models, being sensitive in particular to internal mixing and envelope inflation (N. Langer 2012). To gauge this uncertainty, we calculated additional MESA grids with reduced core overshooting ($\sigma_{\rm ov} = 0.18$; see Section 2 for details). This variant shifts the predicted peak host metallicity from $\simeq$0.2 to $\simeq$0.4 $Z_\odot$, yet it still favors subsolar environments. Consistently, the parameter window for BH+MS mergers tends to increase with metallicity, with such events potentially giving rise to longer, less luminous FBOT-like transients (Section 4.5).

[11] I. Linial & E. Quataert (2024) propose a mechanism to create an extended CSM even around the subset of TDEs with circular orbits (see also C. Xin et al. 2024).

[12] Mass loss driven by waves excited during final stages of nuclear burning decades prior to explosion (S. C. Wu & J. Fuller 2022) could also produce an extended CSM; however, it may be unlikely for this wave-driven CSM to extend to the large radii $\sim 10^{17}$ cm required by LFBOTs (J. Fuller, private communication; Figure 4).

### 4.3. Progenitor Appearance

A unique prediction of DDI scenarios for LFBOTs regards the appearance of the progenitor binary system, which might be detectable through pre-imaging observations. Just prior to dynamical instability, the mass transfer rate onto the BH can approach $\approx 0.1\,M_\odot$ yr$^{-1}$ (Figure 15, right panel), corresponding to $\gtrsim 10^4 - 10^5 \dot{M}_{\rm Edd}$. Although only a portion of this mass is likely to reach the central BH (once disk outflows are taken into account; e.g., R. D. Blandford & M. C. Begelman 1999), the resulting accretion luminosity is likely to be highly super-Eddington, potentially powering ultraluminous X-ray (ULX) emission for observers aligned with the binary orbital axis. The most luminous known ULX sources with $L_{\rm X} \gtrsim 10^{42}$ erg s$^{-1}$ can be observed to distances $d \gtrsim 100$ Mpc (e.g., D. J. Walton et al. 2022) exceeding the nearest LFBOTs such as AT2018cow ($d \simeq 60$ Mpc); unfortunately, $\sim$all-sky complete ULX catalogs to this distance are currently lacking, making such an association difficult in practice. For example, a bright ULX source with $L_{\rm X} = 10^{41}$ erg s$^{-1}$ at a distance of 90 Mpc corresponds to an X-ray flux of $F_{\rm X} \approx 10^{-13}$ erg s$^{-1}$ cm$^{-2}$, which was achieved for no more than 10% of the sky in the XMM-Newton Slew Survey (in soft X-rays; R. D. Saxton et al. 2008). For moderate ULX luminosities $L_{\rm X} \approx 10^{39}$–$10^{40}$ erg s$^{-1}$, this drops down to $F_{\rm X} \approx 10^{-11}$ to $10^{-12}$ erg s$^{-1}$ cm$^{-2}$, which requires deep X-ray imaging (e.g., $\sim$1.6% of the sky in the Chandra Source Catalog; I. N. Evans et al. 2024). Combining scarce sky coverage with the requirement of viewing angles nearly aligned with the binary axis, no more than a few percent of LFBOTs generated by DDI are likely to have detectable X-ray progenitors.

The BH accretion-powered jets from the short-lived mass transfer phases leading up to DDI also inflate energetic compact nebulae ("hypernebulae"), akin to those seen to encase Galactic ULX sources and microquasars, which may be more readily detected through their synchrotron emission by wide-field radio surveys (N. Sridhar & B. D. Metzger 2022). Such nebulae can in principle grow to be several parsecs across, in which case their nebular emission could remain visible as a steady radio source, even after the transient emission from the LFBOT has faded.

As a result of the dusty quasi-spherical outflows from the binary (Section 3.2), emission from the star−BH system may be largely obscured from most viewing angles, resulting in reprocessing of the binary luminosity through the much larger dust photosphere radius $r_{\rm ph} \gtrsim 10^{16}$ cm (Figure 8). The binary emission will therefore be pushed into the mid-infrared, similar to other transients with dust-obscured progenitors (e.g., SN 2008S; J. L. Prieto et al. 2008). To illustrate this, in Figure 11 we plot the H-R diagram positions of donor stars from our binary models taken 1 yr before the BH+HeC merger (blue points) and shift them to their photosphere radius $r_{\rm ph}$ derived from the CSM density profile of each binary (Equation (A13)). The luminosity of the central source hidden by the dust could take on any value between the luminosity of the star ($L_\star$; red points) and the combined luminosity of the star and the BH accretion disk ($L_\star + L_{\rm disk}$; green points), depending on what fraction of $L_{\rm disk}$ is directed toward the observer. Here we take $L_{\rm disk} = L_{\rm Edd}(1 + \ln(\dot{m}_{\rm Edd}))$ (e.g., N. I. Shakura & R. A. Sunyaev 1973), where $L_{\rm Edd}$ is the Eddington luminosity of the BH and $\dot{m}_{\rm Edd} \equiv \dot{M}_{\rm L1}/\dot{M}_{\rm Edd}$ is the Eddington ratio of the mass transfer rate. As a result of their very red colors (likely peaking in the mid-infrared) and the current lack of wide-field surveys at these wavelengths, the progenitors of LFBOTs will likely be challenging to discover even in nearby galaxies.

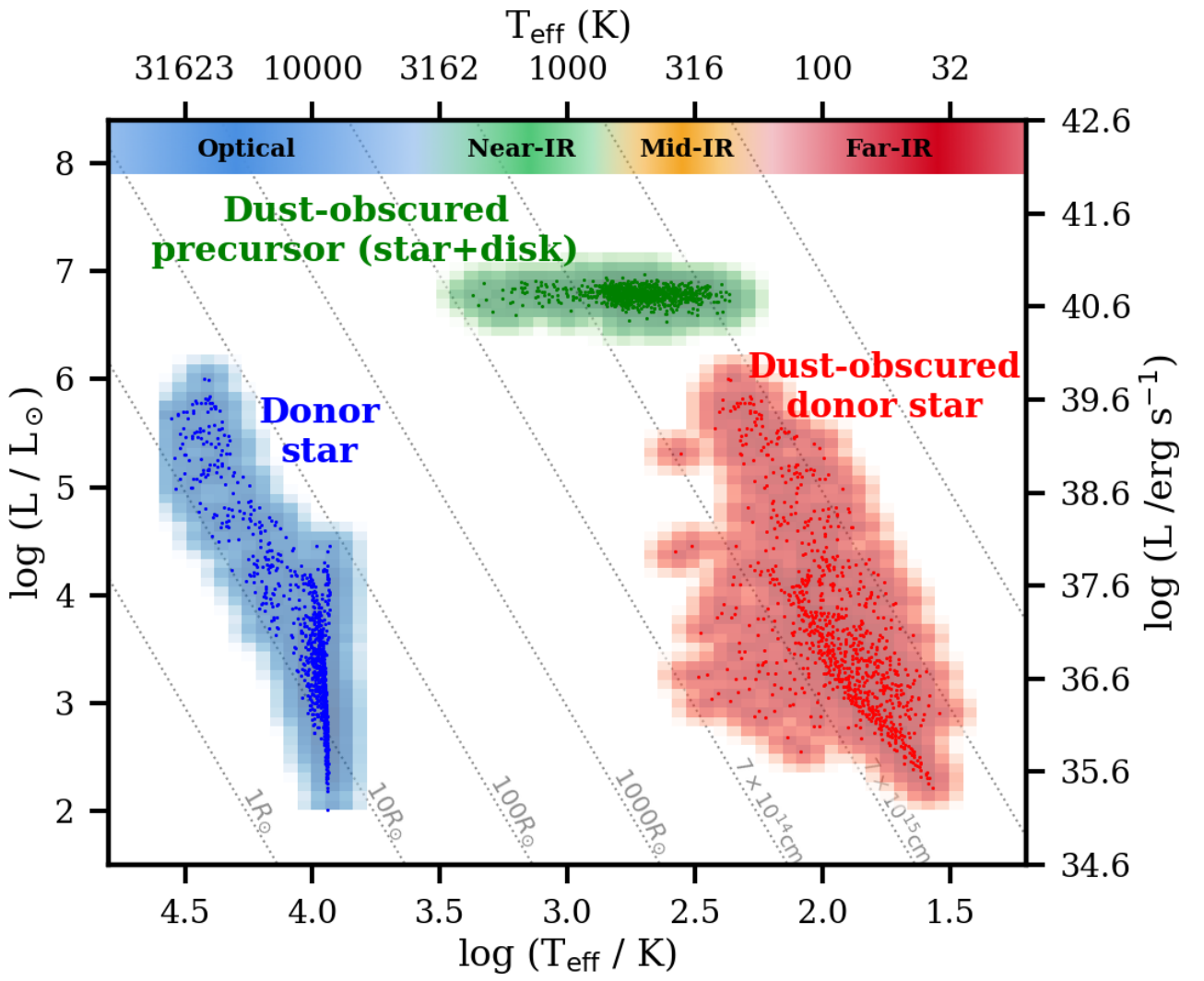

**Figure 11.** The LFBOT progenitors are likely to be dust obscured and therefore to peak in the mid-infrared. Here we plot the H-R diagram positions of unobscured donor stars taken from our binary models 1 yr before the merger (blue points) and shift them to the photosphere radius $r_{\rm ph}$ estimated from their CSM density profile (red points). Green points combine the luminosity of the star with that of the super-Eddington disk $L_{\rm disk} = L_{\rm Edd}(1 + \ln(\dot{m}_{\rm Edd}))$.

A related question is whether LFBOT progenitor binaries might be present in the Milky Way or nearby galaxies. From the measured LFBOT rate of $\lesssim$0.6% of the core-collapse SN rate (A. Y. Q. Ho et al. 2023b) and the Milky Way core-collapse rate of 1/60 yr (K. Rozwadowska et al. 2021), we estimate a Galactic LFBOT rate $R \lesssim 10^{-4}$ yr$^{-1}$. Given also the lifetime $t_{\rm ST} \lesssim 10$ kyr of the stable mass transfer phase (Section 3.1), we would expect only $\sim R t_{\rm ST} \lesssim 1$ such sources in our Galaxy. Indeed, most known high-mass BH X-ray binaries either are undergoing wind-fed accretion or possess much lower mass transfer rates (longer mass transfer durations) than the immediate LFBOT progenitors we consider. Nevertheless, systems such as pulsating ULX source P13 in the nearby Sculptor group galaxy NGC 7793, in which a massive B9 Ia supergiant donor is feeding mass onto an NS companion (M. Bachetti et al. 2014; C. Motch et al. 2014), may evolve into systems similar to those we have modeled.

### 4.4. Connection to Ultralong Gamma-ray Bursts?

Although the late-time X-ray emission from the BH accretion funnel is likely viewing angle dependent (e.g., A. Y. Q. Ho et al. 2022a; Y. Yao et al. 2022a), most emission signatures we have described thus far are quasi-isotropic and hence could be seen by observers located off the binary orbital/disk axis. However, if the BH is rapidly spinning, then it may also power a tightly collimated ultrarelativistic jet via the R. D. Blandford & R. L. Znajek (1977; BZ) process. This has the potential to generate luminous nonthermal emission visible for select observers within the (likely narrow) opening angle of the BZ jet, similar to the geometry of GRBs (C. L. Fryer & S. E. Woosley 1998).

The maximal efficiency of the relativistic jet can be estimated as $\eta_{\rm BZ} \approx 0.1(a_{\bullet}/0.5)^2$ (e.g., A. Tchekhovskoy et al. 2011), where $a_{\bullet}$ is the dimensionless spin of the BH. Substituting $\eta = \eta_{\rm BZ} \sim 0.1$ in Equation (21), we see that a moderately spinning BH could power a jet of luminosity $L_{j,0}(\eta = 0.1) \sim 10^{47}$–$10^{48}$ erg s$^{-1}$ and duration $t_{\rm visc,0} \sim 10^4$–$10^5$ s (Equation (13); Figure 6). Accounting also for the boost in observed luminosity due to relativistic beaming effects, these predicted X-ray/gamma-ray properties for on-axis views broadly overlap those of so-called "ultralong GRBs" of duration $\sim 10^3$–$10^5$ s (e.g., B. Gendre et al. 2013; A. J. Levan et al. 2014). Indeed, the merger of a HeC with an NS was proposed as a model for the ultralong GRB 101225A (the so-called "Christmas burst"; C. C. Thöne et al. 2011).

A recent extreme member of the ultralong class, GRB 250702B (A. J. Levan et al. 2025; B. P. Gompertz et al. 2026; B. O'Connor et al. 2025; J. Carney et al. 2025; E. Neights et al. 2026), produced multiple distinct flares of luminous gamma-ray emission over several hours, emitting an isotropic energy $E \sim 10^{54}$ erg. However, the beaming-corrected energy is likely significantly smaller, $\lesssim 10^{52}$ erg, consistent with $\lesssim 0.1\,M_{\odot}$ of accreted mass for a jet of efficiency $\eta = 0.1$. If this event were a HeC–BH merger (e.g., as proposed by E. Neights et al. 2026), the soft X-ray precursor emission seen 24 hr before the gamma-ray trigger (e.g., B. O'Connor et al. 2025) could arise from the jet shock breakout phase from the nearby CSM (Equation (22)). Unfortunately, because of the bright afterglow and dust extinction in the host galaxy, it is challenging to detect LFBOT emission following GRB 250702B (B. P. Gompertz et al. 2026). Nevertheless, prospects may be better for detecting the dust echo signal in the infrared if an extended CSM surrounds the source (Section 3.6).

Although an ultralong GRB should be included as possible counterparts to HeC–BH mergers, we caution that the opposite is not necessarily true: the requirements of strong magnetic fields and high BH spin may preclude all LFBOT systems from generating powerful relativistic jets. For example, the modest mass accreted by the BH following the HeC disruption, $M_{\rm acc} \sim 0.1\,M_{\odot}$ (Equation (17); Figure 6), precludes significant accretion-induced spin-up, requiring high BH spin prior to the disruption.

### 4.5. Longer-duration Transients from MS+BH Mergers

Our discussion of LFBOT progenitors has focused on systems in which the donor is an evolved star with a well-defined HeC. However, for systems with shorter initial periods, DDI can occur when the donor is still on the MS (BH+MS; blue region in Figure 15). Although the initial plunge of the BH into the donor and the ejection of its outer envelope will likely proceed similarly to those in the evolved case (Section 3.2), the central density of the star the BH encounters as it inspirals, $\rho_c$, will be $\sim$2–3 orders of magnitude smaller than for a HeC. As a result, the disk that forms around the BH will be larger, $R_{d,0} \propto \rho_c^{-1/3}$, and accrete onto the BH over a timescale $t_{\rm visc,0} \propto R_{d,0}^{3/2} \propto \rho_c^{-1/2}$ (Equation (13)), at least an order of magnitude longer than the HeC case.

Thus, rather than the engine activity peaking on a timescale of less than a day (Figure 6), the bulk of the accretion power will be released over a week or longer, similar to the duration of an ordinary SN instead of an FBOT. Furthermore, as a result of both the longer engine duration and the greater wind mass loss from the larger disk (Equation (15)), the peak engine luminosity will typically be smaller by several orders relative to the FBOT case, e.g., $L_{\rm pk} \sim L_{j,0}/100 \sim 10^{44}$ erg s$^{-1}$, still potentially in the range necessary to power luminous SNe or even SLSNe. However, given the weaker power of the jet and the presence of a larger and more massive CSM than in the HeC merger case, it is less clear whether the fast jet from the inner disk winds will be able to successfully break out of the CSM in these systems. Nevertheless, a significant portion of the disk outflow energy will likely be thermalized upon colliding with the surrounding CSM and radiated on a timescale comparable to the ejecta diffusion timescale (J. Dexter & D. Kasen 2013). Our scenario thus predicts a class of DDI-instigated engine-powered hydrogen-rich SNe, which may contribute to the populations of "interacting" Type IIn and/or Type II SLSNe (see also R. A. Chevalier 2012; S. L. Schrøder et al. 2020).

### 4.6. Neutron Star–HeC Mergers

Although we have focused on mergers between evolved massive stars and BHs, a qualitatively similar binary evolution channel can occur for NS accretors (e.g., N. Soker et al. 2019; A. Grichener 2023). Due to a more extreme mass ratio of such systems, the dynamical instability would develop more quickly following the onset of mass transfer but would nonetheless be delayed by hundreds of years (e.g., N. Blagorodnova et al. 2021), producing a somewhat more compact and less massive CSM than predicted in this work for BH accretors (Figure 4). However, the physical picture that leads to the final accretion-driven explosion may more closely resemble that of the NS spiraling into the center of the HeC, rather than it tidally disrupting the core into a disk all at once. Furthermore, the solid surface of an NS results in the accreted gas forming an extended quasi-spherical gas envelope (R. A. Chevalier 1993). By increasing the inner radius of the accretion disk $R_{\rm in}$, this limits the released accretion power $\propto G\dot{M}/R_{\rm in}$ (e.g., L. Combi et al. 2025) relative to the BH case. This suggests that NS–HeC mergers are likely to be less energetic than BH–HeC mergers. Another potential difference relates to the final fate of the system, i.e., whether the entire HeC is accreted or ejected in winds, or whether a portion of the core is able to find hydrostatic balance forming a K. S. Thorne & A. N. Zytkow (1977) envelope around the NS. The disklike X-ray and UV spectra of AT2018cow observed several years after the explosion exclude a spherical remnant and favor a BH accretor (A. Inkenhaag et al. 2023; G. Migliori et al. 2024; M. Winter-Granic & E. Quataert 2025).

## 5. Conclusions

We use MESA binary stellar evolution models coupled with analytic estimates and rapid population synthesis to explore the properties of luminous transients produced by the merger between a BH and the HeC of a massive star, following a DDI that terminates an extended phase of stable mass transfer. This builds on and broadly supports an earlier model for FBOTs developed by B. D. Metzger (2022), who, however, instead focused on HeC/BH mergers that take place with a similarly long delay after a common envelope, starting from a wider initial binary (see also N. Soker et al. 2019; S. L. Schrøder et al. 2020).

The stable mass transfer phase ejects several solar masses of slowly expanding material via outflows from L2, generating a dense extended circumstellar medium (CSM) extending to $\sim 10^{17}$ cm. The predicted CSM density profile broadly matches the inferred environments surrounding LFBOTs (Figure 4). However, we acknowledge uncertainties related to the quantity and geometry of the L2 outflows and their relative importance compared to the BH accretion disk winds along the polar axis of the binary likely probed by the radio/millimeter data. The subsequent dynamical plunge of the BH into the partially stripped donor ejects the remaining stellar envelope at $\sim 10^2 - 10^3$ km s$^{-1}$, creating a compact He-rich CSM ($\gtrsim 10^{13}$ cm), providing a source of shock interaction for the explosion to follow. K26 show that this late dynamical instability occurs for donor star radii spanning a relatively narrow range around $\sim 10 - 30\,R_\odot$ (orbital periods of days; Figure 3), contributing to relatively uniform predictions for the CSM properties across a wide range of initial binaries. Even if this dynamical phase is more gradual than we envision, a compact medium can also be produced by the earliest stages of the HeC merger phase to follow (M22).

Tidal disruption of the HeC forms a thick accretion disk around the BH achieving highly super-Eddington peak accretion rates $\gg 10^{-4}\,M_\odot$ s$^{-1}$, driving both fast ($v \gtrsim 0.3c$) and slow ($v \sim 5000$ km s$^{-1}$) outflows. The jet luminosity reaches $L_j \sim 10^{46}$ erg s$^{-1}$ and powers shock interaction and reprocessing emission broadly consistent with the optical/UV light curves of LFBOTs, including short rise times of a few days set by photon diffusion through the completely low-mass polar region and high peak luminosities ($\gtrsim 10^{44}$ erg s$^{-1}$).

As the CSM environment surrounding the disk is cleared, X-rays from the polar funnel become more readily visible and carry a larger fraction of the engine's power. A portion of the UV/optical luminosity can also be powered by the collision between the slow disk outflows and the even slower $\lesssim 10^3$ km s$^{-1}$ compact CSM. Because the CSM mass from the stripped envelope sometimes exceeds that of the disk outflows, the latter ejecta can be appreciably decelerated by this interaction, particularly in the binary plane. This might create ejecta components that emit narrow lines centered close to zero velocity, such as those seen in optical spectra of AT2024wpp (N. LeBaron et al. 2026). Delayed fallback accretion of this strongly shock-decelerated ejecta onto the central BH might offer an explanation for the late X-ray flare observed from AT2024wpp roughly a month after the explosion (J. Liu et al. 2026).

The disk continues to spread viscously, over several years expanding to large radii $\gtrsim 30\,R_\odot$ as a result of the high angular momentum of the disrupted binary (M22; A. Inkenhaag et al. 2023; G. Migliori et al. 2024; M. Winter-Granic & E. Quataert 2025). Depending on the details of the disk's thermal evolution, this could result in the X-ray light curve steepening when the accretion rate drops enough that the outer disk becomes sub-Eddington and geometrically thin, as well as a potential flattening of the light curve somewhat later after the inner disk closest to the BH becomes sub-Eddington and its radiative efficiency rises (M22).

Shock interaction of the fast jet with the extended CSM generates luminous radio/submillimeter synchrotron emission peaking over several months (a unique characteristic of multiple LFBOTs), corresponding to the propagation time of the fast outflow to the edge of the CSM shell at $\sim 10^{16.5}$ cm. This size of the extended CSM and the steep density drop at $>10^{16}$ cm are natural predictions of binary models that stem from the thermal timescale of massive stars and orbital velocity scaling for the outflow speed. The same slow binary outflow likely formed dust prior to the merger, which was suddenly destroyed by the UV light of the rising transient. However, before dust destruction occurred, enough energy was reprocessed into the infrared to give rise to an "echo" lasting weeks or longer.

Stable mass transfer has received significant interest in recent years as a formation channel for tight binary systems capable of contributing to the population of compact binary mergers detected by LIGO/Virgo/Kagra (E. P. J. van den Heuvel 2017; S. S. Bavera et al. 2021; C. J. Neijssel et al. 2021; P. Marchant et al. 2021; M. Gallegos-Garcia et al. 2021; A. Olejak et al. 2021; L. A. C. van Son et al. 2022; K26). However, for every such binary that successfully detaches at a sufficiently tight orbit to potentially form a gravitational-wave source, a comparable fraction with somewhat different initial conditions will undergo mergers of the type studied here (Figure 9). Within our interpretation, LFBOTs are thus luminous signposts for "failed" gravitational-wave sources. Similar failures that occur in more compact initial binaries with MS donors instead predict a class of longer-duration transients, which may present as luminous Type II SNe (Section 4.5).

## Acknowledgments

We thank the anonymous referee for useful comments and suggestions that helped to improve the paper. We are grateful to Ashley Chrimes for sharing data on the environmental densities of LFBOTs and to Martyna Chruslinska for sharing metallicity-specific star formation models. J.K. acknowledges helpful conversations with Jim Fuller, Jing-Ze Ma, and Lida Oskinova. B.D.M. acknowledges helpful conversations with Kishalay De, Natalie LeBaron, Nayana Alakkal Jagadeeswaran, Viraj Karambelkar, Raffaella Margutti, and Daniel Perley. B.D.M. acknowledges support from NASA ATP (grant No. 80NSSC22K0807), the Fermi Guest Investigator Program (grant No. 80NSSC24K0408), and the Simons Foundation (grant No. 727700). The Flatiron Institute is supported by the Simons Foundation.

## Appendix A
## Additional Model Details

### A.1. Condition for Shell Crossing in the L2 Outflow

Our construction of the CSM density profile assumes that parcels lost through L2 propagate ballistically after ejection with velocity $v_{\rm L2}$ that evolves slowly with the binary orbit. This requires a monotonic mapping between retarded time $t_{\rm ret}$ (measured before the onset of the dynamical instability) and radius. The radius reached by a parcel ejected at $t_{\rm ret}$ is

$$r(t_{\rm ret}) = v_{\rm L2}(t_{\rm ret})\, t_{\rm ret}. \tag{A1}$$

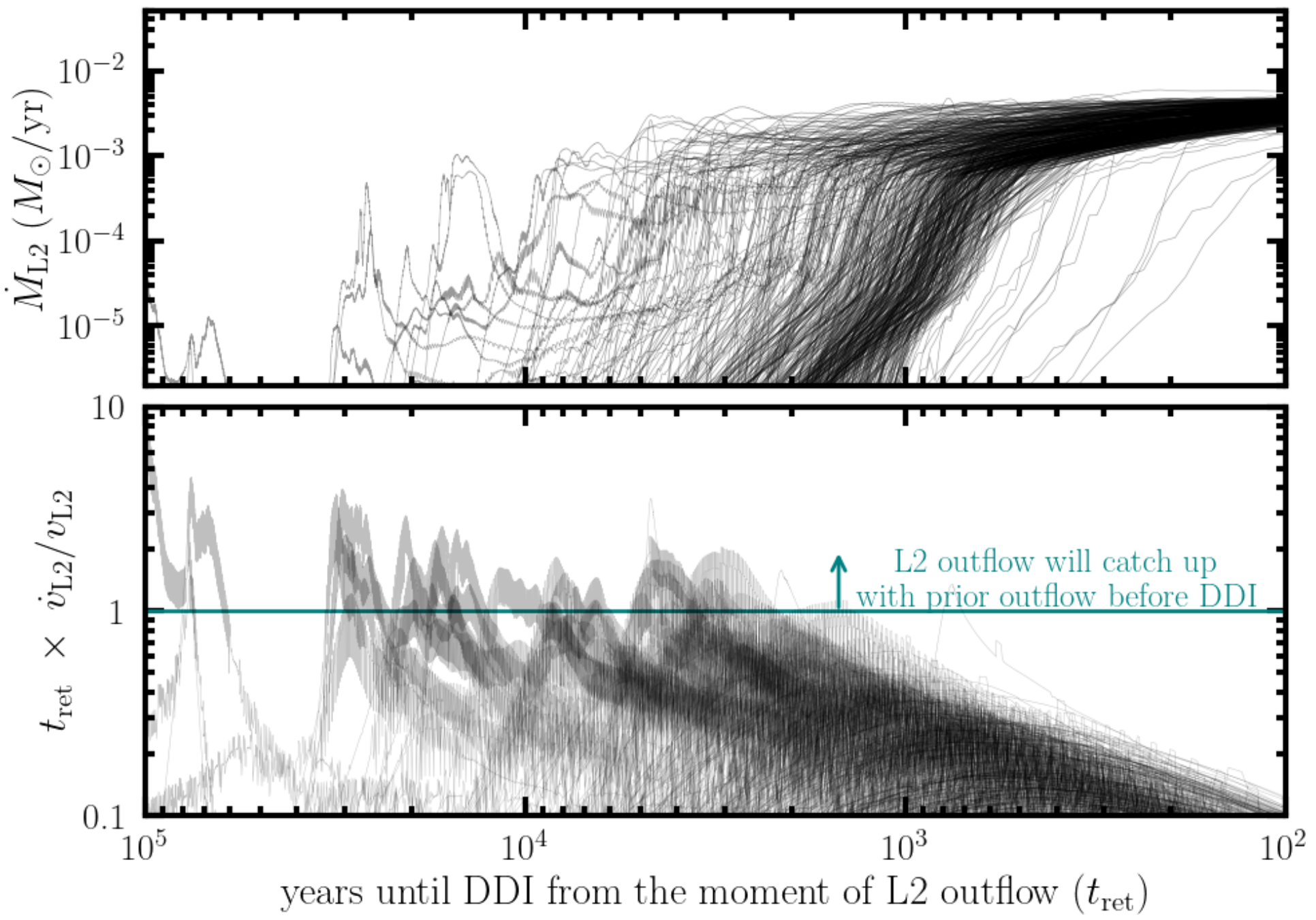

**Figure 12.** Top panel: L2 mass-loss rate as a function of retarded time before the onset of the dynamical instability. Short-timescale fluctuations reflect numerical behavior of the MESA mass transfer solver, while longer-timescale variations arise from the structural response of the donor star to mass loss (e.g., J. Klencki et al. 2022). In a small subset of systems, temporary drops in $\dot{M}_{\rm L2}$ at $t_{\rm ret} \gtrsim 10^4$ yr occur when RLOF begins near the end of the MS and resumes after post-MS expansion. Bottom panel: overtaking factor $t_{\rm ret}\dot{v}_{\rm L2}/v_{\rm L2}$ for the same models. Values above unity indicate that later ejecta overtake earlier outflows. This occurs only in a small subset of systems with long pre-DDI interaction phases.

Shell crossing occurs when this mapping ceases to be monotonic, i.e., when $\frac{dr}{dt_{\rm ret}} = 0$. Differentiating gives $\frac{dr}{dt_{\rm ret}} = v_{\rm L2} + t_{\rm ret}\,\dot{v}_{\rm L2}$, so that later ejecta overtake earlier material whenever

$$t_{\rm ret}\,\frac{\dot{v}_{\rm L2}}{v_{\rm L2}} > 1. \qquad \text{(A2)}$$

Figure 12 shows this overtaking factor computed from our binary evolution models. In nearly all systems the condition above is not satisfied, indicating that the ballistic approximation remains valid throughout most of the pre-DDI mass transfer phase.

Only a small subset of binaries with unusually long ($\gtrsim 10^4$ yr) pre-DDI interaction phases reach $t_{\rm ret}\,\dot{v}_{\rm L2}/v_{\rm L2} > 1$. In these cases, compression and shocks may develop at large radii ($\sim 10^{17}$–$10^{18}$ cm), where interaction with previously ejected material or the ambient interstellar medium could modify the outer CSM structure. Because such systems are rare and the effect occurs only at large distances, it does not affect the characteristic CSM profile shown in Figure 4 or its comparison with LFBOT environments.

### *A.2. Nearby CSM from Dynamical Envelope Removal*

Following DDI, the BH plunges into the remaining envelope with mass $M_{\rm env}$ that is by now enriched in helium. What follows is a dynamical inspiral that is similar to a traditional CE phase. Because the remaining envelope is fully radiative and without as strong a density gradient at the core–envelope boundary as in the case of convective envelopes, we expect the inspiral to continue all the way until the BH merges with the HeC. The energy released by the inspiral ($\Delta E$) will eject part of the envelope $M_{\rm nCSM}$ with a binding energy $E_{\rm bind} < \Delta E$ to form the nearby He-enrich CSM. In Equation (8), we estimate the average velocity of this ejecta $v_{\rm nCSM}$ by assuming that $M_{\rm nCSM} \approx M_{\rm env}$ and neglecting the binding energy term, such that

$$v_{\rm nCSM} \approx \left(\frac{2\Delta E}{M_{\rm env}}\right)^{1/2}. \qquad \text{(A3)}$$

Here we explore a more detailed approach by integrating through the envelope structure taken from a MESA binary model at the point of DDI. Following the BH inspiral from the surface ($r_{\rm surf}$) inward ($r_{\rm in}$), we calculate the binding energy of the outer part of the envelope $r > r_{\rm in}$ as

$$E_{\rm bind}(r_{\rm in}) = -\int_{r_{\rm in}}^{r_{\rm surf}} \left(-\frac{GM(r)}{r} + u(r)\right)\frac{dm}{dr}dr, \qquad \text{(A4)}$$

where $M(r)$ is the star mass enclosed within radius $r$ and $u(r)$ is the specific internal energy at radius $r$ that we assume to fully contribute to the unbinding of the envelope. We compare the binding energy with the energy input from the inspiral $\Delta E$ that we estimate as

$$\Delta E(r_{\rm in}) = -\frac{GM_\bullet M}{2a_{\rm DDI}} + \frac{GM_\bullet M(r_{\rm in})}{2a(r_{\rm in})}, \qquad \text{(A5)}$$

where $M$ and $a_{\rm DDI}$ are the donor mass and orbital separation at the onset of DDI, respectively. Based on the excess energy $\Delta E - E_{\rm bind}$ at a given $r_{\rm in}$, we estimate the velocity of the unbound part of the envelope as

$$v_{\rm nCSM}(r_{\rm in}) = \left[\frac{2(\Delta E - E_{\rm bind})}{M_{\rm env}(r_{\rm in})}\right]^{1/2}, \qquad \text{(A6)}$$

where $M_{\rm env}(r_{\rm in})$ is the mass of the envelope within $r > r_{\rm in}$.

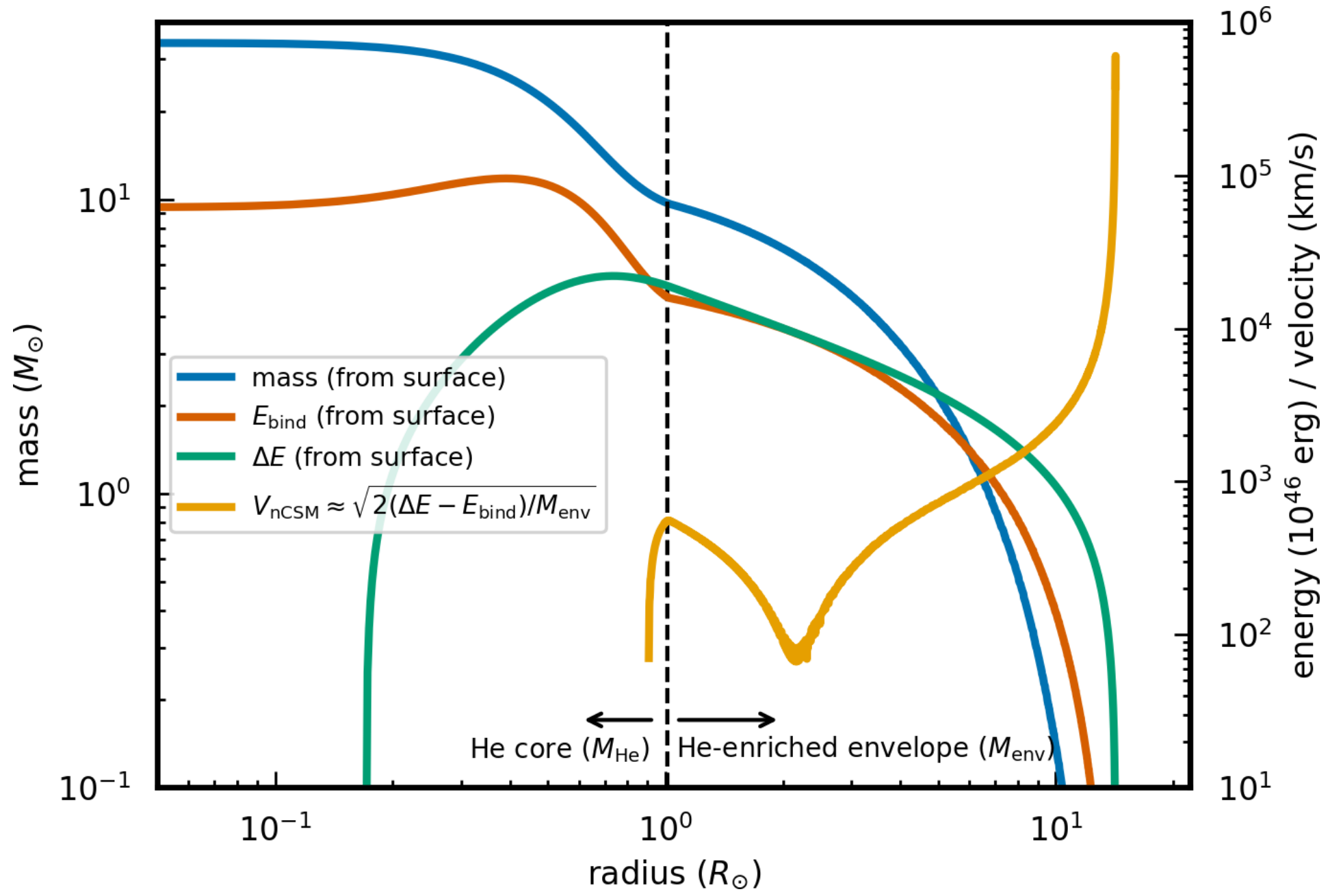

**Figure 13.** The internal profile of a 34 $M_\odot$ donor at the moment of DDI, when its 10 $M_\odot$ BH companion begins to spiral in throughout the remaining He-enriched envelope. The envelope extends from $r_{\rm core} \approx 1\,R_\odot$ to $r_{\rm surf} \approx 14\,R_\odot$. The outer envelope layers are loosely bound, such that the energy input from the inspiral ($\Delta E$) is much greater than the binding energy ($E_{\rm bind}$), leading to fast outflows with $v_{\rm nCSM} \sim 1000$ km s$^{-1}$ that produce a nearby He-rich CSM before the BH eventually plunges into the HeC.

Figure 13 shows $E_{\rm bind}$, $\Delta E$, and $v_{\rm nCSM}$ as a function of radius within the envelope $r_{\rm in}$, for an example binary model with a $\approx 34\,M_\odot$ donor. The envelope extends out to $r_{\rm surf} \approx 14\,R_\odot$ and down to the core–envelope boundary at $r_{\rm core} \approx 1\,R_\odot$. Comparing $E_{\rm bind}$ and $\Delta E$, we find that in the outer half of the envelope ($r \gtrsim 4\,R_\odot$) the assumption of $\Delta E \gg E_{\rm bind}$ is valid and results in $v_{\rm nCSM} \approx 1000$ km s$^{-1}$, broadly supporting our estimate in Equation (6). However, the innermost envelope layers are more tightly bound and hence are likely to be ejected with a comparatively lower velocity $\sim$100 km s$^{-1}$, forming a more compact inner region of the nearby CSM. The uncertainties in the dynamical envelope removal process will affect the total mass of the nearby CSM mass and consequently the predicted peak optical/UV luminosity from the jet−nearby CSM interaction. In our modeling of the transient emission, we crudely account for this by varying that portion of the nearby CSM mass, $M_{\rm sh}$, that interacts with the fast disk outflow (Figure 7).

### A.3. Late-time Decay of Accretion Engine

At late times $t \gg t_{\rm visc,0}$ (Equation (13)), the outer edge of the disk formed from the disrupted HeC will continue to spread outward owing to the redistribution of angular momentum. If the disk outflows carry away only the local specific angular momentum of the disk material, then the outer edge of the disk will grow with time as (e.g., J. K. Cannizzo et al. 1990)

$$R_d \simeq R_{d,0}\left(\frac{t}{t_{\rm visc,0}}\right)^{2/3},\quad t \gg t_{\rm visc,0}. \tag{A7}$$

Over a couple years, the disk will thus spread from its initial radius $R_{d,0} \lesssim R_\odot$ (Equation (12)) by a factor of $\gtrsim$100 (for $t_{\rm visc,0} \lesssim 1$ day) to an outer scale $R_{d,0}(t/t_{\rm visc,0})^{2/3} \gtrsim 30\,R_\odot$, broadly consistent with the large blackbody radius inferred from the UV/optical emission from AT2018cow on a similar timescale (N.-C. Sun et al. 2022, 2023; Y. Chen et al. 2023; A. Inkenhaag et al. 2023).

The accretion rate at radii $r < R_d$ will likewise drop as a power law in time (e.g., B. D. Metzger et al. 2008), viz.,

$$\dot{M} \propto r^p t^{-4(p+1)/3},\quad t \gg t_{\rm visc,0}. \tag{A8}$$

The accretion rate near the outer edge of the disk decays as

$$\dot{M}(R_d) \sim \dot{M}_0\left(\frac{t}{t_{\rm visc,0}}\right)^{-\frac{2(2p+1)}{3}}, \tag{A9}$$

while the accretion rate reaching the BH decays more steeply in time,

$$\dot{M}_\bullet \sim \dot{M}_0\left(\frac{R_{\rm in}}{R_{d,0}}\right)^p\left(\frac{t}{t_{\rm visc,0}}\right)^{-\beta},\quad t \gg t_{\rm visc,0}, \tag{A10}$$

where

$$\beta = \frac{4(p+1)}{3} \approx 2 - 2.7, \tag{A11}$$

and the final equality is for a range $p = 0.5-1$. A more steeply decaying accretion rate (larger $\beta$) is possible if disk outflows carry away greater angular momentum than that of Keplerian rotation (e.g., B. D. Metzger et al. 2008). Taking $R_{\rm in} = 6GM_\bullet/c^2$ and $p = 0.6$ ($\beta = 2.13$), the resulting fast-outflow luminosity obeys

$$L_{\rm jet} \approx \eta\dot{M}_\bullet c^2 \approx 1.5\times10^{45}\ {\rm erg\ s^{-1}}\eta_{-2}\alpha_{0.1}^{-1.13}\left(\frac{M_\bullet}{10\,M_\odot}\right)^{0.03} \times\left(\frac{\theta}{1/3}\right)^{-2.27}\left(\frac{M_{\rm He}}{20\,M_\odot}\right)\left(\frac{R_{d,0}}{R_\odot}\right)^{1.1}\left(\frac{t}{3}\ {\rm days}\right)^{-2.13},\quad t \gtrsim t_{\rm visc,0}, \tag{A12}$$

where we have used Equations (12) and (14). The evolution of $L_{\rm acc}$ is broadly similar in normalization and decay rate to the inferred engine luminosities of LFBOTs (e.g., R. Margutti et al. 2019; N. LeBaron et al. 2026).

### A.4. Dust Echo

For outflow densities typical of those predicted by our models, B. D. Metzger & D. A. Perley (2023) estimate that dust nucleation will occur on radial scales $\gtrsim 10^{15}$ cm from the binary and can result in grain growth up to particle sizes $a \sim 1\ \mu$m. Given the CSM density profiles $n(r)$ predicted by our models, the dust photosphere can be estimated according to

$$\tau_{\rm UV}(>r) = \int_{r_{\rm ph}}^{\infty} n m_p \kappa_{\rm UV} dr = 1, \tag{A13}$$

where

$$\kappa_{\rm UV} = \frac{3}{4}\frac{X_d}{\rho_b a} \approx 20\ {\rm cm}^2\ g^{-1}\frac{X_{d,-2}}{a_{\mu m}} \tag{A14}$$

is the dust opacity in the geometric optics limit to the UV light of the transient, $a_\mu = a/(1\ \mu{\rm m})$, $\rho_b = 3.8$ g cm$^{-3}$ is the bulk density of the dust, and $X_{d,-2} = 10^{-2}X_d$ is the mass fraction of the dust grains. As the transient brightens and its UV luminosity $L_{\rm UV}$ rises, dust will be destroyed by sublimation above the temperature $T_d = T_s \approx 1700$ K (a typical value for silicate grains), at the radius (E. Waxman & B. T. Draine 2000)

$$r_s \approx 4.4 \times 10^{16}{\rm cm}\ a_{\mu{\rm m}}^{-1/2}\left(\frac{L_{\rm UV}}{10^{43}\ {\rm erg\ s^{-1}}}\right)^{1/2} \times \left(\frac{T_s}{1700\ {\rm K}}\right)^{-5/2}. \tag{A15}$$

Once the transient rises to a critical luminosity $L_{\rm UV} = L_{\rm thin}$, dust is sublimated out to its photosphere, i.e., $r_s = r_{\rm ph}$. As long as the transient peaks at a luminosity $L_{\rm pk} \gtrsim L_{\rm thin}$, it will destroy all of the optically thick dust and the bulk of the transient's radiation will not be substantially attenuated. The duration and luminosity of the infrared light echo (from the optically thick dust) can be written in terms of $L_{\rm thin}$ and $r_{\rm ph}$ according to (B. D. Metzger & D. A. Perley 2023; S. Tuna et al. 2025)

$$t_{\rm echo} \simeq \frac{2r_{\rm ph}}{c};\quad L_{\rm echo} \simeq \frac{L_{\rm thin}^{3/2}}{L_{\rm pk}^{1/2}}\left(\frac{t_{\rm pk}}{2t_{\rm echo}}\right), \tag{A16}$$

where $L_{\rm pk}$ (Equation (26)) and $t_{\rm pk} \approx t_d$ (Equation (23)) are the peak luminosity and duration of the transient UV emission, respectively.

## Appendix B
## Population Synthesis Model

The volumetric rate of LFBOTs depends on how frequently BHs pair with massive O or early B-type stars and subsequently experience a DDI. Full stellar evolution calculations of binaries starting from the ZAMS all the way to the final outcomes (such as LFBOTs or binary BH mergers) are still prohibitively expensive for a population study. Instead, we adopt a rapid population synthesis method that evolves ZAMS binaries until the formation of a BH+OB system. The further evolution and the final outcome of these BH+OB binaries are then modeled using our precomputed grids of MESA-detailed binary models (Section B.3). This appendix summarizes the sampling of initial conditions, the rapid binary evolution modeling of systems from ZAMS to the BH+OB stage, and our procedure for mapping the population of BH+OB binaries to final outcomes guided by MESA.

### B.1. Sampling of Initial Conditions

We generate binaries with a primary mass $M_1$ between 10 and 150 $M_\odot$ from the initial mass function of P. Kroupa (2001): $dN/dM = \xi(M) \propto M^{-\alpha_i}$, where $\alpha_1 = 1.3$ for $M/M_\odot \in [0.08, 0.5]$, $\alpha_2 = 2.2$ for $M/M_\odot \in [0.5, 1.0]$, and $\alpha_3 = 2.3$ for $M/M_\odot \in [1.0, 150.0]$. The initial secondary mass $M_2$ is obtained by sampling a mass ratio $q_i = M_2/M_1$ from a uniform distribution between 0.1 and 1.0. Orbital periods are drawn from a distribution $f(\log P) \propto (\log P)^{-0.55}$, derived by H. Sana et al. (2012) based on a spectroscopic monitoring campaign of O-type binaries. While spectroscopic binary detections are the most sensitive for orbital periods of up to several thousand days, we sample $\log(P/{\rm day})$ from a wider range from 0.0 to 7.0 in which massive binaries are known to exist (M. Moe & R. Di Stefano 2017). Orbits are assumed to be circular at the ZAMS.

We adopt a mass-dependent binary fraction following M. Moe & R. Di Stefano (2017), with fixed values at representative masses

| $M_1/M_\odot$ | 0.2 | 0.5 | 1.0 | 3.0 | 10 | 30 |
|---|---|---|---|---|---|---|
| $f_{\rm bin}$ | 0.20 | 0.35 | 0.45 | 0.70 | 0.90 | 0.95 |

$f_{\rm bin}(M_1) =$

and smooth interpolation between points. Outside this range we keep $f_{\rm bin}$ fixed at the boundary values. We neglect higher-order multiples and assume that all nonbinaries are single stars. We simulate five metallicities: $Z/Z_\odot = 0.04$, 0.1, 0.2, 0.4, and 1.0, with $Z_\odot = 0.017$ (N. Grevesse et al. 1996). For each combination of metallicity, accretion efficiency, stability assumption, and SN engine model (see below), we evolve $10^8$ massive binaries.

### B.2. Rapid Binary Evolution from ZAMS to BH+OB

For each massive binary we follow the evolution from the ZAMS until one of the stars (typically the primary) ends its life in core collapse. The computation halts at the moment of a BH formation; we do not model the subsequent interaction between the newborn BH and its companion in the rapid code. If the BH's companion is still on the MS (the majority of cases), the system is recorded as a BH+OB binary and passed to the mapping procedure described in Section B.3.

#### B.2.1. Stellar Evolution

To obtain the evolution of basic stellar properties for both stars in each binary (e.g., masses, radii, wind mass-loss histories), we interpolate along grids of single-star tracks from J. Klencki et al. (2020, 2026). The grids cover 80 ZAMS masses from 10 to 180 $M_\odot$. For masses in between we use smooth interpolation, and we evolve each track to central carbon depletion. The underlying microphysics and mixing assumptions are described in detail in J. Klencki et al. (2020, 2026). Briefly, we adopt step core overshooting with $\alpha_{\rm ov} = 0.35$ (I. Brott et al. 2011), the Ledoux criterion with $\alpha_{\rm sc} = 33$ for semiconvection, and metallicity-dependent stellar winds with mass-loss enhancement near the Eddington limit (see J. Klencki et al. 2022 for implementation details). Tracks are nonrotating, and we neglect tides.

At each time step we update the orbit owing to isotropic wind mass loss and check for RLOF. If a star is stripped of its H-rich envelope during mass transfer and becomes a helium star, we currently do not switch to separate helium-star evolutionary tracks. Instead, from that point to core collapse we estimate further mass loss using the W-R/He-star wide prescription of W. R. Hamann et al. (1995), reduced by a factor of 10 (S. C. Yoon et al. 2006).

### B.2.2. Binary Mass Transfer

Before BH formation, many binaries undergo RLOF. We treat mass transfer as follows. A donor star with a convective envelope experiences stable mass transfer only if the mass ratio $q = M_{\rm accretor}/M_{\rm donor}$ exceeds $q_{\rm crit}^{\rm conv} = 0.8$ at the onset of mass transfer (M. S. Hjellming & R. F. Webbink 1987; K. D. Temmink et al. 2023). Otherwise, a traditional CE phase ensues. We predict the CE outcome using the energy formalism with efficiency parameter $\alpha_{\rm CE} = 1$, including full contribution from the internal and recombination energy as well. We obtain envelope binding energies from our stellar models, calculated as in J. Klencki et al. (2021).

For donors with radiative envelopes we impose two stability requirements: (i) the post-interaction orbit must be wider than $20\,R_\odot$ to avoid a DDI, guided by the fundamental separation limit found in K26; and (ii) the mass ratio at the onset of mass transfer must exceed a critical $q_{\rm crit}^{\rm rad}(P_{\rm RLOF}, \beta)$ to avoid a runaway expansion of the mass-gainer star in response to rapid accretion (C. Schürmann & N. Langer 2024; M. Y. M. Lau et al. 2024; E. Bear & N. Soker 2025). If either condition is violated, the mass transfer episode is treated as a stellar merger. The value of $q_{\rm crit}^{\rm rad}(P_{\rm RLOF}, \beta)$ depends on the orbital periods at the onset of RLOF $P_{\rm RLOF}$ as follows:

$$q_{\rm crit}^{\rm rad}(P_{\rm RLOF}, \beta) = \begin{cases} 0.35, & P_{\rm RLOF} \leqslant 10 \text{ days}, \\ [1ex] q_{10} + w\,(q_{1000}(\beta) - q_{10}), & 10 < P_{\rm RLOF} \leqslant 1000 \text{ days}, \\ [1ex] q_{1000}(\beta), & P_{\rm RLOF} > 1000 \text{ days}, \end{cases}$$

where $q_{10} = 0.35$ and $q_{1000}(\beta)$ depends on the accretion efficiency $\beta$ (see below), whereas $w \equiv \frac{\log(P_{\rm RLOF}) - 1}{2}$ to obtain linear interpolation in $\log P_{\rm RLOF}$ between 10 and 1000 days. The $\beta$ dependence of $q_{1000}(\beta)$ is set by a few constants at small $\beta \leqslant 0.05$ and by linear interpolation in $\beta$ for higher $\beta$ values as follows:

| $\beta$ range | $q_{1000}(\beta)$ |
|---|---|
| $\beta \leq 0.005$ | 0.16 |
| $0.005 < \beta \leq 0.02$ | 0.2 |
| $0.02 < \beta \leq 0.05$ | 0.35 |
| $0.05 < \beta \leq 0.15$ | linear from 0.35 to 0.6 |
| $0.15 < \beta \leq 0.35$ | linear from 0.6 to 0.95 |
| $0.35 < \beta \leq 1.00$ | linear from 0.95 to 1.5 |

For a linear interval with endpoints $(\beta_a, q_a)$ and $(\beta_b, q_b)$, we use $q_{1000}(\beta) = q_a + \frac{\beta - \beta_a}{\beta_b - \beta_a}(q_b - q_a)$. The above expressions were fit to approximate the critical mass ratios for massive radiative donors $q_{\rm crit}^{\rm rad}(P, \beta)$ found in the COMBINE code (C. Schürmann et al. 2025, see their Figures E.6–E.9) based on a detailed study of C. Schürmann & N. Langer (2024). A key feature of this stability model is that the higher the accretion efficiency $\beta$, the less stable the mass transfer.

When the mass transfer is stable, we evolve the orbital separation by integrating $da/dM_{\rm don}$ as a function of the decreasing donor mass (Equation (1) in K26). The orbital evolution depends on the accretion efficiency $\beta$ and the specific angular momentum carried away by the lost matter. For the latter, we consider the "isotropic re-emission" model, i.e., specific angular momentum of the accretor. We consider accretion efficiencies $\beta = 0.1$, 0.3, 0.5, 0.7, and 0.9. Higher values of $\beta$ lead to less stable mass transfer: more binaries merge during the first mass transfer episode, and fewer BH+OB systems form. On the other hand, survivors tend to have OB companions that are several times more massive than the BH, making them promising LFBOT progenitors.

We distinguish between mass transfer events with an MS donor (case A) and those with an evolved donor (cases B/C). For the duration of a mass transfer phase, we assume a thermal timescale for case B/C interactions and a nuclear timescale (until the end of the MS) for case A events. We further assume that post-MS donors (cases B/C) transfer 90% of the hydrogen-rich envelope before mass transfer ceases; this fraction is motivated by detailed mass transfer calculations and ensures that most of the envelope is stripped, leaving the core mass unaffected. For donors still on the MS (case A), we rely on fits to detailed binary models by C. Schürmann et al. (2024), which capture how the donor core retreats in mass during the interaction.

### B.2.3. Remnant Formation and Natal Kicks

We assign remnant masses and types (BH or NS) using three alternative prescriptions: (i) the neutrino-driven explodability criteria of K. Maltsev et al. (2025), which build on E. Laplace et al. (2021) and F. R. N. Schneider et al. (2021) and predict outcomes based on the carbon–oxygen core mass, metallicity, and mass transfer history; (ii) the delayed neutrino-driven explosion model of C. L. Fryer et al. (2012); and (iii) the remnant-mass mapping of S. E. Woosley et al. (2020). In cases where a prescription predicts direct collapse to a BH with negligible mass ejection, we impose a minimum mass loss of 5% of the pre-collapse HeC mass to account for neutrino loses (e.g., A. Vigna-Gómez et al. 2024).

For each prescription, given the remnant mass $M_{\rm remnant}$, we compute the fallback fraction as

$$f_{\rm fb} = \frac{M_{\rm remnant} - M_{\rm protoNS}}{M_{\rm preCC} - M_{\rm protoNS}},$$

where $M_{\rm preCC}$ is the final mass of the collapsing star and $M_{\rm protoNS} = 1.0\,M_\odot$, in analogy to C. L. Fryer et al. (2012). We then draw a natal kick velocity $v_{\rm kick}$ from a Maxwellian with dispersion $\sigma = 265\,{\rm km\,s^{-1}}$ (G. Hobbs et al. 2005) and scale it down by $(1 - f_{\rm fb})$, assuming isotropic directions. If the binary is disrupted owing to the combined effect of sudden mass loss and the kick, it does not contribute to the BH+OB population. The assumption of a fallback-reduced Hobbs distribution results in BH natal kicks spanning a wide range of values, from a few tens to several hundred kilometers per second (left panel of Figure 14). While kicks of up to ∼100 km s$^{-1}$ are loosely supported by the spatial distribution and proper motions of BH low-mass X-ray binaries (e.g., S. Repetto et al. 2012; S. Repetto & G. Nelemans 2015), the magnitude of BH natal kicks remains highly uncertain (S. Popov et al. 2025). Depending on the explosion energy and the amount of fallback, 3D simulations of BH formation predict kicks ranging from several hundred down to only a few kilometers per second in the case of direct collapse or neutrino-driven recoil alone (C. Chan et al. 2018; D. Kresse et al. 2021; H.-T. Janka & D. Kresse 2024; A. Burrows et al. 2024). The recently discovered BH system VFTS 243 provides strong evidence that such weak kicks do sometimes occur, with the natal kick constrained to $\lesssim$10 km s$^{-1}$ (A. Vigna-Gómez et al. 2024). To assess how the assumed BH kick distribution affects the predicted number of LFBOTs in our synthetic population, we also consider a model with constant BH natal kicks of 30 km s$^{-1}$ (Figure 14). In this case, we find that the LFBOT yield increases by a factor of ∼3–5, depending on the adopted explodability criterion.

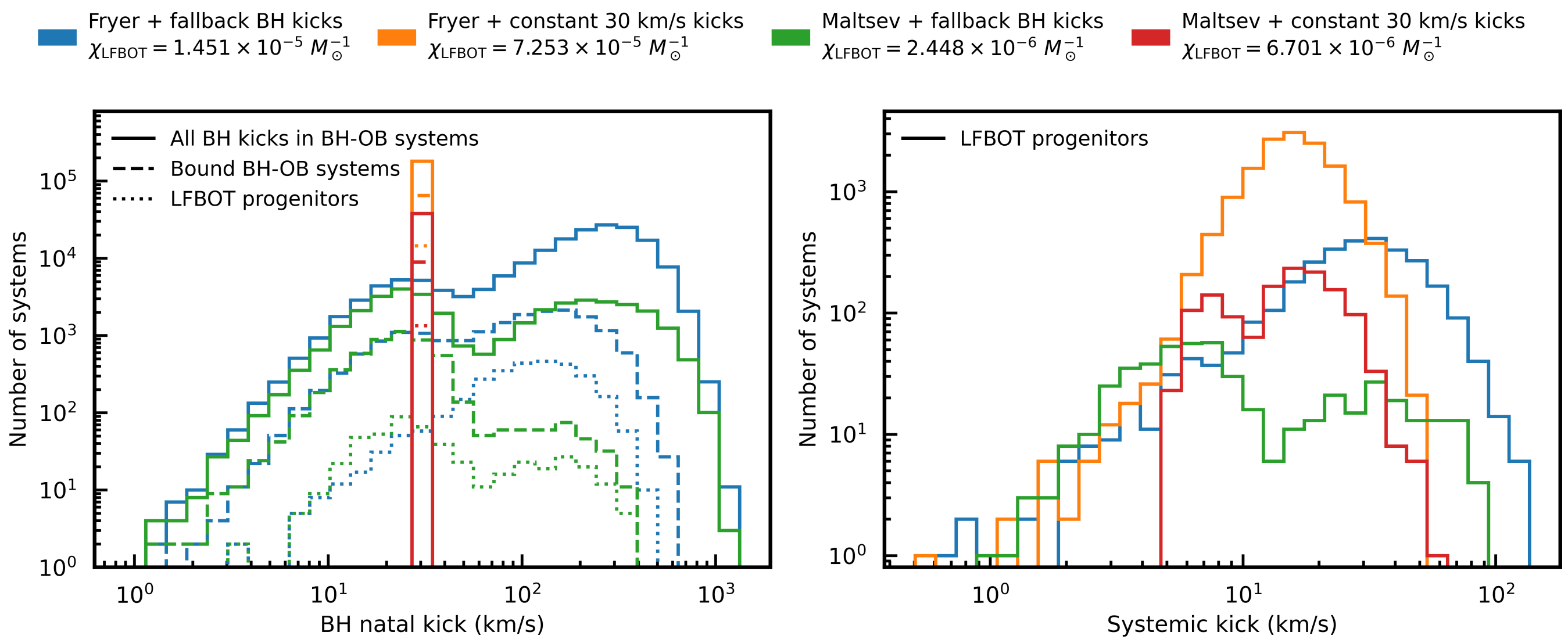

**Figure 14.** Distribution of natal kicks assigned to a BH upon its formation (left) and the resulting systemic kicks of surviving BH−OB systems that evolve to become LFBOTs (right) in our synthetic population model at $0.2\,Z_\odot$ metallicity. Different colors correspond to variations in the assumed BH formation criteria (Fryer delayed or Maltsev) and two assumptions on BH natal kick magnitudes (fallback-reduced Hobbs distribution or constant natal kick of 30 km s$^{-1}$). For each model variation, we calculate the formation efficiency of LFBOTs $\chi_{\rm LFBOT}$, defined as the number of LFBOTs formed per unit of star-formed mass.

### B.3. Mapping BH+OB Systems to the Final Outcomes

Once a BH+OB system forms, we map its properties (mass, mass ratio, orbital period) onto grids of detailed MESA binary models of BH+OB systems. The grids are an extension of those computed in K26 and cover five metallicities (0.04, 0.1, 0.2, 0.4, and 1.0 $Z_\odot$) and two BH masses (5 and 10 $M_\odot$). In addition, we consider two variations in the internal mixing: Type (a) grids adopt step convective core overshooting with $\sigma_{\rm ov} = 0.35$ (following I. Brott et al. 2011) and assume that 25% of the transferred mass is lost through the outer Lagrangian point. Type (b) grids reduce the overshooting parameter to $\sigma_{\rm ov} = 0.18$. The latter choice emulates the possible internal structure of mass gainers: OB companions to BHs often accreted mass in the past and may not have fully rejuvenated (e.g., X.-T. Xu et al. 2025). Lower overshooting yields smaller core masses at a given stellar mass, resulting in more compact MS stars and increasing the fraction of interactions that occur after the donor leaves the MS. DDIs from evolved donors are therefore more frequent in the low-overshooting models, enhancing the predicted LFBOT rate.

The BH+OB grids are organized in the parameter space of varying initial mass ratios $q_{\rm BHOB} = M_\bullet/M_{\rm donor}$ and orbital periods $P_{\rm ini}$ and a fixed BH mass. Each BH+OB system culminates in one of the following final outcomes introduced in Section 2.1: (1) BH+BH mergers, (2) wide BH+BH binaries, (3) BH+HeC mergers (LFBOTs), (4) BH+MS mergers (LFBOTs?), and (5) traditional CE.

The overview of all the grids with the color-coded outcomes is presented in Figures 15 and 16. As it turns out, BH+OB systems with the same outcomes are grouped together. This allows us to fit approximate parameter ranges in the $P_{\rm ini} - q_{\rm BHOB}$ space of BH+OB systems that correspond to each of the five final outcomes (color shadings in Figures 15 and 16). To construct these maps, we trained a supervised classifier on the grid of outcomes of MESA binary models using a support vector classifier (SVC) implemented in the `sklearn` library. The classifier was trained on the ($P_{\rm ini}$, $q_{\rm BHOB}$) coordinates of systems labeled by their final outcomes, with outcome-dependent class weights adopted to improve performance in sparsely sampled regions of the parameter space. The trained classifier was then evaluated on a dense ($P_{\rm ini}$, $q_{\rm BHOB}$) grid to generate the colored outcome regions shown in Figures 15 and 16, and the resulting boundaries were inspected manually to ensure that they provide a satisfactory approximate representation of the outcome domains given the finite sampling of the underlying MESA grids.

We then apply the outcome classifier to assign each of the BH+OB systems predicted by our rapid population synthesis code a MESA-informed final fate. BH+OB binaries with BHs between 3 and 7.5 $M_\odot$ are mapped onto the 5 $M_\odot$ MESA grids, whereas those with BH masses between 7.5 and 20 $M_\odot$ are mapped onto the 10 $M_\odot$ grids. BH+OBs with BHs heavier than ∼20 $M_\odot$ are discarded (the parameter space for LFBOTs from BH+HeC mergers nearly disappears at these high masses).

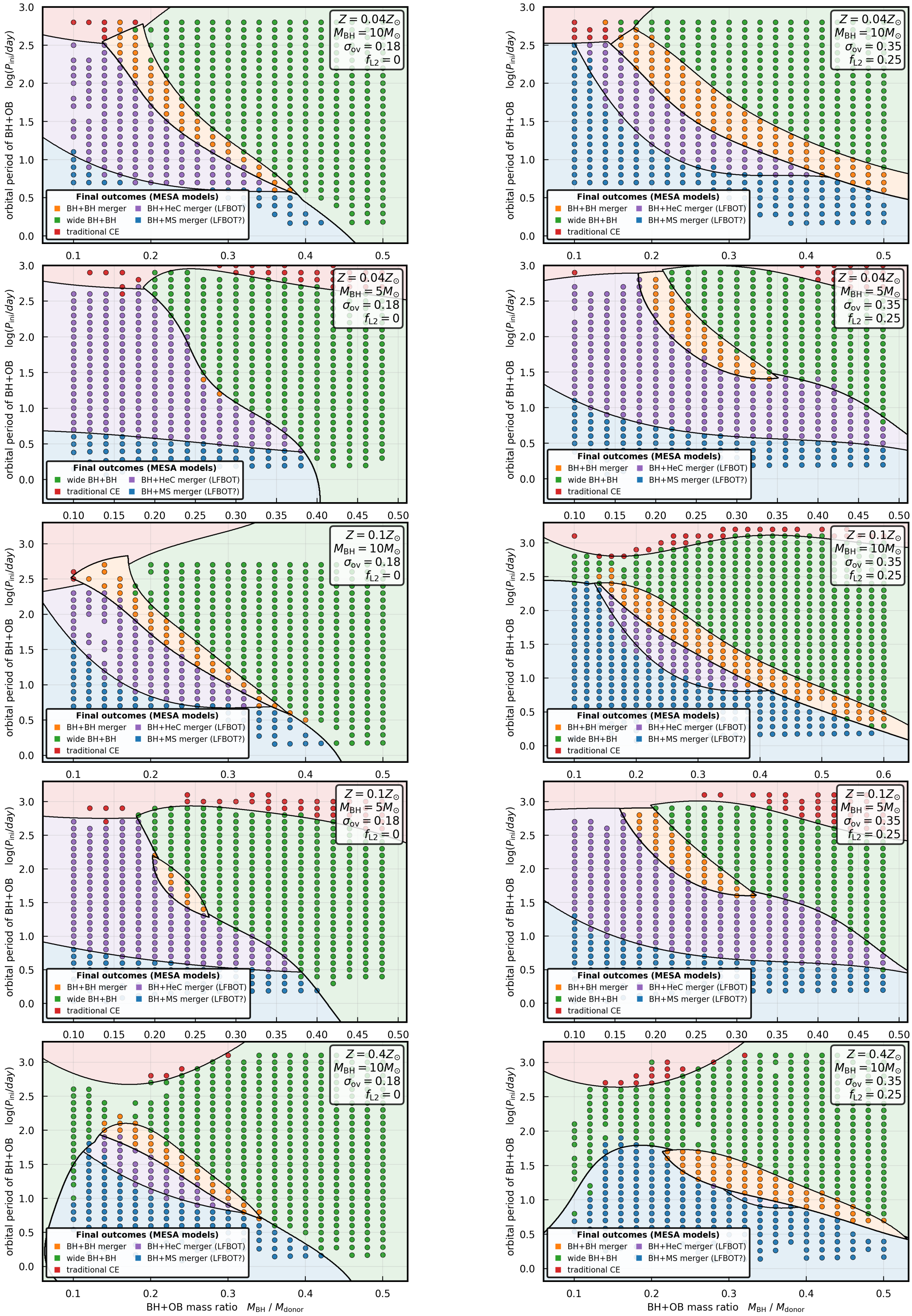

**Figure 15.** Final outcome maps (part 1/2). Each panel shows one grid of BH+star models; see text for details.

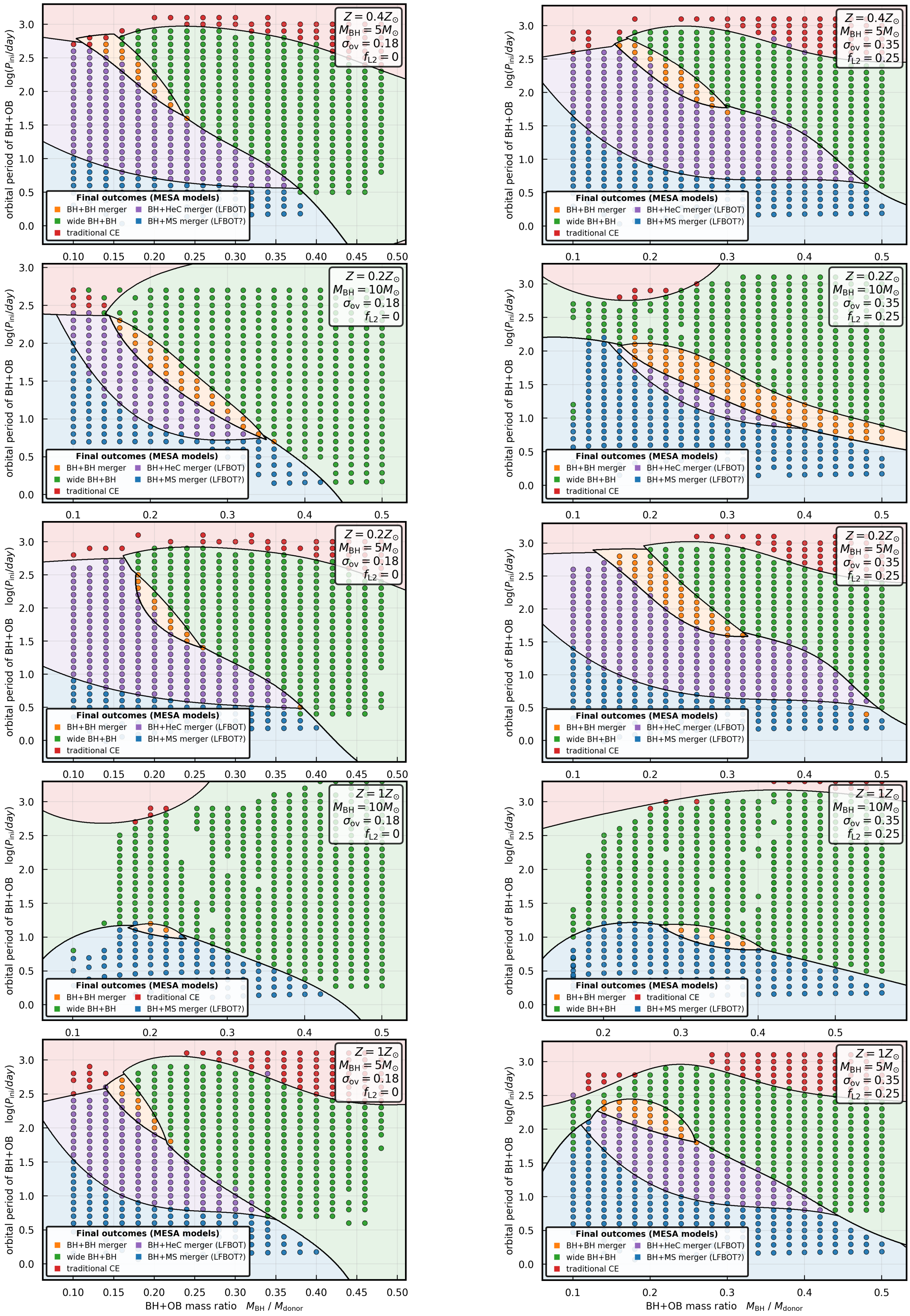

**Figure 16.** Final outcome maps (part 2/2). Each panel shows one grid of BH+star models; see text for details.

### B.4. Event Rate and Cosmic Star Formation History

To convert the predicted number of LFBOT events from our population model into a volumetric event rate, we convolve with a metallicity-specific cosmic star formation history (MS-CSFH). We adopt a suite of 10 MS-CSFH variations from the observation-based framework developed by M. Chruślińska et al. (2019, 2021, 2025), summarized in Figure 17. Because the delay time from ZAMS to an LFBOT is only ∼10 Myr, the formation of the progenitor and the ensuing transient are separated by a negligible fraction of the Hubble time. We therefore evaluate the MS-CSFH in the local volume in which the LFBOTs are observed ($z < 0.3$), where the average SFR density is $\sim 10^{-1.8}\,M_\odot\,\mathrm{yr}^{-1}\,\mathrm{Mpc}^{-3}$ (P. Madau & M. Dickinson 2014). Integrating the metallicity-weighted event fraction over these CSFH models yields the local LFBOT rate. Our fiducial metallicity distribution is consistent with the observed distribution of nearby star-forming galaxies.

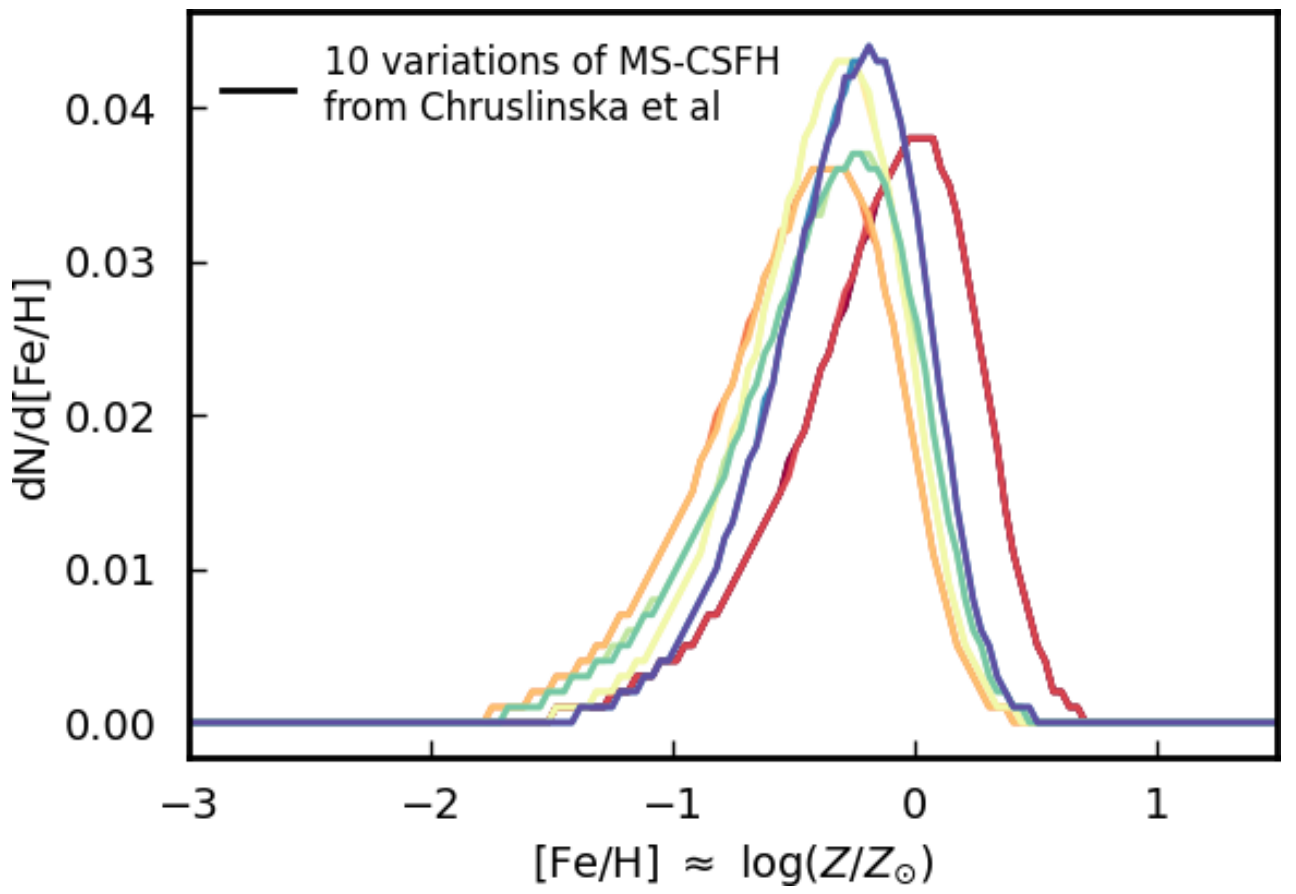

**Figure 17.** Metallicity distribution of star formation in the local Universe ($z < 0.3$), based on the observation-based framework developed by M. Chruślińska et al. (2019, 2021) to derive MS-CSFH. The 10 variations explore uncertainties in the contribution from low-mass galaxies, the absolute abundance scale of gas-phase metallicity, and the oxygen-over-iron sSFR relation (M. Chruślińska et al. 2024).

## ORCID iDs

Jakub Klencki https://orcid.org/0000-0002-7527-5741
Brian D. Metzger https://orcid.org/0000-0002-4670-7509

## References

Abbott, B. P., Abbott, R., Abbott, T. D., et al. 2017, PhRvL, 119, 161101
Arcavi, I., Wolf, W. M., Howell, D. A., et al. 2016, ApJ, 819, 35
Bachetti, M., Harrison, F. A., Walton, D. J., et al. 2014, Natur, 514, 202
Balbus, S. A., & Hawley, J. F. 1998, RvMP, 70, 1
Bavera, S. S., Fragos, T., Zevin, M., et al. 2021, A&A, 647, A153
Bear, E., & Soker, N. 2025, RAA, 25, 025010
Begelman, M. C. 1979, MNRAS, 187, 237
Belczynski, K., Repetto, S., Holz, D. E., et al. 2016, ApJ, 819, 108
Blagorodnova, N., Klencki, J., Pejcha, O., et al. 2021, A&A, 653, A134
Blandford, R. D., & Begelman, M. C. 1999, MNRAS, 303, L1
Blandford, R. D., & Znajek, R. L. 1977, MNRAS, 179, 433
Braun, H., & Langer, N. 1995, A&A, 297, 483
Breivik, K., Coughlin, S., Zevin, M., et al. 2020, ApJ, 898, 71
Bright, J. S., Margutti, R., Matthews, D., et al. 2022, ApJ, 926, 112
Broekgaarden, F. S., Justham, S., de Mink, S. E., et al. 2019, MNRAS, 490, 5228
Brott, I., de Mink, S. E., Cantiello, M., et al. 2011, A&A, 530, A115
Burrows, A., Wang, T., Vartanyan, D., & Coleman, M. S. B. 2024, ApJ, 963, 63
Calderón, D., Pejcha, O., & Duffell, P. C. 2021, MNRAS, 507, 1092
Cannizzo, J. K., Lee, H. M., & Goodman, J. 1990, ApJ, 351, 38
Caprioli, D., & Spitkovsky, A. 2014, ApJ, 783, 91
Carney, J., Andreoni, I., O'Connor, B., et al. 2025, ApJL, 994, L46
Casares, J., Negueruela, I., Ribó, M., et al. 2014, Natur, 505, 378
Chan, C., Müller, B., Heger, A., Pakmor, R., & Springel, V. 2018, ApJL, 852, L19
Chen, C., & Shen, R.-F. 2022, RAA, 22, 035017
Chen, Y., Drout, M. R., Piro, A. L., et al. 2023, ApJ, 955, 43
Chen, Y.-X., Jiang, Y.-F., Goodman, J., & Ostriker, E. C. 2023, ApJ, 948, 120
Chevalier, R. A. 1992, ApJ, 394, 599
Chevalier, R. A. 1993, ApJL, 411, L33
Chevalier, R. A. 1998, ApJ, 499, 810
Chevalier, R. A. 2012, ApJL, 752, L2
Chrimes, A. A., Coppejans, D. L., Jonker, P. G., et al. 2024a, A&A, 691, A329
Chrimes, A. A., Jonker, P. G., Levan, A. J., et al. 2024b, MNRAS, 527, L47
Chrimes, A. A., Jonker, P. G., Levan, A. J., & Mummery, A. 2026, A&A, 706, A327
Chruslinska, M., Belczynski, K., Klencki, J., & Benacquista, M. 2018, MNRAS, 474, 2937
Chruślińska, M., Curti, M., Pakmor, R., et al. 2025, arXiv:2511.15782
Chruślińska, M., Nelemans, G., & Belczynski, K. 2019, MNRAS, 482, 5012
Chruślińska, M., Nelemans, G., Boco, L., & Lapi, A. 2021, MNRAS, 508, 4994
Chruślińska, M., Pakmor, R., Matthee, J., & Matsuno, T. 2024, A&A, 686, A186
Combi, L., Thompson, C., Siegel, D. M., Philippov, A., & Ripperda, B. 2025, ApJ, 987, 71
Coppejans, D. L., Margutti, R., Terreran, G., et al. 2020, ApJL, 895, L23
de Mink, S. E., Langer, N., Izzard, R. G., Sana, H., & de Koter, A. 2013, ApJ, 764, 166
de Mink, S. E., & Mandel, I. 2016, MNRAS, 460, 3545
Dessart, L., Hillier, D. J., Sukhbold, T., Woosley, S. E., & Janka, H. T. 2021, A&A, 656, A61
Detmers, R. G., Langer, N., Podsiadlowski, P., & Izzard, R. G. 2008, A&A, 484, 831
Dexter, J., & Kasen, D. 2013, ApJ, 772, 30
Dominik, M., Belczynski, K., Fryer, C., et al. 2012, ApJ, 759, 52
Drout, M. R., Chornock, R., Soderberg, A. M., Sanders, N. E., et al. 2014, ApJ, 794, 23
Duffell, P. C., & MacFadyen, A. I. 2014, ApJL, 791, L1
Eggleton, P. P. 1983, ApJ, 268, 368
Ercolino, A., Jin, H., Langer, N., & Dessart, L. 2024, A&A, 685, A58
Ercolino, A., Jin, H., Langer, N., & Dessart, L. 2025, A&A, 696, A103
Evans, I. N., Evans, J. D., Martínez-Galarza, J. R., et al. 2024, ApJS, 274, 22
Falk, S. W., & Arnett, W. D. 1977, ApJS, 33, 515
Fox, O. D., & Smith, N. 2019, MNRAS, 488, 3772
Fragos, T., Andrews, J. J., Ramirez-Ruiz, E., et al. 2019, ApJL, 883, L45
Frank, J., King, A., & Raine, D. J. 2002, Accretion Power in Astrophysics (3rd ed.; Cambridge Univ. Press)
Fryer, C. L., Belczynski, K., Wiktorowicz, G., et al. 2012, ApJ, 749, 91
Fryer, C. L., & Woosley, S. E. 1998, ApJL, 502, L9
Fuller, J., & Lu, W. 2022, MNRAS, 511, 3951
Fuller, J., Piro, A. L., & Jermyn, A. S. 2019, MNRAS, 485, 3661
Fulton, M., Chen, T. W., Smartt, S. J., et al. 2024, TNSAN, 206, 1
Gagnier, D., & Pejcha, O. 2023, A&A, 674, A121
Gallegos-Garcia, M., Berry, C. P. L., Marchant, P., & Kalogera, V. 2021, ApJ, 922, 110
Gammie, C. F. 2001, ApJ, 553, 174
Ge, H., Webbink, R. F., Chen, X., & Han, Z. 2015, ApJ, 812, 40
Ge, H., Webbink, R. F., & Han, Z. 2020, ApJS, 249, 9
Gendre, B., Stratta, G., Atteia, J. L., et al. 2013, ApJ, 766, 30
Giacobbo, N., & Mapelli, M. 2018, MNRAS, 480, 2011
Glebbeek, E., Gaburov, E., Portegies Zwart, S., & Pols, O. R. 2013, MNRAS, 434, 3497
Gomez, S., & Grindlay, J. E. 2021, ApJ, 913, 48
Gompertz, B. P., Levan, A. J., Laskar, T., et al. 2026, ApJL, 997, L4
Götberg, Y., de Mink, S. E., Groh, J. H., et al. 2018, A&A, 615, A78
Gottlieb, O., & Metzger, B. D. 2024, ApJL, 974, L9
Gottlieb, O., Tchekhovskoy, A., & Margutti, R. 2022, MNRAS, 513, 3810
Grasberg, E. K., & Nadezhin, D. K. 1976, Ap&SS, 44, 409
Grevesse, N., Noels, A., & Sauval, A. J. 1996, ASPC, 99, 117
Grichener, A. 2023, MNRAS, 523, 221
Grichener, A. 2025, Ap&SS, 370, 11
Groh, J. H., Ekström, S., Georgy, C., et al. 2019, A&A, 627, A24

Gutiérrez, C. P., Mattila, S., Lundqvist, P., et al. 2024, ApJ, 977, 162
Hamann, W. R., Koesterke, L., & Wessolowski, U. 1995, A&A, 299, 151
Hirai, R., & Podsiadlowski, P. 2022, MNRAS, 517, 4544
Hjellming, M. S., & Webbink, R. F. 1987, ApJ, 318, 794
Ho, A. Y. Q., Margalit, B., Bremer, M., et al. 2022a, ApJ, 932, 116
Ho, A. Y. Q., Margalit, B., Bremer, M., et al. 2022b, ApJ, 932, 116
Ho, A. Y. Q., Perley, D. A., Chen, P., Schulze, S., et al. 2023a, Natur, 623, 927
Ho, A. Y. Q., Perley, D. A., Gal-Yam, A., et al. 2023b, ApJ, 949, 120
Ho, A. Y. Q., Perley, D. A., Kulkarni, S. R., Dong, D. Z. J., et al. 2020, ApJ, 895, 49
Ho, A. Y. Q., Phinney, E. S., Ravi, V., et al. 2019, ApJ, 871, 73
Hobbs, G., Lorimer, D. R., Lyne, A. G., & Kramer, M. 2005, MNRAS, 360, 974
Iben, I. J., & Tutukov, A. V. 1984, ApJS, 54, 335
Inkenhaag, A., Jonker, P. G., Levan, A. J., et al. 2023, MNRAS, 525, 4042
Inserra, C. 2019, NatAs, 3, 697
Ivanova, N., Justham, S., Chen, X., et al. 2013, A&ARv, 21, 59
Ivanova, N., & Nandez, J. L. A. 2016, MNRAS, 462, 362
Janka, H.-T., & Kresse, D. 2024, Ap&SS, 369, 80
Janssens, S., Shenar, T., Degenaar, N., et al. 2023, A&A, 677, L9
Jermyn, A. S., Bauer, E. B., Schwab, J., et al. 2023, ApJS, 265, 15
Jiang, J.-a., Yasuda, N., Maeda, K., et al. 2022, ApJL, 933, L36
Justham, S., Podsiadlowski, P., & Vink, J. S. 2014, ApJ, 796, 121
King, A. R., Pringle, J. E., & Livio, M. 2007, MNRAS, 376, 1740
Kitaki, T., Mineshige, S., Ohsuga, K., & Kawashima, T. 2021, PASJ, 73, 450
Klencki, J., Istrate, A., Nelemans, G., & Pols, O. 2022, A&A, 662, A56
Klencki, J., Nelemans, G., Istrate, A., & Chruslinska, M. 2021, A&A, 645, A54
Klencki, J., Nelemans, G., Istrate, A. G., & Pols, O. 2020, A&A, 638, A55
Klencki, J., Podsiadlowski, P., Langer, N., et al. 2026, A&A, 706, A296
Kremer, K., Lu, W., Piro, A. L., et al. 2021, ApJ, 911, 104
Kresse, D., Ertl, T., & Janka, H.-T. 2021, ApJ, 909, 169
Kroupa, P. 2001, MNRAS, 322, 231
Kruckow, M. U., Tauris, T. M., Langer, N., et al. 2016, A&A, 596, A58
Kruckow, M. U., Tauris, T. M., Langer, N., Kramer, M., & Izzard, R. G. 2018, MNRAS, 481, 1908
Kuin, N. P. M., Wu, K., Oates, S., et al. 2019, MNRAS, 487, 2505
Langer, N. 1992, A&A, 265, L17
Langer, N. 2012, ARA&A, 50, 107
Laplace, E., et al. 2021, A&A, 656, A58
Laplace, E., Götberg, Y., de Mink, S. E., Justham, S., & Farmer, R. 2020, A&A, 637, A6
Laplace, E., Schneider, F. R. N., & Podsiadlowski, P. 2025, A&A, 695, A71
Lau, M. Y. M., Hirai, R., Mandel, I., & Tout, C. A. 2024, ApJL, 966, L7
Lau, M. Y. M., Hirai, R., Price, D. J., & Mandel, I. 2022, MNRAS, 516, 4669
Lau, M. Y. M., Hirai, R., Price, D. J., Mandel, I., & Bate, M. R. 2025, A&A, 699, A274
Lazzati, D., Perna, R., Ryu, T., & Breivik, K. 2024, ApJL, 972, L17
LeBaron, N., Margutti, R., Chornock, R., et al. 2026, ApJL, 997, L10
Lechien, T., de Mink, S. E., Valli, R., et al. 2025, ApJL, 990, L51
Leung, S.-C., Blinnikov, S., Nomoto, K., et al. 2020, ApJ, 903, 66
Levan, A. J., Martin-Carrillo, A., Laskar, T., Eyles-Ferris, R. A. J., et al. 2025, ApJL, 990, L28
Levan, A. J., Tanvir, N. R., Starling, R. L. C., et al. 2014, ApJ, 781, 13
Li, J.-Y., Yu, Y.-W., Liu, L.-D., & Xiao, M.-Y. 2025, ApJ, 992, 25
Linial, I., & Quataert, E. 2024, ApJ, 974, 67
Liu, J., Wang, B., Wang, X., et al. 2026, arXiv:2602.20523
Liu, Z., Malyali, A., Krumpe, M., et al. 2023, A&A, 669, A75
Lombardi James, C. J., Warren, J. S., Rasio, F. A., Sills, A., & Warren, A. R. 2002, ApJ, 568, 939
Lu, W., Fuller, J., Quataert, E., & Bonnerot, C. 2023, MNRAS, 519, 1409
Lyman, J. D., Galbany, L., Sánchez, S. F., et al. 2020, MNRAS, 495, 992
MacLeod, M., & Loeb, A. 2020, ApJ, 895, 29
MacLeod, M., Ostriker, E. C., & Stone, J. M. 2018, ApJ, 863, 5
Madau, P., & Dickinson, M. 2014, ARA&A, 52, 415
Maeder, A. 1987, A&A, 178, 159
Mahy, L., Sana, H., Shenar, T., et al. 2022, A&A, 664, A159
Maltsev, K., Schneider, F. R. N., Mandel, I., et al. 2025, A&A, 700, A20
Mandel, I., & Broekgaarden, F. S. 2022, LRR, 25, 1
Mandel, I., & de Mink, S. E. 2016, MNRAS, 458, 2634
Marchant, P., Langer, N., Podsiadlowski, P., et al. 2017, A&A, 604, A55
Marchant, P., Langer, N., Podsiadlowski, P., Tauris, T. M., & Moriya, T. J. 2016, A&A, 588, A50
Marchant, P., Pappas, K. M. W., Gallegos-Garcia, M., et al. 2021, A&A, 650, A107
Margalit, B. 2022, ApJ, 933, 238
Margalit, B., & Metzger, B. D. 2016, MNRAS, 461, 1154
Margalit, B., Quataert, E., & Ho, A. Y. Q. 2022, ApJ, 928, 122
Margutti, R., Metzger, B. D., Chornock, R., et al. 2019, ApJ, 872, 18
Martínez-Sebastián, C., Simón-Díaz, S., Jin, H., et al. 2025, A&A, 693, L10
Matthews, D., Margutti, R., Metzger, B. D., et al. 2023, RNAAS, 7, 126
Maund, J. R., Höflich, P. A., Steele, I. A., et al. 2023, MNRAS, 521, 3323
Metzger, B. D. 2022, ApJL, 937, L12
Metzger, B. D., & Perley, D. A. 2023, ApJ, 944, 74
Metzger, B. D., Piro, A. L., & Quataert, E. 2008, MNRAS, 390, 781
Metzger, B. D., & Stone, N. C. 2016, MNRAS, 461, 948
Migliori, G., Margutti, R., Metzger, B. D., et al. 2024, ApJL, 963, L24
Miller-Jones, J. C. A., Bahramian, A., Orosz, J. A., et al. 2021, Sci, 371, 1046
Miszuda, A., Guo, Z., & Townsend, R. H. D. 2025, A&A, 702, A203
Moe, M., & Di Stefano, R. 2017, ApJS, 230, 15
Motch, C., Pakull, M. W., Soria, R., Grisé, F., & Pietrzyński, G. 2014, Natur, 514, 198
Nakar, E., & Sari, R. 2010, ApJ, 725, 904
Narayan, R., & Yi, I. 1995, ApJ, 452, 710
Nayana, A. J., & Chandra, P. 2021, ApJL, 912, L9
Nayana, A. J., Margutti, R., Wiston, E., et al. 2025, ApJL, 993, L6
Neights, E., Burns, E., Fryer, C. L., et al. 2026, MNRAS, 545, staf2019
Neijssel, C. J., Vinciguerra, S., Vigna-Gómez, A., et al. 2021, ApJ, 908, 118
Nicholl, M., Srivastav, S., Fulton, M. D., Gomez, S., Huber, M. E., et al. 2023, ApJL, 954, L28
Nugent, A. E., Villar, V. A., Metzger, B. D., et al. 2026, arXiv:2603.23597
O'Connor, B., Gill, R., DeLaunay, J., et al. 2025, ApJL, 994, L17
Ofek, E. O., Ozer, L., Konno, R., et al. 2025, ApJ, 993, 76
Olejak, A., Belczynski, K., & Ivanova, N. 2021, A&A, 651, A100
Olejak, A., Klencki, J., Xu, X.-T., et al. 2024, A&A, 689, A305
Ondratschek, P. A., Röpke, F. K., Schneider, F. R. N., et al. 2022, A&A, 660, L8
Orosz, J. A., Steeghs, D., McClintock, J. E., et al. 2009, ApJ, 697, 573
Ouyed, R. 2025, arXiv:2506.20540
Paczynski, B. 1976, IAUS, 73, 75
Park, J., Caprioli, D., & Spitkovsky, A. 2015, PhRvL, 114, 085003
Pasham, D. R., Ho, W. C. G., Alston, W., Remillard, R., et al. 2022, NatAs, 6, 249
Pavlovskii, K., & Ivanova, N. 2015, MNRAS, 449, 4415
Pavlovskii, K., Ivanova, N., Belczynski, K., & Van, K. X. 2017, MNRAS, 465, 2092
Paxton, B., Bildsten, L., Dotter, A., et al. 2011, ApJS, 192, 3
Paxton, B., Cantiello, M., Arras, P., et al. 2013, ApJS, 208, 4
Paxton, B., Marchant, P., Schwab, J., et al. 2015, ApJS, 220, 15
Paxton, B., Smolec, R., Schwab, J., et al. 2019, ApJS, 243, 10
Pejcha, O., Metzger, B. D., & Tomida, K. 2016a, MNRAS, 455, 4351
Pejcha, O., Metzger, B. D., & Tomida, K. 2016b, MNRAS, 461, 2527
Pejcha, O., Metzger, B. D., Tyles, J. G., & Tomida, K. 2017, ApJ, 850, 59
Pellegrino, C., Howell, D. A., Vinkó, J., et al. 2022, ApJ, 926, 125
Perets, H. B., Li, Z., Lombardi, J. C. J., & Milcarek, S. R. J. 2016, ApJ, 823, 113
Perley, D. A., Ho, A. Y. Q., McGrath, Z., et al. 2026, MNRAS, 549, stag678
Perley, D. A., Ho, A. Y. Q., Yao, Y., et al. 2021, MNRAS, 508, 5138
Perley, D. A., Mazzali, P. A., Yan, L., et al. 2019, MNRAS, 484, 1031
Perna, R., Lazzati, D., & Cantiello, M. 2018, ApJ, 859, 48
Phinney, E. S., & Hansen, B. M. S. 1993, ASPC, 36, 371
Piro, A. L. 2015, ApJL, 808, L51
Piro, A. L., & Lu, W. 2020, ApJ, 894, 2
Podsiadlowski, P., Joss, P. C., & Hsu, J. J. L. 1992, ApJ, 391, 246
Podsiadlowski, P., Rappaport, S., & Pfahl, E. D. 2002, ApJ, 565, 1107
Pols, O. R. 1994, A&A, 290, 119
Popov, S., Müller, B., & Mandel, I. 2025, NewAR, 101, 101734
Prentice, S. J., Maguire, K., Smartt, S. J., et al. 2018, ApJL, 865, L3
Prieto, J. L., Kistler, M. D., Thompson, T. A., et al. 2008, ApJL, 681, L9
Pursiainen, M., Childress, M., Smith, M., et al. 2018, MNRAS, 481, 894
Pursiainen, M., Killestein, T. L., Kuncarayakti, H., et al. 2025, MNRAS, 537, 3298
Quataert, E., & Kasen, D. 2012, MNRAS, 419, L1
Ramachandran, V., Oskinova, L. M., Hamann, W. R., et al. 2022, A&A, 667, A77
Renzo, M., & Götberg, Y. 2021, ApJ, 923, 277
Repetto, S., Davies, M. B., & Sigurdsson, S. 2012, MNRAS, 425, 2799
Repetto, S., & Nelemans, G. 2015, MNRAS, 453, 3341
Rivera Sandoval, L. E., Maccarone, T. J., Corsi, A., et al. 2018, MNRAS, 480, L146
Röpke, F. K., & De Marco, O. 2023, LRCA, 9, 2
Rozwadowska, K., Vissani, F., & Cappellaro, E. 2021, NewA, 83, 101498

Sadowski, A., & Narayan, R. 2016, MNRAS, 456, 3929
Sana, H., de Mink, S. E., de Koter, A., et al. 2012, Sci, 337, 444
Saxton, R. D., Read, A. M., Esquej, P., et al. 2008, A&A, 480, 611
Scherbak, P., Lu, W., & Fuller, J. 2025, ApJ, 990, 172
Schneider, F. R. N., Izzard, R. G., Langer, N., & de Mink, S. E. 2015, ApJ, 805, 20
Schneider, F. R. N., Podsiadlowski, P., & Laplace, E. 2024, A&A, 686, A45
Schneider, F. R. N., Podsiadlowski, P., & Müller, B. 2021, A&A, 645, A5
Schrøder, S. L., MacLeod, M., Loeb, A., Vigna-Gómez, A., & Mandel, I. 2020, ApJ, 892, 13
Schürmann, C., & Langer, N. 2024, A&A, 691, A174
Schürmann, C., Langer, N., Kramer, J. A., et al. 2024, A&A, 690, A282
Schürmann, C., Xu, X. T., Langer, N., et al. 2025, A&A, 704, A219
Sen, K., Renzo, M., Jin, H., et al. 2026, ApJ, 1000, 2
Sevilla, C., Ho, A. Y. Q., Nayana, A. J., Schulze, S., et al. 2026, arXiv:2601.18926
Shakura, N. I., & Sunyaev, R. A. 1973, A&A, 24, 337
Shenar, T., Sana, H., Mahy, L., et al. 2022, NatAs, 6, 1085
Soker, N., Grichener, A., & Gilkis, A. 2019, MNRAS, 484, 4972
Somalwar, J. J., Ravi, V., Margutti, R., et al. 2025, ApJ, 995, 228
Song, H. F., Meynet, G., Maeder, A., Ekström, S., & Eggenberger, P. 2016, A&A, 585, A120
Spitkovsky, A. 2006, ApJL, 648, L51
Sravan, N., Marchant, P., & Kalogera, V. 2019, ApJ, 885, 130
Sridhar, N., & Metzger, B. D. 2022, ApJ, 937, 5
Strubbe, L. E., & Quataert, E. 2009, MNRAS, 400, 2070
Sun, N.-C., Maund, J. R., Crowther, P. A., & Liu, L.-D. 2022, MNRAS, 512, L66
Sun, N.-C., Maund, J. R., Shao, Y., & Janiak, I. A. 2023, MNRAS, 519, 3785
Taam, R. E., & Spruit, H. C. 2001, ApJ, 561, 329
Tanikawa, A., Yoshida, T., Kinugawa, T., Takahashi, K., & Umeda, H. 2020, MNRAS, 495, 4170
Tauris, T. M., & van den Heuvel, E. P. J. 2023, Physics of Binary Star Evolution. From Stars to X-ray Binaries and Gravitational Wave Sources (Princeton Univ. Press)
Tchekhovskoy, A., Metzger, B. D., Giannios, D., & Kelley, L. Z. 2014, MNRAS, 437, 2744
Tchekhovskoy, A., Narayan, R., & McKinney, J. C. 2011, MNRAS, 418, L79
Teboul, O., & Metzger, B. D. 2023, ApJL, 957, L9
Temmink, K. D., Pols, O. R., Justham, S., & Blagorodnova, N. 2025, A&A, 703, A121
Temmink, K. D., Pols, O. R., Justham, S., Istrate, A. G., & Toonen, S. 2023, A&A, 669, A45
The LIGO Scientific Collaborationthe Virgo Collaborationthe KAGRA Collaboration, et al. 2025, arXiv:2508.18082
Thöne, C. C., de Ugarte Postigo, A., Fryer, C. L., et al. 2011, Natur, 480, 72
Thorne, K. S., & Zytkow, A. N. 1977, ApJ, 212, 832
Tsuna, D., & Lu, W. 2025, ApJ, 986, 84
Tuna, S., & Metzger, B. D. 2023, ApJ, 955, 125
Tuna, S., Metzger, B. D., Jiang, Y.-F., & White, C. 2025, ApJ, 989, 27
Uno, K., & Maeda, K. 2020, ApJL, 905, L5
van Dalen, J. N. D., Levan, A. J., Jonker, P. G., et al. 2025, ApJL, 982, L47
van den Heuvel, E. P. J. 1976, IAUS, 73, 35
van den Heuvel, E. P. J. 2017, JApA, 38, 45
van Son, L. A. C., de Mink, S. E., Renzo, M., et al. 2022, ApJ, 940, 184
Vetter, M., Röpke, F. K., Schneider, F. R. N., et al. 2024, A&A, 691, A244
Vetter, M., Röpke, F. K., Schneider, F. R. N., et al. 2025, A&A, 698, A133
Vigna-Gómez, A., Neijssel, C. J., Stevenson, S., et al. 2018, MNRAS, 481, 4009
Vigna-Gómez, A., Willcox, R., Tamborra, I., et al. 2024, PhRvL, 132, 191403
Vinkó, J., Yuan, F., Quimby, R. M., et al. 2015, ApJ, 798, 12
Wagg, T., Johnston, C., Bellinger, E. P., et al. 2024, A&A, 687, A222
Walton, D. J., Mackenzie, A. D. A., Gully, H., et al. 2022, MNRAS, 509, 1587
Wang, C., Langer, N., Schootemeijer, A., et al. 2020, ApJL, 888, L12
Wang, Z. Y., Pastorello, A., Cai, Y. Z., et al. 2025, A&A, 700, A156
Waxman, E., & Draine, B. T. 2000, ApJ, 537, 796
Wei, D., Schneider, F. R. N., Podsiadlowski, P., et al. 2024, A&A, 688, A87
Winter-Granic, M., & Quataert, E. 2025, arXiv:2512.09017
Woosley, S. E., Sukhbold, T., & Janka, H. T. 2020, ApJ, 896, 56
Wu, S. C., & Fuller, J. 2022, ApJ, 930, 119
Xiang, D., Wang, X., Lin, W., et al. 2021, ApJ, 910, 42
Xin, C., Haiman, Z., Perna, R., Wang, Y., & Ryu, T. 2024, ApJ, 961, 149
Xu, X.-T., Langer, N., Klencki, J., Wang, C., & Li, X.-D. 2025, arXiv:2512.20054
Yao, Y., Ho, A. Y. Q., Medvedev, P. J. N. A., et al. 2022a, ApJ, 934, 104
Yao, Y., Lu, W., Guolo, M., et al. 2022b, ApJ, 937, 8
Yoon, S. C., & Langer, N. 2005, A&A, 443, 643
Yoon, S. C., Langer, N., & Norman, C. 2006, A&A, 460, 199
Yuan, F., & Narayan, R. 2014, ARA&A, 52, 529
Zapartas, E., de Mink, S. E., Justham, S., et al. 2019, A&A, 631, A5
Zhang, W., Shu, X., Chen, J.-H., et al. 2022, RAA, 22, 125016